\pdfoutput=1
\RequirePackage{ifpdf}
\ifpdf 
\documentclass[pdftex]{sigma}
\else
\documentclass{sigma}
\fi

\usepackage{subfigure}

\numberwithin{equation}{section}

\begin{document}


\renewcommand{\thefootnote}{$\star$}

\newcommand{\arXivNumber}{1603.00172}

\renewcommand{\PaperNumber}{068}

\FirstPageHeading

\ShortArticleName{Exact Renormalisation Group Equations and Loop Equations for Tensor Models}

\ArticleName{Exact Renormalisation Group Equations\\ and Loop Equations for Tensor Models\footnote{This paper is a~contribution to the Special Issue on Tensor Models, Formalism and Applications. The full collection is available at \href{http://www.emis.de/journals/SIGMA/Tensor_Models.html}{http://www.emis.de/journals/SIGMA/Tensor\_{}Models.html}}}

\Author{Thomas KRAJEWSKI~$^\dag$ and Reiko TORIUMI~$^\ddag$}

\AuthorNameForHeading{T.~Krajewski and R.~Toriumi}

\Address{$^\dag$~Aix Marseille Universit\'e, Universit\'e de Toulon, CNRS, CPT,\\
\hphantom{$^\dag$}~UMR 7332, 13288 Marseille, France}
\EmailD{\href{mailto:thomas.krajewski@cpt.univ-mrs.fr}{thomas.krajewski@cpt.univ-mrs.fr}}

\Address{$^\ddag$~Department of Physics, University of California Berkeley, USA}
\EmailD{\href{mailto:torirei@gmail.com}{torirei@gmail.com}}

\ArticleDates{Received February 29, 2016, in f\/inal form July 05, 2016; Published online July 14, 2016}

\Abstract{In this paper, we review some general formulations of exact renormalisation group equations and loop equations for tensor models and tensorial group f\/ield theories. We illustrate the use of these equations in the derivation of the leading order expectation values of observables in tensor models. Furthermore, we use the exact renormalisation group equations to establish a suitable scaling dimension for interactions in Abelian tensorial group f\/ield theories with a closure constraint. We also present analogues of the loop equations for tensor models.}

\Keywords{tensor models; group f\/ield theory; large $N$ limit; exact renormalisation equation}

\Classification{81T17; 81T18}

{\small \tableofcontents}

\renewcommand{\thefootnote}{\arabic{footnote}}
\setcounter{footnote}{0}

\section{Introduction}

Tensor models are higher-dimensional generalisations of matrix models that have been introduced more than twenty f\/ive years ago \cite{Ambjorn:1990ge}. The aim was to reproduce in dimension $D>2$ the successes of matrix models in providing a theory of random geometries. This is of high interest in multiple f\/ields of physics, especially in discrete approaches to quantising gravity. The basic idea is that a suitable integral over a rank~$D$ tensor~$T$ is the partition function of a theory of random triangulations in dimension~$D$,
\begin{gather}
\int_{\text{rank $D$}\atop\text{ tensors}} d\mu_{C}(T)\exp\{-{V}(T)\}
\quad=\quad\sum_{\text{graph ${\cal G}$ $\Leftrightarrow$}
\text{ dimension $D$ triangulations}}{\cal A}({\cal G}).\label{tensor:eq}
\end{gather}
In this Feynman graph expansion, $d\mu_{C}$ is a Gau{\ss}ian measure on tensors, $V$ an interaction potential and ${\cal A}$ the resulting weight associated to the graph ${\cal G}$. By a certain duality procedure, these graphs are in one to one correspondence with triangulations of dimension~$D$. Thus random tensors provide a natural way to construct models of random geometries in dimension~$D$.

However, the combinatorics of tensor models turned out to be much more intricate than the matrix model one. Therefore, progress in this f\/ield remained very slow until a couple of years ago, when a new impetus arose thanks to the introduction of coloured tensor models by Gurau~\cite{Gurau:2009tw}. This allowed him to a derive for the f\/irst time a $1/N$ expansion~\cite{Gurau:2011xq}, where $N$ is the size of the tensor and prove Gau{\ss}ian universality at large~$N$~\cite{Gurau:2011kk}. This breakthrough was followed by many other works in the f\/ield, ref\/ining the $1/N$ expansion or adding new degrees of freedom that can be interpreted as matter living on the triangulations. Finally, a double scaling limit has been obtained: when some parameters in~$V$ are taken to their critical values together with~$N\rightarrow\infty$, the series over triangulations reproduces a~continuum limit, see~\cite{Bonzom:2014oua} and~\cite{Dartois:2013sra}. We refer to~\cite{Rivasseau:2013uca} for a recent overview of random tensors.

A second important breakthrough occurred when Ben Geloun and Rivasseau constructed the f\/irst renormalisable group f\/ield theory~\cite{BenGeloun:2011rc}. Group f\/ield theories~\cite{Oriti:2014uga} are specif\/ic quantum f\/ield theories def\/ined, in dimension~$D$, over~$D$ copies of a group often taken to be $\text{SU}(2)$ or several copies of $\text{U}(1)$. As far as the combinatorics is concerned, the group f\/ield $\Phi(g_{1},\dots,g_{D})$ is analogous to the tensor $T_{i_{1},\dots,i_{D}}$. Thus, a functional integral over ${\Phi}$ generates a sum over triangulations weighted by some group integrals. When the potential is suitably chosen, these weights reproduce the spin foam amplitudes in loop quantum gravity. We refer the reader to the monographs by Rovelli \cite{Rovelli:2004tv} and Thiemann \cite{Thiemann:2007zz} for an account of loop quantum gravity. Then, group f\/ield theory provides a prescription on how to sum these amplitudes. This is crucial in order to investigate some properties of the theory, like triangulation independence or summation over topologies. Moreover, if available, a double scaling limit could be interpreted a continuum limit, including a sum over topologies. This interplay between the various amplitudes also plays a key role in the renormalisation of group f\/ield theories: the renormalised theory is constructed recursively using the combinatorics of subgraphs, thus the amplitude of smaller triangulations. Let us note that the model renormalised in~\cite{BenGeloun:2011rc} is based on a abelian group of the type $\text{U}(1)^{d}$. This has been followed by the renormalisation of other abelian models. A further decisive step was made by Carrozza, Oriti and Rivasseau in~\cite{Carrozza:2013wda}, when a rank three model based on $\text{SU}(2)$ with closure constraint has been renormalised. Such a model would open the possibility, if def\/ined nonperturbatively, to construct a quantum theory of gravity in $D=3$ that includes topology changes. Finally, let us mention that the study of $D=4$ models with a Barrett--Crane or Engle--Peirera--Rovelli--Livine/Freidel--Krasnov vertex is currently under scrutiny by various researchers. These models include the simplicity constraint as well as the closure constraint, the former being essential in making contacts with general relativity in $D=4$.

\looseness=1
In this paper, we aim at revisiting these two breakthroughs in light of the exact renorma\-li\-sation group equation, in Polchinski's formulation~\cite{Polchinski:1983gv}. The latter is a functional dif\/ferential equation that allows to implement Wilson's ideas in a general setting in quantum f\/ield theory. It translates into a f\/low equation for all the couplings of the theory, hence the name ``exact'', that prescribes how the couplings should be modif\/ied in order to take into account the ef\/fects of the high energy modes that have been integrated. As a consequence, the evolution equation allows to classify the various couplings according to their degree of relevancy. This is what is done here: In the tensor model case, we identify the dominant couplings in the large $N$ limit while for group f\/ield theories we identify the relevant couplings that govern the high spin limit.

Moreover, it allows to trade the functional integral formalism for a functional dif\/ferential equation, from which many physical consequences can be drawn without having to resort to Feynman graphs evaluation. This is precisely the point of view adopted here: while the fundamental results presented above have been obtained by expanding \eqref{tensor:eq} in Feynman graphs, these results can also be derived within the exact renormalisation group equation. The latter is an equation for the ef\/fective vertices that collects a sum over Feynman graphs. From a~geo\-metrical perspective, the ef\/fective vertices correspond to boundary triangulations and collect all triangulations corresponding to a f\/ixed boundary.

However, our approach remains perturbative in nature since no attempt has been made to def\/ine the quantities we manipulate as analytic functions of the parameters of the potential. Recently, various nonperturbative results have been obtained~\cite{Gurau:2013pca} by summing up the perturbative series. Such an approach is called ``constructive f\/ield theory'' and we refer the reader to book~\cite{Rivasseau:1991ub} and the recent review~\cite{Gurau:2014vwa} for a detailed account of the f\/ield theory case. Alternatively, the exact renormalisation group equation in Wetterich's formulation can be solved using a truncation as will be brief\/ly reviewed in Section~\ref{truncations:sec}. Wetterich's equation is equivalent at the perturbative level to Polchinski's one but more suitable for truncations. In this respect, let a mention the review by Carrozza~\cite{Carrozza:2016vsq} in this special issue, dealing with exact renormalisation group equation, with an emphasis on the Wetterich equation.

As a functional dif\/ferential equation to be satisf\/ied by the ef\/fective theory, exact renormalisation group equations bear some common features with the Schwinger--Dyson equations. More in general, the invariance of the integral~\eqref{tensor:eq} under change of variables translate into a set of constraints on the latter that can be expressed as dif\/ferential operators with respect to the parameters appearing in the potential. In the matrix model case, these equations are known as the ``loop equations'' and play a prominent role, especially in deriving the double scaling limit and in establishing some connection to integrable hierarchies. For tensor models, analogues of these equations have been derived in~\cite{Gurau:2012ix} and applied to the large~$N$ limit in~\cite{Bonzom:2012cu}. The potential of these equations for tensor models has not yet been completely explored and it is possible that they lead to signif\/icant results, as in the matrix model case.

It is important to emphasize the rationale of our study: Polchinski's equation and Schwinger--Dyson equations are exact relations that are satisf\/ied by the ef\/fective action and the generating functions of invariant observables. In our context, we work at the perturbative level only, considering these quantities as formal series in the couplings. Then, the aforementioned equations lead to nontrivial relations between these quantities, collecting a sum over Feynman graphs, or equivalently, a~sum over space-time triangulations.

Our paper is organised as follows. In Section~\ref{section2}, we present the general framework of tensor models and tensorial group f\/ield theories. Section~\ref{section3} is devoted to the exact renormalisation group equation. First, we introduce Wilson's ideas and formulate Polchinski's equation in the context of quantum f\/ield theory. Then, we apply this formalism to tensor models. As a~f\/irst application, we derive the scaling properties of the ef\/fective couplings which lead to an alternative proof of Gurau's universality~\cite{Gurau:2011kk}. These equations are also formulated for group f\/ield theories in a~general setting. Then, we apply them to propose a~scaling dimension for all the couplings in the case of abelian group f\/ield theories with closure constraints. As a byproduct, we recover the list of f\/ive renormalisable theories f\/irst identif\/ied in~\cite{Carrozza:2013wda}. We close this section by a~formulation of the f\/low equation for spin networks and a~brief discussion of the nonperturbative results based on Wetterich's equation. Sections~\ref{section4} deals with loop equations. We start by expressing reparametrisation invariance in quantum f\/ield theory and then formulate the analogue of the loop equations for tensor models and stress the analogies between tensors and matrices. Finally, we illustrate the use of the loop equations in deriving the values of the leading order expectation values, following~\cite{Bonzom:2012cu}.

\section{Tensor models}\label{section2}

 Matrix models turn out to be ubiquitous in mathematics and physics. For instance, they appear in combinatorics of maps and statistical physics on random triangulated surfaces. Tensor models aim at providing a higher-dimensional generalisation of matrix models, with a particular view towards generation of random triangulations of spaces of dimension $D>2$. Thus, the basic variables we consider are rank $D$ complex tensors $T_{i_{1},\dots,i_{D}}$ and its complex conjugate $\overline{T}_{i_{1},\dots,i_{D}}$, with indices $i_{k}\in \{1,\dots,N \}$, the integer $N$ being the size of the tensor. Let us stress that the models considered here involve a tensor and its complex conjugate, treated as independent variables. Furthermore, these tensors are not required to fulf\/il any symmetry relations upon permutations of their indices. For instance, in the $D=2$ case, they correspond to nonhermitian matrices.

There is a natural transformation of a rank $D$ complex tensor under the direct product of $D$ copies of the group $\text{U}(N)$
\begin{gather}
T_{i_{1},\dots,i_{D}}\rightarrow \!\sum_{j_{1},\dots,j_{D}}\!U^{1}_{i_{1}j_{1}}\cdots U^{D}_{i_{D}j_{D}}T_{j_{1},\dots,j_{D}},
\qquad
\overline{T}_{i_{1},\dots,i_{D}}\rightarrow \!\sum_{j_{1},\dots,j_{D}}\!\overline{U}^{1}_{i_{1}j_{1}}\cdots \overline{U}^{D}_{i_{D}j_{D}}\overline{T}_{j_{1},\dots,j_{D}},\!\!\!\!\label{unitary:eq}
\end{gather}
with $(U^{1},\dots,U^{D})\in \text{U}(N)\times\cdots\times \text{U}(N)$. This large transformation group is available because we do not impose any symmetry under permutations of indices, otherwise two indices that can be permuted must transform in the same way under the unitary group, so that the symmetry is reduced from $\text{U}(N)\times \text{U}(N)$ down to~$\text{U}(N)$. Finally, let us also note that it is possible to let the various indices vary in dif\/ferent ranges, $i_{k}\in \{1,\dots,N_{k} \}$. Then, unitary transformations correspond to the group $\text{U}(N_{1})\times\cdots\times \text{U}(N_{k})$. Although useful from a combinatorial point of view, we do not consider this generalisation here and stick to the case $N_{1}=\cdots =N_{k}=N$.

 \subsection{Invariant potentials for random tensors}

 In analogy with random matrices, we construct a theory of random tensors, we have to def\/ine a certain probability measure on the space of tensors. The latter is of the form
 \begin{gather*}
\frac{d\mu_{C}(T,\overline{T})\exp\{-V(T,\overline{T})\}}{Z},
 \end{gather*}
where $d\mu_{C}(C,\overline{T})$ is a Gau{\ss}ian measure of covariance~$C$, $V(T,\overline{T})$ the interaction potential and~$Z$ the partition function def\/ined as
 \begin{gather*}
 Z=\int d\mu_{C}(T,\overline{T})\exp\{-V(T,\overline{T})\}.
 \end{gather*}
 Then, the expectation value of an observable, taken to be any function ${\cal O}(T,\overline{T})$ of the components of the tensors, is def\/ined as
\begin{gather*}
\langle{\cal O}\rangle=
\frac{\int d\mu_{C}(T,\overline{T}){\cal O}(T,\overline{T})\exp\{-V(T,\overline{T})\}}{Z}.
 \end{gather*}
 Unless otherwise stated, we restrict our attention to invariant random tensors. This means that $d\mu_{C}(T,\overline{T})$, ${\cal O}(T,\overline{T})$ and $V(T,\overline{T})$ have to be invariant under the action of $\text{U}(N)\times\cdots\times \text{U}(N)$, see~\eqref{unitary:eq}.

\looseness=-1 Polynomial invariants of the components of the tensors are constructed using particular graphs which are called $D$-bubbles. A $D$-bubble is a~bipartite $D$-coloured graph. This means that:
 \begin{itemize}\itemsep=0pt
 \item There are two types of vertices, black ones $\bullet$ and white ones $\circ$.
 \item An edge can connect only a black vertex to a white vertex.
 \item At any vertex there are exactly $D$ incident edges.
 \item Each edge is decorated by a colour in $\{1,\dots,D\}$ in such a way that the colours of the $D$ edges incident to any vertex are all dif\/ferent.
 \end{itemize}

\begin{figure}[t]\centering
\begin{subfigure}[Dipole graph (Gau{\ss}ian measure).]{
\parbox{5cm}{\includegraphics[width=4cm]{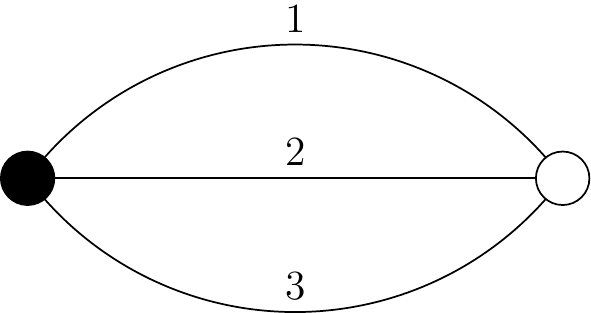}}
\hspace{2cm}
$\sum_{i_{a}}\overline{T}_{i_{1},i_{2},i_{3}}
T _{i_{1}i_{2}i_{3}}
$}
\end{subfigure}

\begin{subfigure}[Degree 6 interaction.]{
\parbox{4cm}{\includegraphics[width=4cm]{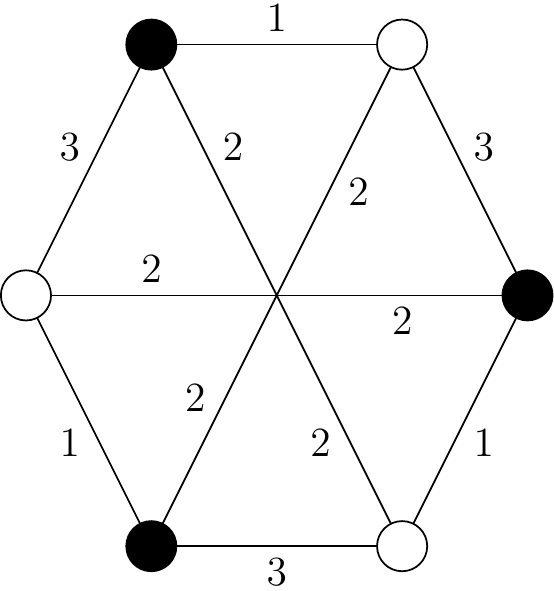}}
\hspace{1cm}
$
\sum_{i_{a},\,j_{b},\,k_{c}}
\overline{T}_{i_{1}i_{2}i_{3}}\overline{T}_{j_{1}j_{2}j_{3}}\overline{T}_{k_{1}k_{2}k_{3}}
T_{i_{1}k_{2}j_{3}}T_{j_{1}i_{2}k_{3}}T_{k_{1}j_{2}i_{3}}
$}
\end{subfigure}
\caption{Some bubble invariants.}\label{invariantexample:fig}
\end{figure}

 The invariant associated to a $D$-bubble is def\/ined by assigning a tensor $T$ to a white vertex, a~tensor $\overline{T}$ to a black vertex. We identify the indices $i_{k}$ in $T$ and $\overline{i}_{k}$ in $\overline{T}$ whenever they are connected by a~line of colour $k$ and summing over all tensor indices. In analogy with the trace invariants of matrix models, such an invariant is written as
 \begin{gather*}
 \operatorname{Tr}_{\cal B}(T,\overline{T})=
 \sum_{i_{e}\atop\text{edge indices}}
 \prod_{\overline{v}\atop \text{black vertices }}\overline{T}_{\overline{i}_{\overline{v},1},\dots,\overline{i}_{\overline{v},D}}
 \prod_{{v}\atop \text{white vertices }}{T}_{{i}_{{v},1},\dots,{i}_{{v},D}}
 \prod_{e\atop\text{edges}}
 \delta_{i_{e},\overline{i}_{\overline{v}(e),c(e)}}
 \delta_{i_{e},i_{v(e),c(e)}},
 \end{gather*}
 where $e$ is an edge between a white vertex $v(e)$ and a black vertex $\overline{v}(e)$ and $c(e)$ its colour. For example, the invariants associated to two 3-bubbles are given in Fig.~\ref{invariantexample:fig}.

Let us note that we do not restrict ourselves to connected bubbles, though the invariant of a disconnected bubble factorizes over its connected components. In the particular case $D=2$, connected bubble invariants are def\/ined by a single even integer, the number~$n$ of edges, and are just traces of power of the matrix~$MM^{\dagger}$, $\operatorname{Tr}_{{\cal B}}(M,\overline{M})=\operatorname{Tr}(MM^{\dagger})^{n/2}$. In this case, non connected bubbles correspond to product of traces, also known as multitrace operators.

A polynomial invariant potential for a rank $D$ tensor is def\/ined by summing over a f\/inite set of $D$-bubbles, each weighted by a complex number~$\lambda_{\cal B}$, called its coupling constant,
\begin{gather}
V(T,\overline{T})=\sum_{{\cal B}}\frac{\lambda_{\cal B}}{C_{\cal B}} \operatorname{Tr}_{\cal B}(\overline{T},T).
\label{invariantpotential:eq}
\end{gather}
$C_{\cal B}$ is a combinatorial factor introduced for later convenience. It is the cardinal of the group of transformation of edges and vertices that preserve the bubble. In this paper, we mostly stay at a formal level, i.e., we expand the functional integral~\eqref{tensor:eq} in power series of the couplings using Gau{\ss}ian integration. In this case, the couplings can be any complex numbers but if we require the partition function and the expectation values to be real, then we have to take the couplings to be real. At a~nonperturbative level, the main dif\/f\/iculty lies in the fact that the trace invariants $\operatorname{Tr}_{\cal B}(\overline{T},T)$ are not necessarily positive, except for a few special cases, like the quartic models studied in~\cite{Gurau:2013pca}. When these invariant are positive, positive couplings lead to a~well def\/ined integral. In the general case, it may be useful to consider purely imaginary couplings in such a way that the exponential of the interaction term remain bounded.

There is a single quadratic trace invariant up to a factor. It is associated to the dipole, a~bubble made of two vertices connected by $D$ edges, see Fig.~\ref{invariantexample:fig}. Its trace invariant reads
\begin{gather*}
\operatorname{Tr}_{\text{dipole}}(T,\overline{T})
=\sum_{i_{1}\cdots i_{D}\atop \overline{i}_{1},\dots,\overline{i}_{D}}
\overline{T}_{\overline{i}_{1},\dots,\overline{i}_{D}}T_{i_{1}\cdots i_{D}}=\overline{T}\cdot T.
\end{gather*}
It is used to construct the normalised Gau{\ss}ian measure
\begin{gather*}
d\mu_{C}(T,\overline{T})={\cal N}\exp\big\{{-}\big(t^{-1}\overline{T}\cdot T\big)\big\} dTd\overline{T},
\end{gather*}
 where ${\cal N}$ is a normalisation factor and $dTd\overline{T}$ the standard Lebesgue measure on the space of components of the tensors $T$ and $\overline{T}$. Then, the covariance of the Gau{\ss}ian measure is simply $C_{i_{1},\dots,i_{D}|\overline{i}_{1},\dots,\overline{i}_{D}}=t \delta_{i_{1}\overline{i}_{1}}\cdots \delta_{i_{D}\overline{i}_{D}}$, which means that
\begin{gather}
\int d\mu_{C}(T,\overline{T})
T_{i_{1}\dots i_{D}}
\overline{T}_{\overline{i}_{1}\dots \overline{i}_{D}}
=t\,\delta_{i_{1}\overline{i}_{1}}\cdots \delta_{i_{D}\overline{i}_{D}}\label{covariance:eq}
\end{gather}
for positive $t$.

By the same token, invariant observables are also labelled by $D$-bubbles,
\begin{gather*}
{\cal O}(T,\overline{T})=\sum_{{\cal B}}\frac{{\cal O}_{\cal B}}{C_{\cal B}} \operatorname{Tr}_{\cal B}(\overline{T},T).
\end{gather*}
Thus, we reduce the computation of the expectation value of an observable to that of an invariant bubble. Moreover, bubble interactions may be used to generate expectations of bubble observables,
\begin{align}
\langle\operatorname{Tr}_{\cal B}(\overline{T},T)\rangle & =
\frac{1}{Z}
\frac{\partial }{\partial J_{B}}
\int d\mu_{C}(T,\overline{T})\exp\left\{-V(T,\overline{T})+
\sum_{{\cal B}'}{{J}_{\cal B'}} \operatorname{Tr}_{{\cal B}'}(\overline{T},T)
\right\}\Bigg|_{J_{{\cal B}'}=0}\nonumber \\
& =\frac{\partial }{\partial J_{B}}\log Z
\big|_{J_{{\cal B}'}=0}.
\label{bubbleexpectation:eq}
\end{align}
Alternatively, any bubble observable can be expanded in powers of the components of the tensors. Thus, the basic object to compute is the expectation value of a product of components
\begin{gather}
\big\langle T_{i_{1,1}\dots i_{1,D}}\cdots
T_{i_{n,1}\dots i_{n,D}} \quad
\overline{T}_{\overline{i}_{1,1}\dots \overline{i}_{1,D}}\cdots
\overline{T}_{\overline{i}_{n,1}\dots \overline{i}_{n,D}}
\big\rangle.\label{correlation:eq}
\end{gather}
In analogy with quantum f\/ield theory, these expectation values are referred to as correlation functions. Because of $\text{U}(1)$ invariance, correlation functions vanish if there is not an equal number of $T$ and $\overline{T}$.

Although we restrict ourselves to invariant tensor models, in the more general case of noninvariant tensors, the sum over bubbles invariants in~\eqref{invariantpotential:eq} may be replaced by
\begin{gather}
V(T,\overline{T})= \sum_{{\cal B}}\prod_{\text{edges}}
\sum_{i_{e},\, \overline{i}_{e}\atop \text{edge indices}}
\frac{\lambda_{\cal B}\left\{i_{e},\overline{i}_{e}\right\}}{C_{\cal B}}\nonumber\\
\hphantom{V(T,\overline{T})=}{}\times
\prod_{\overline{v}\atop \text{black vertices }}\overline{T}_{\overline{i}_{\overline{v},1},\dots,\overline{i}_{\overline{v},D}}
 \prod_{{v}\atop \text{white vertices }}{T}_{{i}_{{v},1},\dots,{i}_{{v},D}}
 \prod_{e\atop\text{edges}}
 \delta_{\overline{i}_{e},\overline{i}_{\overline{v}(e),c(e)}}
 \delta_{i_{e},i_{v(e),c(e)}}.\label{noninvariantpotential:eq}
\end{gather}
The invariant potential \eqref{invariantpotential:eq} is recovered for
\begin{gather*}
\lambda_{\cal B}\left\{i_{e},\overline{i}_{e}\right\}=\lambda_{\cal B}
\prod_{e\,\text{edges}}\delta_{\overline{i}_{e},\overline{i}_{e}}.
\end{gather*}
The motivation behind the choice \eqref{noninvariantpotential:eq} lies in the fact that it is covariant under the group $\text{U}(N)\times\cdots\times U(N)$: it is invariant under simultaneous transformations of $\lambda_{\cal B}\left\{i_{e},\overline{i}_{e}\right\}$ and the tensors $T$ and $\overline{T}$. Moreover, this general construction renders the transition to group f\/ield theory more transparent.

 \subsection{Perturbative expansion}

Let us now give the basic features of the expansion of an arbitrary bubble observable as a formal power series in the bubble couplings ${\lambda}_{\cal B}$. Using \eqref{bubbleexpectation:eq}, we reduce such a computation to that of the partition function, considered as a function of the bubble couplings,
\begin{gather*}
Z=\int d\mu_{C}(T,\overline{T})
\exp\left\{-\sum_{{\cal B}}\frac{\lambda_{\cal B}}{C_{\cal B}}\,\operatorname{Tr}_{\cal B}(\overline{T},T)\right\}.
\end{gather*}
To proceed, we expand the exponential and integrate over the tensors using Wick's theorem with the covariance~\eqref{covariance:eq}. The latter can be formulated as the computation of the Gau{\ss}ian expectation value of the product of $n$ pairs of a tensor and its complex conjugate
\begin{gather*}
\int d\mu_{C}(T,\overline{T}) \
T_{i_{1,1}\dots i_{1,D}}\cdots
T_{i_{n,1}\dots i_{n,D}} \
\overline{T}_{\overline{i}_{1,1}\dots \overline{i}_{1,D}}\cdots
\overline{T}_{\overline{i}_{n,1}\dots \overline{i}_{n,D}} \\
\qquad {}= t\sum_{\sigma\in\mathfrak{S}_{n}}
\prod_{1\leq k\leq D}\delta_{i_{1,k},\overline{i}_{\sigma(1),k}}
\cdots
\delta_{i_{n,k},\overline{i}_{\sigma(n),k}}.
\end{gather*}
The sum over all permutations $\mathfrak{S}_{n}$ is nothing but the familiar sum over all pairings, with the constraint that a tensor is paired with its complex conjugate. The Gau{\ss}ian expectation value of a product of $n$ tensors $T$ with $\overline{n}$ tensors $\overline{T}$ vanishes when $n\neq\overline{n}$.

\begin{figure}[t]
\centering
\begin{subfigure}[Generic vertices.]{
\parbox{6cm}{\includegraphics[width=6cm]{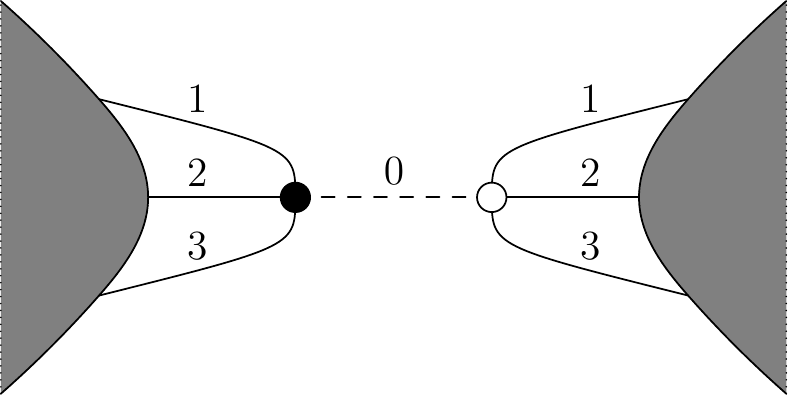}}$\qquad\rightarrow\qquad$
\parbox{4cm}{\includegraphics[width=4cm]{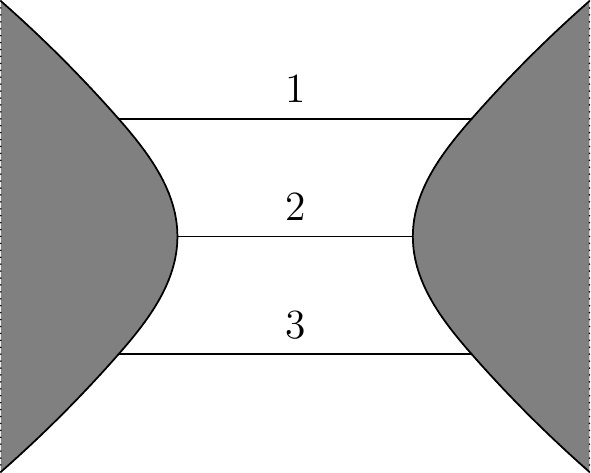}}
}
\end{subfigure}\hspace{2cm}
\begin{subfigure}[Connected vertices.]{\qquad\qquad
\parbox{6cm}{\includegraphics[width=6cm]{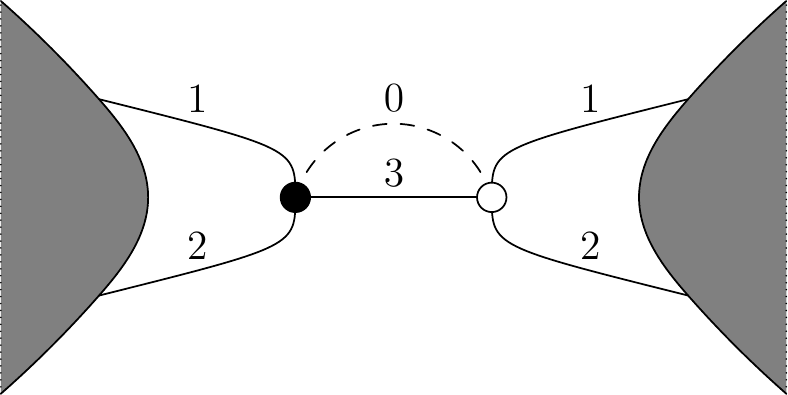}}$\qquad\rightarrow\qquad$
\parbox{4cm}{\includegraphics[width=4cm]{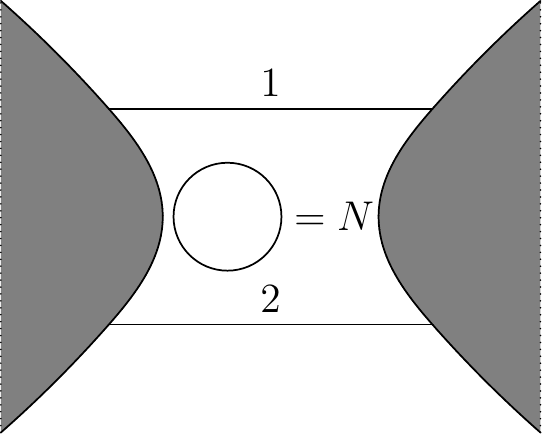}}
}
\end{subfigure}

\caption{Contraction of a pair of vertices.}\label{paircontraction:fig}
\end{figure}

Applying Wick's theorem to any monomial obtained by expanding the potential amounts to summing over all contractions of~$T$ with~$\overline{T}$. Any such contraction corresponds to the contraction of a pair of vertices of dif\/ferent colours in a bubble, def\/ined as follows. Consider a bubble ${\cal B}$, which may be disconnected, as is the case if it arises from two dif\/ferent bubble couplings or from a single bubble coupling associated with a disconnected bubble. Choose a white vertex~$v$ and black vertex~$\overline{v}$ in~${\cal B}$ and remove them, leaving a set of coloured half edges. Then, reattach the half edges respecting the colours. The resulting bubble is denoted by ${\cal B}/{\overline{v}v}$. It is convenient to materialise the choice of the pair of vertices in~${\cal B}$ before contraction by an extra colour~0 edge, represented by a dashed line, see Fig.~\ref{paircontraction:fig}. Note that if the pairs of vertices are already connected by~$k$ edges in ${\cal B}$, then the pair contraction yields $k$ circles, so that we have an extra factor of~$N^{k}$ resulting from the summation over the~$k$ indices.

\begin{figure}[t] \centering
\includegraphics[width=11.5cm]{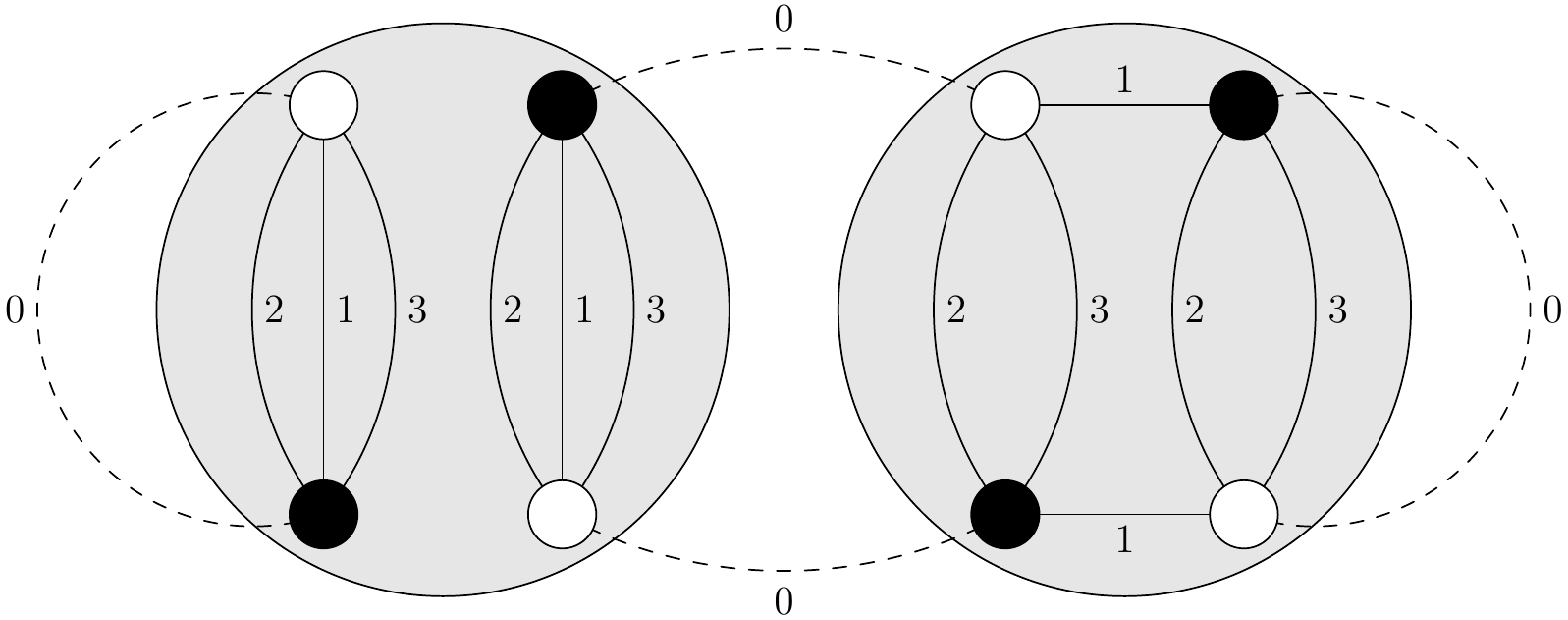}
\caption{A 0-connected bubble with color $0$ edges and two 3-bubbles.}
\label{graph:fig}
\end{figure}

Accordingly, the perturbative expansion of the partition function can be represented as
a~sum over $(D{+}1)$-bubbles (see Fig.~\ref{graph:fig} for an example of such a bubble)
\begin{gather*}
Z=\int d\mu_{C}(T,\overline{T})
\exp\left\{-\sum_{{\cal B}}\frac{\lambda_{\cal B}}{C_{\cal B}}\,\operatorname{Tr}_{\cal B}(\overline{T},T)\right\}=
\sum_{{\cal G}\atop (D{+}1)\text{-bubbles}}
\frac{N^{F({\cal G})}t^{E(G)}}{C_{\cal G}}
\prod_{{\cal B}\subset{\cal G}}(-\lambda_{\cal B}).
\end{gather*}
This expression is obtained by contracting all pairs of vertices, thus yielding $(D{+}1)$-bubbles once we represent the pair contractions by a~colour~0 line. The pair contractions can be used, in any order, to successively reduce the bubbles to a disjoint union of circles, each of them yielding a factor of~$N$. On the bubble ${\cal G}$, it can be viewed as the total number of faces with colour $0$. Recall that a face of colours $0$ and $k$ is a connected subgraph of ${\cal G}$ obtained by removing all colours except~$0$ and~$k$. $E({\cal G})$ is the number of colour 0 edges of ${\cal G}$. The bubbles ${\cal B}\subset{\cal G}$ are simply the $D$-bubbles we started with. On the example of Fig.~\ref{graph:fig}, the 3-bubble in the potentials have been drawn on a shaded disc, there is one disconnected 3-bubble (two dipoles) considered as a single interaction in the potential, whose contribution to the potential is
\begin{gather*}
\frac{1}{2}\lambda_{\includegraphics[width=0.5cm]{Krajewski-Fig01a}\includegraphics[width=0.5cm]{Krajewski-Fig01a}}\big(\overline{T}\cdot T\big)^{2}.
\end{gather*}
The 4-bubble of Fig.~\ref{graph:fig} is 0-connected but not connected.

In many cases, it is useful to consider the logarithm of~$Z$. The latter is expanded over $0$-connected bubbles,
\begin{gather}
\log Z=\sum_{{\cal G}\atop 0\text{-connected}\,(D{+}1)\text{-bubbles}}
\frac{N^{F({\cal G})}t^{E({\cal G)}}}{C_{\cal G}}
\prod_{{\cal B}\subset{\cal G}}(-\lambda_{\cal B}),\label{logexpansion:eq}
\end{gather}
where a bubble is said to be 0-connected if the graph (with only colour 0 lines) obtained by contracting all the bubbles ${\cal B}\subset {\cal G}$ is connected. A~bubble may be 0-connected without being connected if some of the bubbles ${\cal B}\subset {\cal G}$ are disconnected.

\begin{figure}[t]
\centering
\parbox{5cm}{\includegraphics[width=5cm]{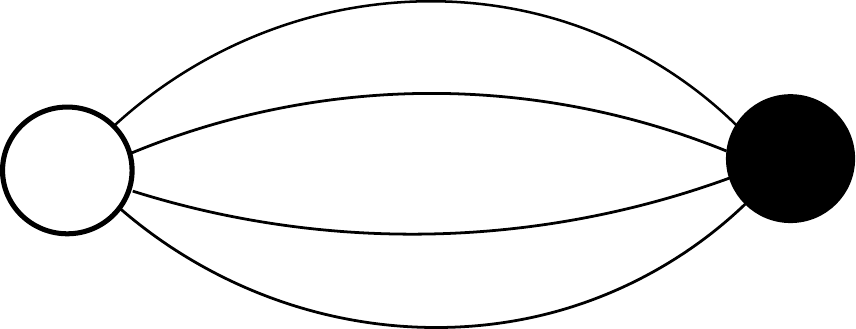}}\quad$\Leftrightarrow$\quad\parbox{5cm}{\includegraphics[width=5cm]{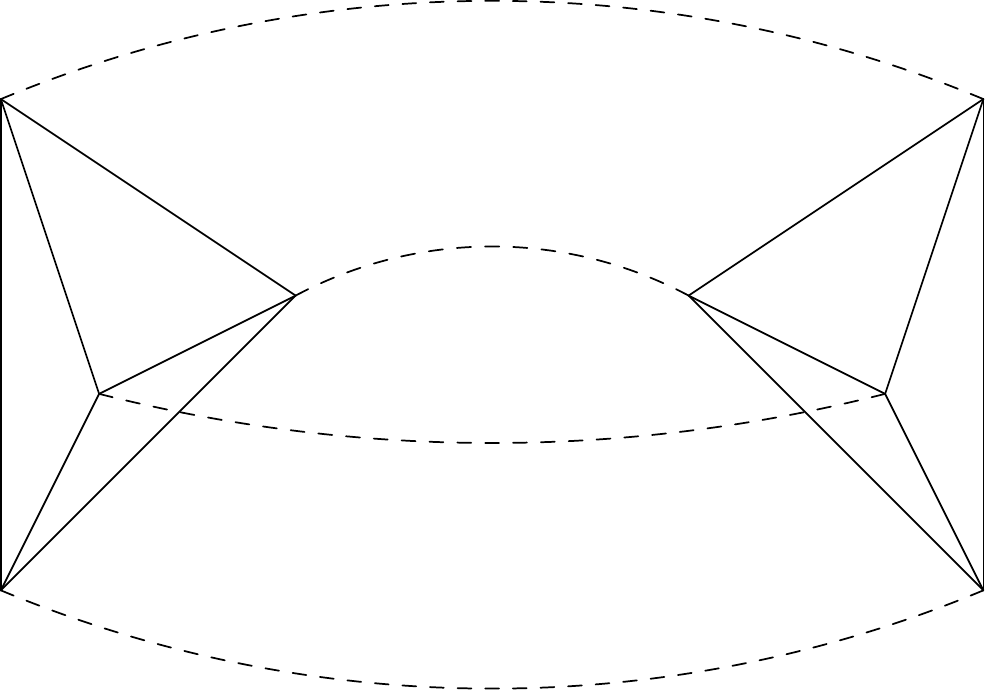}}
\caption{Triangulation of a 3d sphere with 2 tetrahedra.}
\label{dipoletriangulation:fig}
\end{figure}

This expansion can be given a geometrical interpretation that may be used in discretised approaches to quantum gravity and random geometry. Each
bubble ${\cal G}$ with $D+1$ colors {represents} a coloured triangulation of a space of dimension~$D$. Its simplices of dimension~$D$ are the vertices of~${\cal G}$ attached along their $D-1$ faces following the edges of~${\cal G}$. Thus, the expansion of the logarithm of the partition function~\eqref{logexpansion:eq} provides a sum over coloured triangulations. If all the bubbles in the potential are connected, then~$\log Z$ provides us with a sum over connected triangulations. For instance, the dipole graph represents a triangulation of a sphere by~2 simplices, see Fig.~\ref{dipoletriangulation:fig}.

Triangulations with boundaries are represented by correlation functions, see~\eqref{correlation:eq}. $(D{+}1)$-bubbles contributing to a correlation with~$n$ tensors~$T$ and $\overline{T}$ have $2n$ colour 0 external legs. Such a~bubble~${\cal B}$ gives rise to the boundary bubble $\partial{\cal B}$. Vertices of $\partial{\cal B}$ are external legs of~${\cal B}$. They are connected by an edge of colour~$k$ if and only if the two external legs belong to a face of colour~$0k$ in~${\cal B}$. Recall that in a coloured tensor models a face of colours $i_{1}\dots i_{k}$ in a bubble is a connected component of the subgraph made of edges of colours $i_{1}\dots i_{k}$ in ${\cal B}$.

\subsection{Tensorial group f\/ield theories} \label{GFTgeneral:sec}

\begin{figure}[t]
\centering
\parbox{4.5cm}{\includegraphics[width=4.0cm]{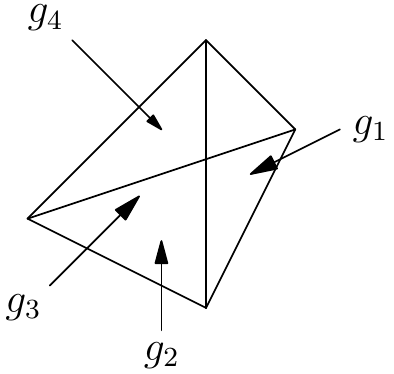}}
\caption{The group f\/ield as a tetrahedron.}
\label{tetrahedrongroup:fig}
\end{figure}

Tensorial group f\/ield theories are particular quantum f\/ield theories def\/ined on $D$ copies of a~group, usually $\text{SU}(2)$, $\text{U(1)}^{d}$, ${\mathbb R}^{d}$ or $\text{SL}(2,{\mathbb C})$. They are constructed in analogy with tensor models, we replace tensors by functions of $D$ copies of the group,
\begin{gather*}
T_{i_{1},\dots,i_{D}}\rightarrow
\Phi(g_{1},\dots,g_{D}),\qquad
\overline T_{\overline{i}_{1},\dots,\overline{i}_{D}}\rightarrow
\overline\Phi(\overline g_{1},\dots,\overline g_{D}).
\end{gather*}
Note that $\overline{g}$ is not the complex conjugate of $g$ but an independent group element that is an argument of the complex conjugate f\/ield $\overline\Phi$. In quantum gravity applications, $D$ denotes the space-time dimension and the group is the Lorentz group, the euclidian group or their universal covers. The group f\/ield $\Phi(g_{1},\dots,g_{D})$ represents a $D-1$ simplex and the group variables $g_{1},\dots,g_{D}$ are associated with its $D$ $(D{-}1)$-faces, see Fig.~\ref{tetrahedrongroup:fig}.

\begin{figure}[t]
\centering
\parbox{6cm}{\includegraphics[width=5.2cm]{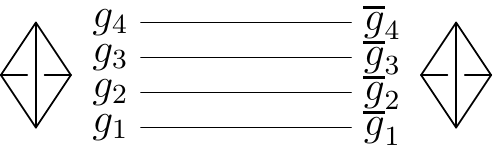}}
\caption{Propagation of a tetrahedron.}\label{tetrahedrongroup:fig+}
\end{figure}

In many models, the group elements are related to the normals of the tetrahedron. Then, it is assumed that they obey a closure constraint. The closure condition is implemented as an invariance of the f\/ield under left translation,
\begin{gather}
\Phi(g_{1},\dots,g_{D})=
\Phi(hg_{1},\dots,hg_{D}),\qquad
\overline\Phi(\overline g_{1},\dots,\overline g_{D})=
\overline\Phi(\overline{h}\overline g_{1},\dots,\overline{h}\overline g_{D}).\label{closure:eq}
\end{gather}
for every group elements $h$ and $\overline{h}$. The covariance is def\/ined by the Gau{\ss}ian integration
\begin{gather*}
\int d\mu_{C}(\Phi,\overline{\Phi})\,\Phi(g_{1},\dots,g_{D})\overline{\Phi}(\overline{g}_{1},\dots,\overline{g}_{D})=
C\big(g_{1}\overline{g}^{-1}_{1},\dots,g_{D}\overline{g}^{-1}_{D}\big),
\end{gather*}
and the integration of $\Phi\,\Phi$ and $\overline{\Phi}\,\overline{\Phi}$ vanishes. Because of the closure constraint~\eqref{closure:eq}, the co\-va\-riance obeys
\begin{gather*}
C\big(\big\{g_{e}\overline{g}_{e}^{-1}\big\}\big)=
C\big(\big\{hg_{e}\overline{g}_{e}^{-1}\overline{h}^{-1}\big\}\big).
\end{gather*}
Note that it is also invariant under a global translation for every colour, since it only depends on products $g_{e}\overline{g}_{e}^{-1}$. The covariance propagates a tetrahedron, in analogy with the quantum f\/ield theory propagation of a particle, see Fig.~\ref{tetrahedrongroup:fig+}.

A suitable covariance may be constructed using the heat kernel on the group $H_{\alpha}$,
\begin{gather}
C_{\Lambda,\Lambda_{0}}\big(\big\{g_{e}\overline{g}_{e}^{-1}\big\}\big)=\int_{\frac{1}{\Lambda_{0}^{2}}}^{\frac{1}{\Lambda^{2}}}d\alpha \int dhd\overline{h}
\prod_{1\leq i\leq D}H_{\alpha}\big(hg_{i}\overline{g}_{i}^{-1}\overline{h}^{-1}\big).
\label{covarianceGFT:eq}
\end{gather}
$\Lambda$ and $\Lambda_{0}$ are cut-of\/fs that have to be introduced to avoid divergences. Group integrations are performed using the Haar measure and implement the required invariances.

As for the tensor models, the general interaction potential can also be expanded over $D$-bubbles. However, in this case there is no reason to impose unitary invariance. Therefore, we use the general form of the bubble couplings~\eqref{noninvariantpotential:eq}, replacing tensor indices by group elements,
\begin{gather}
V(\Phi,\overline{\Phi})=
\sum_{{\cal B}}\frac{1}{C_{\cal B}}
\int\prod_{e} dh_{e} \prod_{v,i} d{g_{v,i}} \prod_{\overline{v},
\overline{i}} d\overline{g}_{\overline{v},\overline{i}} \lambda_{\cal B} ( \{h_{e} \} )\prod_{e}\delta\big(h_{e}^{-1}g_{v(e),c(e)}(\overline{g}_{\overline{v}(e),c(e)})^{-1}
\big) \nonumber\\
\hphantom{V(\Phi,\overline{\Phi})=}{}\times
\prod_{\overline{v}\atop \text{black vertices }}\hspace{-0.5cm}\overline{\Phi}(\overline{g}_{\overline{v},1},\dots,\overline{g}_{\overline{v},D})
 \prod_{{v}\atop \text{white vertices }}\hspace{-0.5cm}{\Phi}(g_{{v},1},\dots,g_{{v},D}),
\label{GFTpotential:eq}
\end{gather}
where $\delta$ is the Dirac distribution on the group, def\/ined by $\int dg\delta (g)f(g)=f(1)$. Note that we have imposed translation invariance, so that the couplings only depend on the products $g_{v(e),c(e)}(\overline{g}_{\overline{v}(e),c(e)})^{-1}$

Then, the perturbative expansion yields a sum over triangulations of dimension $D$, weighted by certain graph amplitudes. By a suitable choice of the bubble couplings $ \lambda_{\cal B}(\{h_{e}\})$, one can show that this amplitude can be made equal to a spin foam amplitude. Indeed, a spin foam is 2-complex and can be constructed from the $(D{+}1)$-bubbles by adding a 2-cell to any colour~$0k$ face, for every~$k$.

\subsection{Couplings in the spin network basis}

The relation to spin foam models can be made more precise if we formulate the theory in terms of spins $j$ (labelling representations of $\text{SU}(2)$) and intertwiners $\iota$ (invariant tensors) in the case of $\text{SU}(2)$. Then, we expand the bubble couplings using the Peter--Weyl theorem,
\begin{gather*}
\lambda_{\cal B} ( \{h_{e} \} )=
\sum_{j_{e}\,\text{spins}\atop
i_{v}\, \text{intertwiners}} \lambda_{\cal B} ( \{j_{e},\iota_{v}\} )
\prod_{e} d_{j_{e}}\\
{}\times
\prod_{e\,\text{edges}}\Bigg\{\!\!
\sum_{m_{v,k},\, \overline{m}_{\overline{v},k}\atop\text{magnetic indices}}\!\!\!
(\iota_{v(e)})_{m_{v(e),1}\dots m_{v(e),D}}
{\cal D}^{j_{e}}_{m_{v(e),c(e)},
\overline{m}_{\overline{v}(e),c(e)}}(h_{e})
(\overline{\iota}_{\overline{v}(e)})_{\overline{m}_{\overline{v}(e),1}\dots \overline{m}_{v(e),D}}
\Bigg\}.
\end{gather*}
${\cal D}_{\overline{m}m}^{j}(h)$ are the Wigner matrices with $d_{j}=2j+1$ the dimension of the spin $j$ representation and the intertwiner appear because of the closure constraint~\eqref{closure:eq}. The intertwiners are invariant tensors in the tensor product of representation of edges incident to a vertex, following the ordering provided by the colours. They are chosen so that they def\/ine a orthonormal basis in this tensor product. The new couplings $\lambda_{\cal B} (\{j_{e},\iota_{v}\})$ depend on the choice of a spin for every edge and of an intertwiner for every vertex. Such a representation theoretic data def\/ines a spin nertwork~$s$, so that the couplings can be simply indexed by a spin network. For example, for the dipole graph, the spin network formulation yields
\begin{gather*}
\lambda_{\includegraphics[width=0.4cm]{Krajewski-Fig01a}}(h_{1},h_{2},h_{3})= \sum_{j_{1},\,j_{2},\,j_{3},\,\iota}
\lambda_{\includegraphics[width=0.4cm]{Krajewski-Fig01a}}(j_{1},j_{2},j_{3},\iota)\\
\hphantom{\lambda_{\includegraphics[width=0.4cm]{Krajewski-Fig01a}}(h_{1},h_{2},h_{3})=}{}\times
\sum_{m_{1},\,m_{2},\,m_{3}\atop
\overline{m}_{1},\,\overline{m}_{2},\, \overline{m}_{3}}
 \iota_{m_{1}m_{2}m_{3}}
D_{m_{1}\overline{m}_{1}}^{j_{1}}(h_{1})
D_{m_{2}\overline{m}_{2}}^{j_{2}}(h_{2})
D_{m_{3}\overline{m}_{3}}^{j_{3}}(h_{3})
\overline{\iota}_{\overline{m}_{1}\overline{m}_{2}\overline{m}_{3}}.
\end{gather*}
This can be considered as a non-Abelian Fourier transform,
\begin{gather*}
\text{position space: $h_{e}$}\qquad
\leftrightarrow\qquad
\text{momentum space: $j_{e}$, $\iota_{v}$.}
\end{gather*}
Moreover, it is important to realise that the existence of the intertwiner between representations labelled by spins puts some constraints on these spins. For instance, for three spins~$j_{1}$,~$j_{2}$,~$j_{3}$, there exists an intertwiner (and it is then unique up to normalisation) if and only if the triangular inequality $|j_{1}-j_{2}|\leq j_{3}\leq j_{1}+j_{2}$ is satisf\/ied.

\section{Exact renormalisation group equations}\label{section3}

\subsection{Wilson's ideas in quantum f\/ield theory}

\label{Wilson:sec}

Let us start with a heuristic overview of Wilson's idea in quantum f\/ield theory, deferring a more precise formulation till the next section. Consider a quantum f\/ield theory with an action $S_{0}$ and a UV cut-of\/f~$\Lambda_{0}$. The basic object of interest is the generating functional of correlation functions,
\begin{gather*}
Z[J]=\int [D\phi]_{\Lambda_{0}}\exp \{-S_{0}[\phi]+J\phi \},
\end{gather*}
where $[D\phi]_{\Lambda_{0}}$ is a measure that integrates only over f\/ields with momenta~$\lesssim\Lambda_{0}$.

Wilson's idea amounts to performing this integration not in a single step but in a series of partial integrations. The cutof\/f~$\Lambda_{0}$ is then reduced to a lower cutof\/f~$\Lambda$ by integrating the modes between~$\Lambda$ and~$\Lambda_{0}$. The low energy physics, for instance correlation functions for f\/ields at momenta~$\lesssim\Lambda$ is then left unchanged. Denoting by~$\phi_{0}$ the original f\/ield with modes less than~$\Lambda_{0}$, $\phi$ a f\/ield with modes less than~$\Lambda$ and by $\chi$ the remaining degrees of freedom between~$\Lambda$ and~$\Lambda_{0}$,
\begin{gather*}
\phi_{0}=\phi +\chi.
\end{gather*}
This separation of modes translates into the following equation for the generating functional
\begin{align*}
Z[J]&=\int [D\phi_{0}]_{\Lambda_{0}}\exp \{-S_{0}[\phi_{0}]+J\phi_{0} \} \\
&=\int [D\phi]_{\Lambda} [D\chi]_{\Lambda,\Lambda_{0}}\exp \{-S_{0}[\phi+\chi]+J\phi+J\chi \}\\
&=\int [D\phi]_{\Lambda} \exp \{-S_{\Lambda}[\phi]+J\phi \},
\end{align*}
where we have used $J\chi=0$ since we are only interested in the low energy correlation functions. $S_{\Lambda}$ is the Wilsonian ef\/fective action at scale $\Lambda$ def\/ined by
\begin{gather*}
S_{\Lambda}[\phi]= -\log\int [D\chi]_{\Lambda,\Lambda_{0}}\exp\{-S_{0}[\phi+\chi]\}.
\end{gather*}
This procedure leads to a f\/low equation for the ef\/fective action
\begin{gather}
\Lambda\frac{dS_{\Lambda}}{d\Lambda}=\beta(\Lambda,S_{\Lambda}),\label{flowS:eq}
\end{gather}
with the initial condition $S_{\Lambda_{0}}=S_{0}$.

The physics of this equation is best captured by expanding the ef\/fective action in local operators, which are monomials of the f\/ields and their derivatives. Writing symbolically $O[\partial^{d},\phi^{n}]$ such a monomial with~$n$ occurrences of the f\/ield and $d$ derivatives, the action reads, in a space-time of dimension~$D$,
\begin{gather*}
S_{\Lambda}[\phi]=\sum_{i\atop
\text{local operators}}\int d^{D}x\,
g_{i}(\Lambda) O\big[\partial^{d_{i}},\phi^{n_{i}}\big].
\end{gather*}
The f\/low equation for the ef\/fective action~\eqref{flowS:eq} translates into a system of equations for the couplings,
\begin{gather*}
\Lambda\frac{dg_{i}}{d\Lambda}=\beta_{i}(\Lambda, \{g_{j} \}).
\end{gather*}
It is convenient to write this equation in terms of dimensionless couplings $u_{i}(\Lambda)$, related to~$g_{i}(\Lambda)$ by
\begin{gather*}
g_{i}(\Lambda)=\Lambda^{\delta_{i}}u_{i}(\Lambda),
\end{gather*}
where $\delta_{i}$ is the canonical dimension of the local operator $i$. Assuming translational invariance, this dimension is
\begin{gather*}
\delta_{i}=D-n_{i}\left(\frac{D}{2}-1\right)-d_{i}.
\end{gather*}
Dimensional analysis allows us to write the f\/low equation for dimensionless couplings as
 \begin{gather*}
\Lambda\frac{du_{i}}{d\Lambda}= -\delta_{i}u_{i}+ \beta_{i}( \{u_{j} \}).
\end{gather*}
 Couplings are classif\/ied according to the sign of their dimension:
\begin{itemize}\itemsep=0pt
\item relevant couplings with $\delta_{i}>0$,
\item irrelevant couplings with $\delta_{i}<0$,
\item marginal couplings with $\delta_{i}=0$.
\end{itemize}
In the vicinity of the Gau{\ss}ian f\/ixed point $u^{\ast}_{i}=0$ (i.e., in perturbation theory),
the physical consequences of this equation are the following:

 \begin{itemize}\itemsep=0pt

 \item Universality: Whatever the irrelevant couplings at scale $\Lambda_{0}$ are (provided $u_{i}(\Lambda_{0})$ are bounded, so that $g_{i}(\Lambda_{0})\sim \Lambda_{0}^{\delta_{i}}$ goes to zero for large $\Lambda_{0}$), when $\Lambda_{0}\rightarrow\infty$, the couplings at scale $\Lambda$ are attracted towards a manifold that can be parametrized by the relevant and marginal couplings only.
 \item Renormalisability: When $\Lambda_{0}\rightarrow \infty$ with $\Lambda$ f\/ixed, the relevant and marginal couplings at scale $\Lambda$ diverge. Perturbative renormalisation amounts to imposing their values at $\Lambda$ by a~measurement. Then, all the other operators can be computed at scale~$\Lambda$ in terms of the relevant and marginal operators at scale~$\Lambda$, in the limit $\Lambda_{0}\rightarrow\infty$.
\end{itemize}

This provides a rationale for renormalisation: When renormalisable couplings (i.e., couplings with $\delta_{i}\geq 0$) are available, the latter dominate the low energy behaviour in the limit $\Lambda_{0}\rightarrow\infty$. Low energy physics can be formulated in terms of low energy parameters. Of course, when no such couplings exist, for instance if they are forbidden by symmetries, the limit $\Lambda_{0}\rightarrow$ cannot be taken.

These statements can be rigorously proved in perturbation theory, see~\cite{Polchinski:1983gv}. Obviously, a~nonperturbative study is required, especially when marginal couplings are involved or when no renormalisable couplings are available. It may also be that the theory is attracted towards a~nontrivial f\/ixed point. Finally, let us mention that these ideas also play a prominent role in constructive physics. The latter aims at a~nonperturbative construction of the theory based on a~multiscale analysis of the Feynman graphs~\cite{Rivasseau:1991ub}.

 \subsection{Exact renormalisation group equations}

 In order to understand how the previous ideas can be implemented in tensor models and tensorial group f\/ield theories, a more precise formulation is required. Within this section, we remain in the general setting of a quantum f\/ield theory. In order to keep the notations as simple and universal as possible, we write f\/ields as~$\phi^{i}$, where $i$ may be a discrete index, a~space-time index or a momentum. We separate the action into a quadratic part and an interaction poten\-tial~$V_{0}[\phi]$. The quadratic part is included into a Gau{\ss}ian measure of covariance $C(t,t_{0})$ which we assume to be written as an integral
\begin{gather}
C(t,t_{0})=\int_{t_{0}}^{t}ds \,K(s).\label{integralC:eq}
\end{gather}
The interacting part of ef\/fective action is def\/ined as
\begin{gather}
V(t,\phi)=-\log \int d\mu_{C(t,t_{0})}(\chi)\exp\{-V_{0}(t,\phi+\chi\},\label{effectiveV:eq}
\end{gather}
with the initial condition $V(t_{0},\phi)=V_{0}[\phi]$ since $C(t_{0},t_{0})=0$ so that there is no integration.

\begin{figure}[t]
\centering
${\displaystyle \frac{\partial}{\partial t}}$\,\,\parbox{2cm}{\includegraphics[width=2cm]{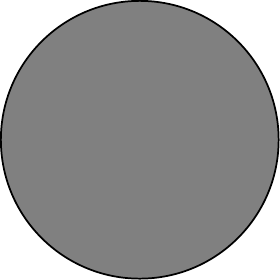}}
\quad=\quad-${\displaystyle \frac{1}{2}}$
\quad\parbox{4cm}{\includegraphics[width=4cm]{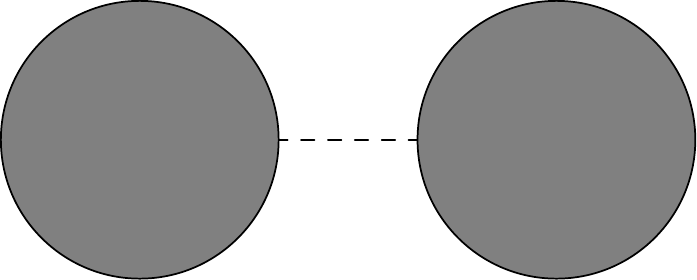}}\quad+${\displaystyle \frac{1}{2}}$\quad
\parbox{3cm}{\includegraphics[width=3cm]{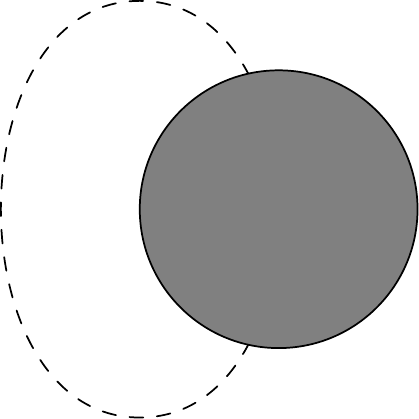}}

\caption{A graphical interpretation of the Polchinski's equation.}\label{Polchinskigeneral:fig}
\end{figure}

The ef\/fective interaction $V(t,\phi)$ obey the semi-group relation, see~\cite{ZinnJustin:1989mi},
\begin{align*}
Z&=
\int d\mu_{C(t,t_{0})}[\phi]\exp\{-V_{0}(t,\phi)\} \\
&=\int d\mu_{C(t,t_{1})}[\phi]
\int d\mu_{C(t_{1},t_{0})}(\chi)
\exp\{-V_{0}(\phi+\chi)\} \\
&=\int d\mu_{C(t,t_{1})}[\phi]\exp\{-V(t_{1},\phi)\}.
\end{align*}
Then, the integration over an inf\/initesimal shell with $t_{1}=t-dt$ leads to the following dif\/ferential equation, known as Polchinski's exact renormalisation group equation,
\begin{gather}
\frac{\partial V}{\partial t}=
\frac{1}{2}\sum_{i,j}
K_{ij}(t)\left(-\frac{\partial V}{\partial \phi^{i}}
\frac{\partial V}{\partial \phi^{j}}+\frac{\partial ^{2}V}{\partial \phi^{i}\partial\phi^{j}}
\right).\label{Polchinskigeneral:eq}
\end{gather}
At the perturbative level, the ef\/fective potential~\eqref{effectiveV:eq} can be expanded over Feynman graphs \mbox{using} the background f\/ield technique. $\phi$ dependent vertices are introduced by expanding \linebreak \mbox{$V_{0}(\phi+\chi)$} in powers of~$\chi$. Then, Feynman graphs are generated by integrating over $\chi$. Then, the f\/low equation provides a recursive construction of these graphs by successively adding edges. The f\/irst term corresponds to a tree edge (between two connected components) and the second one to a loop edge (within the same connected component), see Fig.~\ref{Polchinskigeneral:fig}.

While Polchinski's equation is a powerful tool in the perturbative case, in nonperturbative investigations (for instance search for nontrivial f\/ixed points using truncations), it is often more convenient to sum the contributions of the trees. This is achieved by introducing the Legendre transform~$\Gamma$ of~$V$ with respect to some sources. Alternatively $\Gamma$ can be def\/ined in analogy with~\eqref{effectiveV:eq} by
\begin{gather*}
\Gamma(t,\phi)=-\log \int d\mu_{C(t,t_{0})}(\chi)\exp\left\{-V_{0}(t,\phi+\chi)+\frac{\partial\Gamma}{\partial \phi}\cdot\chi\right\}.
\end{gather*}
It obeys the following dif\/ferential equation, known as Wetterich's equation, which we write symbolically as
\begin{gather}
\frac{\partial\Gamma}{\partial t}=\frac{1}{2}
\operatorname{Tr}\left[C^{-1}KC^{-1}\left(1+C\frac{\partial^{2}\Gamma}{\partial\phi\partial\phi}C
\right)^{-1} \right],\label{Wetterich:eq}
\end{gather}
which has to be understood as a geometrical series for the matrix products
\begin{gather*}
\sum_{j,k}
C_{ij}\frac{\partial^{2}\Gamma}{\partial\phi_{j}\partial\phi_{k}}C_{kl}.
\end{gather*}
Wetterich's equation is particularly useful in the search for f\/ixed points. A f\/ixed point is an ef\/fective action such $\beta(S_{\ast})=0$. Because $\Lambda\frac{dS}{d\Lambda}=\beta(S)$, a f\/ixed point action describes a scale invariant theory. These f\/ixed points play an important role in statistical physics. In the case of the $O(N)$ invariant model in three dimensions, the long distance behavior at a second order phase transition is governed by a infrared stable f\/ixed point (i.e., with only few divergent directions), known as the Wilson--Fisher f\/ixed point. In this case the ultraviolet cut-of\/f $\Lambda_{0}$ is f\/ixed because it is nothing but the inverse lattice spacing. The vicinity of the critical point is described by this f\/ixed point in the long distance limit $\Lambda\ll\Lambda_{0}$. On the other side, ultraviolet stable f\/ixed point may allow the large cut-of\/f limit $\Lambda_{0}\rightarrow\infty$ in theories of fundamental interactions. Asymptotically free theories provide natural examples with a Gau{\ss}ian f\/ixed point but more general f\/ixed points have been found in some simple models, like the $D=3$ Gross--Neveu model, see for instance~\cite{Braun:2010tt}. Then, physical theories are parametrised by the few coordinates on the stable manifold. Such a scenario, known as "asymptotic safety", has been proposed by Weinberg and may allow to construct a~quantum theory of gravity with only f\/ield theoretical degrees of freedom. We refer the reader to the reviews
 \cite{Bagnuls:2000ae,Polonyi:2001se,Rosten:2010vm} for more complete overviews of these renormalisation group equations.

The f\/ield theory case presented in the previous section is recovered by taking $t=\log(\Lambda/\Lambda_{\text{ref}})$, where $\Lambda_{\text{ref}}$ is an immaterial reference scale that disappears from the equations. Then, $C_{t_{0},t}=C_{\Lambda_{0},\Lambda}$ smoothly integrates out modes between $\Lambda$ and $\Lambda_{0}$. In momentum space, it can be written as
\begin{gather*}
C(\Lambda_{0},\Lambda, p,q)=
\delta(p+q)\int_{\Lambda}^{\Lambda_{0}}
\frac{d\Lambda'}{\Lambda'}
K(\Lambda',p).
\end{gather*}
Sources can be included in $V$ for convenience since we assume that they decouple from the high energy behaviour. When formulated in momentum space, the f\/low equation~\eqref{Polchinskigeneral:eq}
for a~translationally invariant f\/ield theory takes the form
\begin{gather*}
\Lambda\frac{dS}{d\Lambda}=
\frac{1}{2}\int d^{D}p\,
K(\Lambda,p)\left(
-\frac{\delta V}{\delta\phi(p)}
\frac{\delta V}{\delta\phi(-p)}
+\frac{\delta^{2}V}{\delta\phi(p)\delta\phi(-p)}
\right).
\end{gather*}
Expanding the ef\/fective interaction in powers of the f\/ield
\begin{gather*}
V[\Lambda,\phi]=\sum_{n}\frac{1}{n!}\int dp_{1}\cdots dp_{n}\delta(p_{1}+\cdots+ p_{n})G_{n}(\Lambda,p_{1},\dots,p_{n})\phi(p_{1})\cdots\phi(p_{n})
\end{gather*}
yields the system of equations
\begin{gather}
\frac{\partial G_{n}}{\partial \Lambda}(\Lambda,p_{1},\dots,p_{n})=
\frac{1}{2}\int d^{D}p\, K(\Lambda,p) \bigg(
\sum_{i=0}^{n}
\big\{-G_{i+1}(\Lambda,p,p_{1},\dots,p_{i})\nonumber\\
\hphantom{\frac{\partial G_{n}}{\partial \Lambda}(\Lambda,p_{1},\dots,p_{n})=}{}
\times G_{n-i+1}(\Lambda,-p,p_{i+1},\dots,p_{n})
+G_{n+2}(\Lambda,p,-p,p_{1},\dots,p_{n})\big\}\bigg).
\label{ERGEcorrelation:eq}
\end{gather}
Then, dimensional analysis can be done by further expanding $G_{n}$ in powers of the momenta.

 The next sections are devoted to an implementation of these ideas in the context of tensor models and tensorial group f\/ield theories, following our presentation in~\cite{Krajewski:2015clk}.

 \subsection{Polchinski's equation for invariant tensor models}

 Let us f\/irst consider an analogue of the Polchinski's exact renormalisation group equation \eqref{Polchinskigeneral:eq} for random tensors. We consider a covariance $C(t,t_{0})$ such that~\eqref{integralC:eq}.
The ef\/fective potential is def\/ined as in~\eqref{effectiveV:eq},
 \begin{gather*}
V(t,T,\overline{T})=-\log \int d\mu_{C(t,t_{0})}(
T',\overline{T}')
\exp\big\{{-}V_{0}\big(t,T+T',\overline{T}+\overline{T}'\big)\big\}.
\end{gather*}
Since we treat $T$ and $\overline{T}$ as independent variables, the f\/low equation for the ef\/fective poten\-tial~\eqref{Polchinskigeneral:eq} takes the form
\begin{gather}
\frac{\partial V}{\partial t}=
\sum_{i_{1},\dots,i_{D}\atop \overline{i}_{1},\dots,\overline{i}_{D}}
K_{i_{1},\dots,i_{D}|\overline{i}_{1},\dots,\overline{i}_{D}}(t)\left(-\frac{\partial V}{\partial T_{i_{1},\dots,i_{D}}}
\frac{\partial V}{\partial \overline{T}_{\overline{i}_{1},\dots,\overline{i}_{D}}}+\frac{\partial ^{2}V}{\partial T_{i_{1},\dots,i_{D}}\partial\overline{T}_{\overline{i}_{1},\dots,\overline{i}_{D}}}
\right).\label{Polchinskitensor:eq}
\end{gather}
In order to preserve unitary invariance~\eqref{unitary:eq}, the covariance has to be diagonal, see~\eqref{covariance:eq}. In this section, we restrict our attention to
\begin{gather*}
C_{i_{1},\dots,i_{D}|\overline{i}_{1},\dots,\overline{i}_{D}}(t,t_{0})=\mathfrak{K}
(t-t_{0})\prod_{1\leq k\leq D} \delta_{i_{k},\overline{i}_{k}},
\end{gather*}
with $\mathfrak{K}$ independent of $t$ and $t_{0}$. Although simple, such a covariance is not really in accordance with the spirit of Wilson's ideas. It would be better to choose a covariance that gradually reduces the size of the tensors by integrating over some of its components. For simplicity, we restrict ourselves to an invariant integration. We refer to~\cite{Eichhorn:2013isa,Eichhorn:2014xaa} for an application of a~reduction of the size to matrix models suing Wetterich's equation.

In the unitary invariant case, we expand the interaction potential over bubbles as in \eqref{invariantpotential:eq}
 \begin{gather}
V(t,T,\overline{T})=\sum_{{\cal B}}\frac{\lambda_{\cal B}(t)}{C_{\cal B}} \operatorname{Tr}_{\cal B}(\overline{T},T),
\label{couplingstensor:eq}
\end{gather}
with couplings that depend on the parameter $t$ controlling the f\/low.

\begin{figure}[t] \centering
 \includegraphics[width=15cm]{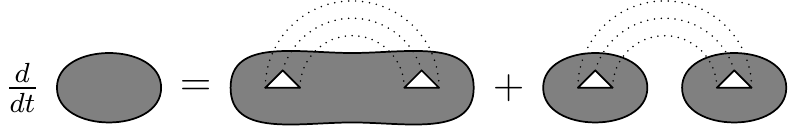}
 \caption{Dual interpretation of the f\/low equation for $D=3$.}
 \label{Polchinski3d:fig}
 \end{figure}

Polchinski's exact renormalisation group equation for tensors \eqref{Polchinskitensor:eq} involves the operations of deriving $V$ with respect to $T$ and $\overline{T}$. At the level of bubble couplings, they respectively amount to removing a white vertex $v$ and a black vertex $\overline{v}$. Then, the contraction with the derivative of the covariance $K$ connects the resulting edges respecting the colours. Therefore, when acting on bubble couplings in $V$,
\begin{gather*}
\sum_{i_{1},\dots,i_{D}\atop \overline{i}_{1},\dots,\overline{i}_{D}}
K_{i_{1},\dots,i_{D}|\overline{i}_{1},\dots,\overline{i}_{D}}(t)\left(-\frac{\partial V}{\partial T_{i_{1},\dots,i_{D}}}
\frac{\partial V}{\partial \overline{T}_{\overline{i}_{1},\dots,\overline{i}_{D}}}+\frac{\partial ^{2}V}{\partial T_{i_{1},\dots,i_{D}}\partial\overline{T}_{\overline{i}_{1},\dots,\overline{i}_{D}}}
\right)
\end{gather*}
 is equivalent to a contraction of a pair of vertices, either on two independent couplings (f\/irst term) or on the same coupling (second term). At the dual level of triangulation, the f\/low equation can be represented graphically as in Fig.~\ref{Polchinski3d:fig}.

 Equation \eqref{Polchinskitensor:eq} translates into a system of equations for the couplings $\lambda_{\cal B}$. To explicit this system, it is useful to introduce the inverse operation of a~pair contraction, which is called a~cut. A~$k$-cut~$c$ in a bubble ${\cal B}$ is def\/ined as a~subset of edges $\{e_{1},\dots,e_{k}\}$ of ${\cal B}$ with dif\/ferent colours. The cut bubble ${\cal B}_{c}$ is the bubble obtained from ${\cal B}$ by cutting the $k$ edges $\{e_{1},\dots,e_{{ k}}\}$ into half-edges, attaching to them a new black~$\overline{v}$ and a new white vertex~$v$ and joining~$v$ and~$\overline{v}$ by $D-k$ edges carrying the colors not in $\{e_{1},\dots,e_{k}\}$. Then, it follows that ${\cal B}/{v\overline{v}}={\cal B}_{c}$. The notion of $k$-cut with $k\neq D$ is necessary to deal with the contraction of a pair that is connected by $D-k$ edges.

This ensures that ${\cal B}_{c}$ is a bubble with $D$ colours. In particular, if $c$ is a 0-cut, ${\cal B}_{c}$ is just the disjoint union of ${\cal B}$ with a dipole. A~1-cut on an edge~$e$ is just the insertion on~$e$ of a pair of vertices joined by $D-1$ edges carrying the colours dif\/ferent from that of $e$.
\begin{figure}[t]\centering
\begin{subfigure}[A 3-cut.]{
\parbox{4cm}{\includegraphics[width=4cm]{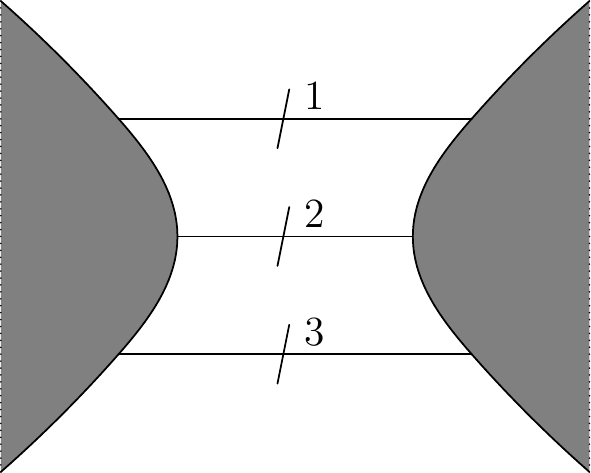}}
$\quad\rightarrow\quad$\parbox{6cm}{\includegraphics[width=6cm]{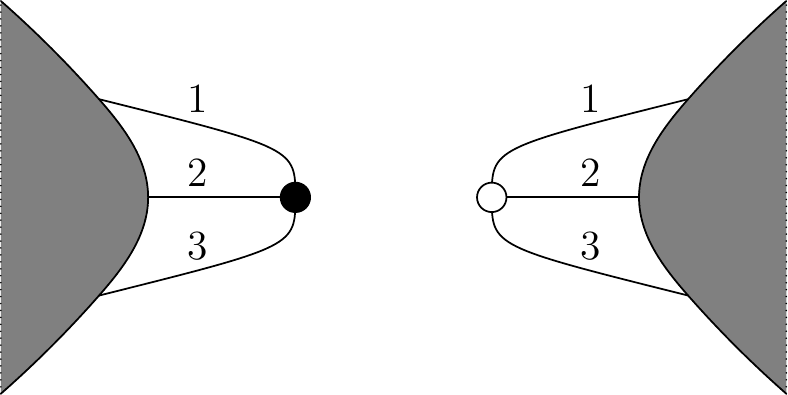}}}
\end{subfigure}
\begin{subfigure}[A 2-cut.]{
\parbox{4cm}{\includegraphics[width=4cm]{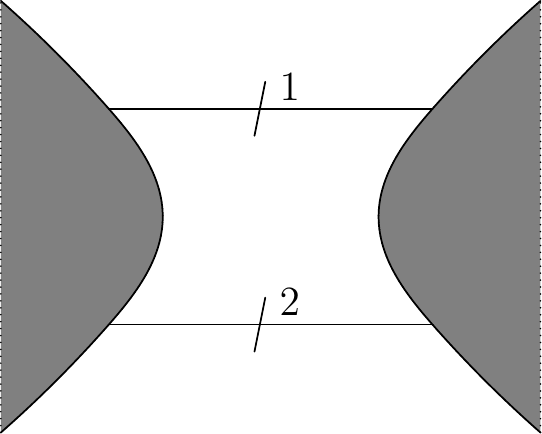}}
$\quad\rightarrow\quad$\parbox{6cm}{\includegraphics[width=6cm]{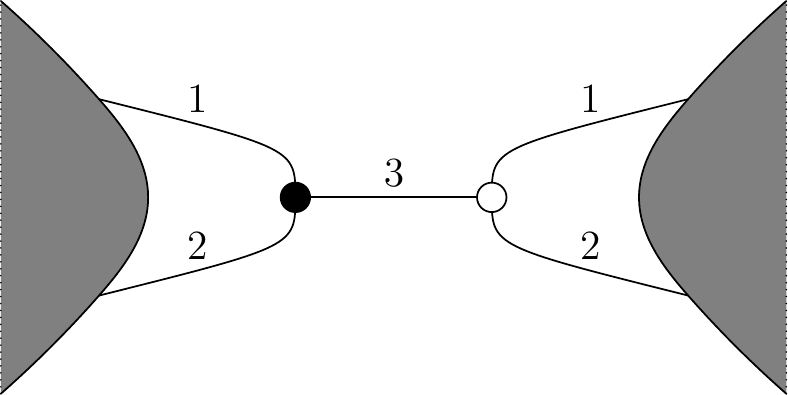}}}
\end{subfigure}
\caption{Examples of cut operations.}
\end{figure}
Substituting the bubble expansion \eqref{couplingstensor:eq} into the f\/low equation~\eqref{Polchinskitensor:eq}, we arrive at the system of dif\/ferential equations governing the bubble couplings. Thus, the evolution of the couplings reads
\begin{gather*}
\frac{\partial\lambda_{\cal B}}{\partial t}=\mathfrak{K}\sum_{k=0}^{D}
\sum_{k\text{-cut } c}
N^{D-k}
\lambda_{B_{c}}-
\mathfrak{K}\sum_{D\text{-cut }c\atop
{\kappa({\cal B}_{c})>\kappa({\cal B})}
}
\sum_{{\cal B}',{\cal B}''\atop
{\cal B}_{c}={\cal B}'\cup {\cal B}'',\, v\in{\cal B}',\, \overline{v}\in{\cal B}''}
\lambda_{\cal B'}
\lambda_{\cal B''},
\end{gather*}
In this equation, the sum runs over $k$-cuts $c$ for $0\leq k\leq D$. The second term involves a~sum\-mation over $D$-cut that increases the number of connected components $\kappa({\cal B})$ of ${\cal B}$ and over ways of writing ${\cal B}_{c}$ as a disjoint union of a bubble~${\cal B}'$ containing $v$ and a bubble ${\cal B}''$ containing~$\overline{v}$. Finally, let us emphasize that even if the initial potential~$V_{0}$ does not contain nonconnected bubbles, the latter are generated by the loop-like term in the f\/low equation. Furthermore, they play a key role in the large~$N$ limit.

 \subsection[Large $N$ melonic universality]{Large $\boldsymbol{N}$ melonic universality} \label{melonic:sec}

 In order to study the large $N$ limit, we make a change of variables in the couplings,
 \begin{gather*}
 {\lambda}_{{\cal B}}(t)=N^{\delta(\cal B)} u_{\cal B}(t).
 \end{gather*}
 $\delta({\cal B})$ has to be determined in such a way that the rescaled variables $u_{{\cal B}}$ are of order unity in~$N$. The latter play the role of the dimensionless variables in the Wilsonian approach to quantum f\/ield theory, with $\delta$ analogous to the canonical dimension of a coupling. This amounts to parametrising the potential as
 \begin{gather*}
V(t,T,\overline{T})=\sum_{{\cal B}}\frac{N^{\delta(\cal B)}u_{\cal B}(t)}{C_{\cal B}}\,\operatorname{Tr}_{\cal B}(\overline{T},T).
\end{gather*}
This change of variables factorizes the leading order behaviour of the bubble couplings in the large $N$ limit.

Since the covariance involves the dipole and the latter could as well be considered as an interaction, invariance under shifts of part of the covariance into the interaction requires that the constant $\mathfrak{K}$ be written as $\mathfrak{K}=N^{\delta(\includegraphics[width=0.5cm]{Krajewski-Fig01a})}$. In a more general discussion, a multiplicative factor that admits a~f\/inite limit when $N\rightarrow\infty$ could be included. For our purposes, this is irrelevant and we choose this constant to be~1. In the renormalisation group language, this number can be considered as a f\/inite, cut-of\/f independent wave function renormalisation and can be removed by a constant rescaling of the f\/ields.

The couplings $u_{{\cal B}}$ are f\/inite at large $N$ if the obey a renormalisation group equation involving only negative powers of~$N$,
\begin{gather*}
\frac{\partial u_{{\cal B}}}{\partial t}=
\beta_{0} ( \{u_{\cal B '} \} )+\frac{1}{N}
\beta_{1} ( \{u_{\cal B'} \} )+\frac{1}{N^{2}}\beta_{2} ( \{u_{\cal B'} \} )+\cdots,
\end{gather*}
so that in the large $N$ limit the couplings $u_{{\cal B}}$ are f\/inite and solely determined by~$\beta_{0}$. To de\-ter\-mine~$\delta({\cal B})$ such that this is the case, let us assume that it is of the form
\begin{gather*}
\delta({\cal B})=\alpha+\beta\kappa({\cal B})+\gamma v({\cal B}),
\end{gather*}
where $\kappa({\cal B})$ is the number of connected components of the bubble ${\cal B}$ and $v({\cal B})$ its total number of vertices. It is not necessary to introduce the number of edges since $e({\cal B})=\frac{D}{2}v({\cal B})$, but more ref\/ined numbers could be used, like the number of faces. Then, the renormalisation group equation reads
\begin{gather*}
\frac{\partial u_{\cal B}}{\partial t} =\sum_{k=0}^{D}
\sum_{k\text{-cut } c}
N^{\delta({\cal B}_{c})+D-k-\delta({\cal B})
-\delta(\includegraphics[width=0.5cm]{Krajewski-Fig01a})}
u_{B_{c}}\\
\hphantom{\frac{\partial u_{\cal B}}{\partial t} =}{} -
\sum_{D\text{-cut }c\atop
{\kappa({\cal B}_{c})>\kappa({\cal B})}
}
\sum_{{\cal B}',{\cal B}''\atop
{\cal B}_{c}={\cal B}'\cup {\cal B}'',\, v\in{\cal B}',\, \overline{v}\in{\cal B}''}
N^{\delta({\cal B}')+\delta({\cal B}'')-\delta({\cal B})-\delta(\includegraphics[width=0.5cm]{Krajewski-Fig01a})}
u_{\cal B'}
u_{\cal B''},
\end{gather*}
Note that for the dipole $\delta(\includegraphics[width=0.5cm]{Krajewski-Fig01a})=\alpha+\beta+2\gamma$.
Then, the tree-like term (second term in the previous equation) is independent of $N$ because $v({\cal B})=v({\cal B}')+v({\cal B}'')-2$ and $\kappa({\cal B})=\kappa({\cal B}')+\kappa({\cal B}'')-1$. The power of~$N$ in the loop-like term (f\/irst term in the previous equation) can be written as
\begin{align}
\delta({\cal B}_{c})+D-k-\delta({\cal B})
-\delta(\includegraphics[width=0.5cm]{Krajewski-Fig01a})&=D-\alpha-k-\beta\kappa({\cal B},c),
\end{align}
where $\kappa({\cal B},c)$ is def\/ined as the number of connected components of ${\cal B}$ containing edges of the cut, except for a
{$D$-cut} that increases the number of connected components of ${\cal B}$, in which case $\kappa({\cal B},c)=0$. This implies that $\kappa({\cal B}_{c})=\kappa({\cal B})-\kappa({\cal B},c)+1$. This exponent is negative if we choose $\alpha=D$ and $\beta=-1$ with $\gamma$ arbitrary\footnote{We thank Nicolas Dub for pointing this to us.}, but other choices are possible, see the discussion in~\cite{Bonzom:2015axa}. Then, the renormalisation group equation is
\begin{gather}
\frac{\partial u_{\cal B}}{\partial t}=\sum_{k=0}^{D}
\sum_{k\text{-cut } c}
N^{\kappa({\cal B},c)-k}
u_{B_{c}}-
\sum_{D\text{-cut }c\atop
{\kappa({\cal B}_{c})>\kappa({\cal B})}
}
\sum_{{\cal B}',{\cal B}''\atop
{\cal B}_{c}={\cal B}'\cup {\cal B}'',\,v\in{\cal B}', \,\overline{v}\in{\cal B}''}
u_{\cal B'}
u_{\cal B''}. \label{rescaledflow:eq}
\end{gather}
We always have $k\geq\kappa({\cal B},c)$ and $k= \kappa({\cal B},c)$ if and
only if all edges of the cut belong to dif\/ferent connected components of ${\cal B}$. In particular, this holds for any 0-cut and any 1-cut. The evolution equations for low order bubble couplings are given in Appendix~\ref{example3:sec} for $D=3$ and in Appendix~\ref{example4:sec} for $D=4$.

Since the exponent of $N$ in~\eqref{rescaledflow:eq} is negative, the limit $N\rightarrow\infty$ can be taken. For matrices ($D=2$), the large~$N$ limit is equivalent to the equation derived in~\cite{Gurau:2010ts}.

 In this regime, only $D$-cuts that disconnect the graph or $k$-cuts with each edge in distinct connected components (including the 0-cuts) survive. Staring with the empty bubble coup\-ling~$u_{\varnothing}$, that corresponds to $\log Z$, only melonic bubbles are generated when solving iteratively~\eqref{rescaledflow:eq} in powers of $t-t_{0}$. Recall that a~bubble~${\cal M}$ is said to be melonic if, for every white vertex~$v$, there is a black vertex $\overline{v}$ such that the removal of~$v$ and $\overline{v}$ increases the number of connected components by~$D-1$.

\begin{figure}[t]\centering

\begin{subfigure}[Melonic bubbles.]{
\parbox{2cm}{\includegraphics[width=2cm]{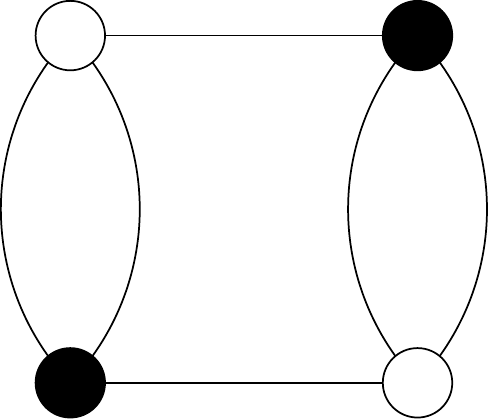}}
\quad
\parbox{2cm}{\includegraphics[width=2cm]{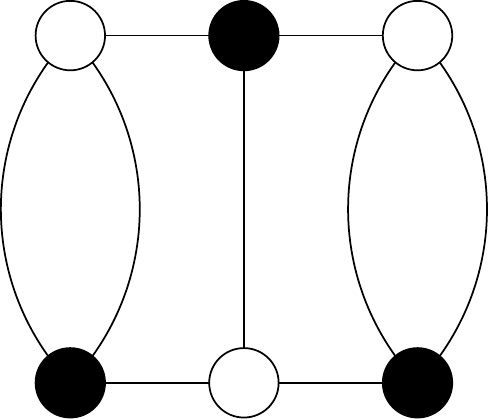}}
\quad
\parbox{2cm}{\includegraphics[width=2cm]{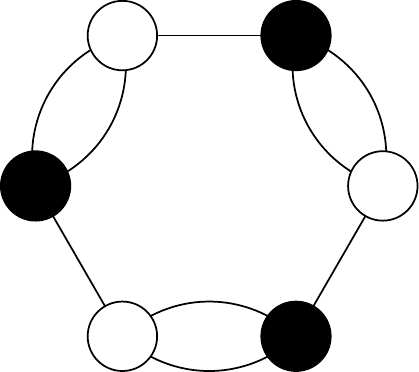}}}
\end{subfigure}\qquad\qquad
\begin{subfigure}[Non melonic bubbles.]
{\parbox{2cm}{\includegraphics[width=2cm]{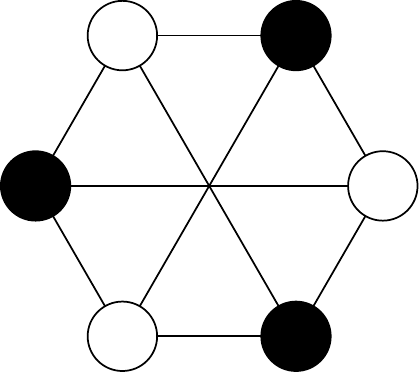}}
\quad
\parbox{2cm}{\includegraphics[width=2cm]{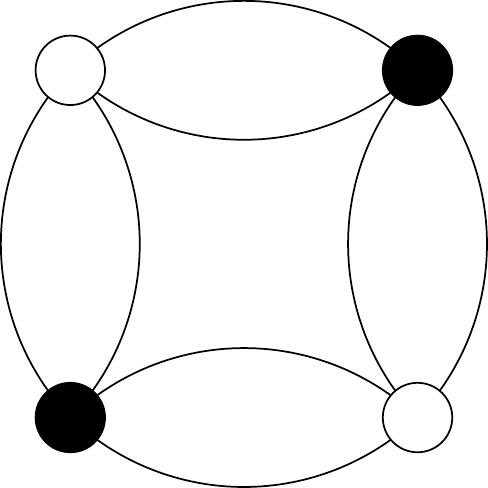}}}
\end{subfigure}
\caption{Melonic and nonmeloinc bubbles for $D=3$ and $D=4$.}
\label{melon:fig}
\end{figure}

Consequently, there is a kind of melonic universality: Whatever the initial couplings $u_{\cal B}(t_{0})$ are, the large $N$ partition function only depends on the melonic couplings $u_{\cal M}(t_{0})$. Thus, as far as the partition function is concerned, nonmelonic bubbles are irrelevant. The partition function is all we need to compute the expectation value of a bubble since
\begin{gather*}
\big\langle \operatorname{tr}_{\cal B}(T,\overline{T})\rangle =
-\frac{1}{N^{D-\kappa({\cal B})}}\frac{\partial }{\partial u_{\cal B}(t_{0})}
\log Z =N^{\kappa({\cal B})}
\frac{\partial u_{\varnothing}(t)}{\partial u_{\cal B}(t_{0})},
\end{gather*}
since $\log Z=-N^{D} u_{\varnothing}(t)$. If ${\cal B}$ is melonic, the derivative $\frac{\partial u_{\varnothing}(t)}{\partial u_{\cal B}(t_{0})}$ is of order unity. If~${\cal B}$ is not melonic, then it is suppressed by negative powers $N$. Taking the second derivative of $\log Z$ shows that the expectation value of a product of melonic bubbles factorizes at leading order. The actual computation of this expectation value is performed in Section~\ref{melonicdominance:sec}. Note that the melonic universality holds for any $D$, including the matrix model case $D=2$, for which all interactions are melonic. However, the Gau{\ss}ian ansatz presented in Section~\ref{melonicdominance:sec} does not hold for matrices. These techniques also allow to show the large $N$ Gau{\ss}ian nature of the correlation functions for random tensors, see~\cite{Krajewski:2015clk}.

Finally, let us emphasize that our treatment is perturbative: The partition and the expectation values of observables are considered as formal power series in the initial couplings $u_{\cal B}(t_{0})$ and the large $N$ limits holds order by order. It is not excluded that small $1/N$ contributions add up and alter the scaling exponents at a~nonperturbative level. This could happen, for instance, by resuming contributions to the dipole graph, thus leading to a nontrivial wave function renormalisation.

\subsection{Group f\/ield theory}\label{GFT:sec}

We consider now tensorial group f\/ield theories, analogous to tensor models, provided we replace rank~$D$ tensors by functions over~$D$ copies of a~group~$G$, \begin{gather*}
T_{i_{1},\dots,i_{D}}\rightarrow
\Phi(g_{1},\dots,g_{D}),\qquad
\overline{T}_{\overline{i}_{1},\dots,\overline{i}_{D}}\rightarrow
\overline\Phi(\overline g_{1},\dots,\overline g_{D}).
\end{gather*}
As already emphasized, $\overline{g}$ is not the complex conjugate of $g$ but an independent group element that is an argument of the complex conjugate f\/ield $\overline\Phi$. The covariance includes both a UV cut of\/f~$\Lambda_{0}$ and an IR one~$\Lambda$ (see~\eqref{covarianceGFT:eq}), and depends on the~$D$ products of group elements~$g_{i}\overline{g}_{i}^{-1}$,
\begin{gather*}
C\big(\Lambda,\Lambda_{0},\big\{g_{i}\overline{g}_{i}^{-1}\big\}\big)=\int_{\frac{1}{\Lambda_{0}^{2}}}^{\frac{1}{\Lambda^{2}}}d\alpha K\big(\alpha,\big\{g_{i}\overline{g}_{i}^{-1}\big\}\big),
\end{gather*}
so that
\begin{gather*}
\Lambda\frac{\partial C}{\partial\Lambda}\big(\Lambda,\Lambda_{0},\big\{g_{i}\overline{g}_{i}^{-1}\big\}\big)=
-\frac{2}{\Lambda^{2}}
 K\left(\frac{1}{\Lambda^{2}},\big\{g_{i}\overline{g}_{i}^{-1}\big\}\right).
\end{gather*}
In the following general discussion, $K(\alpha,\{g_{i}\overline{g}_{i}^{-1}\})$ is left unspecif\/ied. It is often convenient to take it as a product of $D$ heat kernels averaged over the group. The general interaction can be expanded in terms of bubble couplings (see~\eqref{GFTpotential:eq}),
\begin{gather*}
V(\Lambda,\Phi,\overline{\Phi})=
\sum_{{\cal B}}\frac{1}{C_{\cal B}}
\int\prod_{e} dg_{v(e),c(e)}d\overline{g}_{\overline{v}(e),c(e)} \\
{}\times \lambda_{\cal B}\big(\Lambda,\big\{
g_{v(e),c(e)}(\overline{g}_{\overline{v}(e),c(e)})^{-1}
\big\}\big)
\!\!\!\! \prod_{\overline{v}\atop \text{black vertices }} \!\!\!\! \overline{\Phi}(\overline{g}_{\overline{v},1},\dots,\overline{g}_{\overline{v},D})
 \!\!\!\!\prod_{{v}\atop \text{white vertices }} \!\!\!\! {\Phi}(g_{{v},1},\dots,g_{{v},D}),
\end{gather*}
where the bubble couplings only depend on $
g_{v(e),c(e)}\overline{g}^{-1}_{\overline{v}(e),c(e)}$ because of the global translation inva\-riance.

For group f\/ield theories, Polchinski's exact renormalisation group equation reads (see the tensor model analogue \eqref{Polchinskitensor:eq})
\begin{gather}
\Lambda\frac{\partial V}{\partial \Lambda}=-\frac{2}{\Lambda^{2}}
\int\prod_{1\leq i\leq D}dg_{i}d\overline{g}_{i}\nonumber\\
\hphantom{\Lambda\frac{\partial V}{\partial \Lambda}=}{}\times
K\left(\frac{1}{\Lambda^{2}},\big\{g_{i}\overline{g}_{i}^{-1}\big\}\right)\left(-
\frac{\delta V}{\delta \Phi( \{g_{i} \})}\frac{\delta V}{\delta \overline{\Phi}( \{\overline{g}_{\overline{i}} \})}+\frac{\delta^{2}V}{\delta{\Phi}( \{{g}_{i} \})\delta\overline{\Phi}( \{\overline{g}_{i} \})}\right).
\label{PolchinskiGFT:eq}
\end{gather}
Let us determine the system of equations satisf\/ied by the bubble couplings $\lambda_{\cal B}$. These involve $k$-cuts $c$ on the edges of ${\cal B}$. Recall that a $k$-cut is a subset of $k$ edges of distinct colours, with $i_{1},\dots,i_{D-k}$ the colours of the edge not in the cut. We cut these edges, attach two new vertices $\overline{v}_{c}$ and $v_{c}$ and complete the graph by connecting these vertices by $D-k$ edges $e_{i_{1}},\dots,e_{i_{k}}$ carrying the colours not in the cut. We denote by ${\cal B}_{c}$ the resulting bubble and, if $k=D$ and if ${\cal B}_{c}$ has one more connected component than ${\cal B}$, ${\cal B}^{'}_{c}$ and ${\cal B}^{''}_{c}$ are any bubbles such that ${\cal B}_{c}={\cal B}^{'}_{c}\cup {\cal B}^{''}_{c}$ with $v_{c}\in{\cal B}^{'}_{c}$ and $\overline{v}_{c}\in{\cal B}^{''}_{c}$.

Then, inserting the expansion of $V$ in terms of bubble couplings \eqref{GFTpotential:eq}
 in the exact renormalisation group equation \eqref{PolchinskiGFT:eq} and identifying the contribution of a bubble on both sides lead to
 \begin{gather}
\Lambda\frac{\partial\lambda}{\partial \Lambda} \big(\big\{ g_{v(e),c(e)}(\overline{g}_{\overline{v}(e),c(e)})^{-1}\big\}\big)=
-\frac{2}{\Lambda^{2}}
\int\prod_{1\leq i\leq D}dg_{i}d\overline{g}_{i}K\left(\frac{1}{\Lambda^{2}},\big\{g_{i}\overline{g}_{i}^{-1}\big\}\right)\nonumber\\
\hphantom{\Lambda\frac{\partial\lambda}{\partial \Lambda} \big(\big\{ g_{v(e),c(e)}(\overline{g}_{\overline{v}(e),c(e)})^{-1}\big\}\big)=}{}\times
\Bigg\{
\sum_{0\leq k\leq D}
\sum_{c}
\lambda_{{\cal B}_{c}}\big(\Lambda,\big\{
g_{v(e),c(e)}(\overline{g}_{\overline{v}(e),c(e)})^{-1}\big\}_{e\in{\cal B}}, g_{i},\overline{g}_{i}\big)\nonumber\\
\hphantom{\Lambda\frac{\partial\lambda}{\partial \Lambda} \big(\big\{ g_{v(e),c(e)}(\overline{g}_{\overline{v}(e),c(e)})^{-1}\big\}\big)=}{}
-\!\!\!
\sum_{c \text {$D$-cut}\atop \text{ disconnecting ${\cal B}$}}
\sum_{{\cal B}_{c}^{'}\cup{\cal B}_{c}^{''}}\!\!
\lambda_{{\cal B}_{c}^{'}}
\big(\Lambda,\big\{
g_{v(e),c(e)}(\overline{g}_{\overline{v}(e),c(e)})^{-1}\big\}_{e\in{\cal B}_{c}^{'}}, g_{i}\big)\nonumber\\
\hphantom{\Lambda\frac{\partial\lambda}{\partial \Lambda} \big(\big\{ g_{v(e),c(e)}(\overline{g}_{\overline{v}(e),c(e)})^{-1}\big\}\big)=}{}
\times
\lambda_{{\cal B}_{c}^{''}}
\big(\Lambda,\big\{
g_{v(e),c(e)}(\overline{g}_{\overline{v}(e),c(e)})^{-1}\big\}_{e\in{\cal B}_{c}^{''}}, \overline{g}_{i}\big)
\Bigg\}.
\label{PolchinskiGFTbubble:eq}
\end{gather}
This expression is not very illuminating, but interesting consequences can be drawn from it in specif\/ic models. In the next two sections, we investigate the case of Abelian models with closure constraints and $\text{SU}(2)$ models in the spin network basis.

\subsection{Scaling and renormalisation for Abelian models with closure constraint}

In this section, we consider Abelian models based on the group $\text{U(1)}^{d}$. Elements of $\text{U(1)}$ are written as $g=\exp{2\text{i}\pi\frac{\theta}{L}}$ with $\theta\in[0,L]$, with $L$ being a length. The covariance is of heat kernel type
\begin{gather*}
C(\Lambda, \Lambda_{0},\{\theta_{i}-\overline{\theta}_{i}\})
=
\int_{\frac{1}{\Lambda_{0}^{2}}}^{\frac{1}{\Lambda^{2}}}d\alpha
\sum_{ \{p_{i}\}\in{\frac{1}{L}\mathbb Z}^{dD}}
\exp\left(-\left\{\alpha\sum_{1\leq i\leq D}p_{i}^{2}+\text{i}\sum p_{i}(\theta_{i}-\overline\theta_{i})\right\}\right)
\delta_{\sum p_{i},0},
\end{gather*}
where we have enforced the condition $\sum_{i} p_{i}=0$ so that the closure constraint is fulf\/illed.

The exact renormalisation group equation written in terms of bubble couplings \eqref{PolchinskiGFTbubble:eq} is
\begin{gather}
\Lambda\frac{\partial \lambda_{\cal B}(\left\{p_{e}\right\})}{\partial \Lambda}
=
-\frac{2}{\Lambda^{2}}\sum_{k=0}^{D}
\sum_{k\text{-cut } c}
\sum_{ \{p_{l} \}_{l\notin c}}
\exp\left(-\frac{1}{\Lambda^{2}}\left\{
\sum_{i=1}^{D}p_{i}^{2}
\right\}\right)
\lambda_{B_{c}}
(\{p_{e}\}_{e\in{\cal B}},
\{p_{l}\}_{l\notin c}
)\delta_{\sum\limits_{i=1}^{D}p_{i},0}
\nonumber
\\
{} +\frac{2}{\Lambda^{2}}\!\!\sum_{D\text{-cut }c\atop
\kappa({\cal B}_{c})>\kappa({\cal B})}
\sum_{{\cal B}',{\cal B}''\atop
{{\cal B}_{c}={\cal B}'\cup {\cal B}'',\atop v\in{\cal B}',\, \overline{v}\in{\cal B}''}}\!\!\!\!\!
\exp\left(\!-\frac{1}{\Lambda^{2}}\left\{
\sum_{i=1}^{D}p_{i}^{2}
\right\}\!\right)
\lambda_{\cal B'}(\{p_{e}\}_{e\in {\cal B'}})
\lambda_{\cal B''}
(\{{p}_{e}\}_{e\in {\cal B}''})
\delta_{\sum\limits_{i=1}^{D}p_{i},0}.\!\!\!\label{AbelianERGE:eq}
\end{gather}
This equation is the group f\/ield theory analogue of \eqref{ERGEcorrelation:eq} in the framework of quantum f\/ield theory. Note that because of the closure constraint, the bubble couplings are only def\/ined on subspaces of momentum space such that the sum of momenta incident to any vertex is~0. We further assume $L$ to be large enough ($1/L\ll \Lambda$) so that momenta can be treated as continuous variables. Then we can trade sums for integrals
\begin{gather*}
\sum_{p} \rightarrow L^{d}\int dp\qquad
\text{and}\qquad
\delta_{\sum{p_{i}},0} \rightarrow L^{-d}\delta\left(\sum{p_{i}}\right).
\end{gather*}
Otherwise, a more precise but cumbersome analysis can be performed using the Poisson resummation. Finally, we also allow derivatives of the bubble couplings with respect to momenta and impose rotational invariance in momentum space. This last hypothesis is fulf\/illed if we assume that the initial couplings are invariant under rotation since the propagator is.

To understand the large $\Lambda$ behaviour, we repeat the steps followed in Section~\ref{melonic:sec}. Let us introduce the analogue of dimensionless bubble couplings $u_{\cal B}$ def\/ined by
\begin{gather*}
\lambda_{\cal B}(\{p_{e}\},\Lambda)=
\Lambda^{\delta({\cal B})}u_{\cal B}(\{q_{e}\},\Lambda) \qquad\text{with}\qquad p_{e}={\Lambda}{q_{e}}.
\end{gather*}
$\delta({\cal B})$ is not the dimension as would be derived by rescaling all lengths, $L$ included. It is merely a scaling dimension for large $\Lambda$ that encodes the dominant behaviour of~$\lambda_{{\cal B}}$.

{\allowdisplaybreaks
In terms of these new couplings, the exact renormalisation group equation~\eqref{AbelianERGE:eq} is
\begin{gather*}
\Lambda\frac{\partial u_{\cal B}(\{q_{e}\})}{\partial \Lambda}
=-\delta({\cal B})\,u_{\cal B}(\{q_{e}\})+\sum_{e}q_{e}\frac{\partial u_{\cal B}(\{q_{e}\})}{\partial q_{e}}
-2\sum_{k=0}^{D}
\sum_{0\leq k\leq D\atop c\,k\text{-cut } }
\frac{\Lambda^{\delta({\cal B}_{c})-\delta({\cal B})-2+d(D-k)-d}}{L^{d-d(D-k)}}\\
{}\times \int \prod_{l\notin c}dq_{l}
\exp\left\{-\left(
\sum_{i}q_{i}^{2}\right)\right\}
u_{B_{c}}
(\{q_{e}\}_{e\in{\cal B}},
\{q_{l}\}_{l\notin c}
) \delta\left(\sum_{i}q_{i}\right)
\\
{}+2\sum_{D\text{-cut }c\atop
\kappa({\cal B}_{c})>\kappa({\cal B})}
\sum_{
{\cal B}_{c}={\cal B}'\cup {\cal B}'',\atop v\in{\cal B}',\, \overline{v}\in{\cal B}''}\!\!\!
 \frac{
\Lambda^{\delta({\cal B}')+\delta({\cal B}'')-\delta({\cal B})-2}}{L^{d}}
\exp\left\{-\left(
\sum_{i}q_{i}^{2}\right)\right\}
u_{\cal B'}(\{q_{e}\}_{e\in {\cal B'}})
u_{\cal B''}
(\{{q}_{e}\}_{e\in {\cal B}''}).
\end{gather*}
Suppose that this equation can be written with only negative powers of $\Lambda$,
\begin{gather*}
\Lambda\frac{\partial u_{{\cal B}}}{\partial \Lambda}=
\beta_{0}(\{u_{\cal B}'\})+\frac{1}{\Lambda}
\beta_{1}(\{u_{\cal B}'\})+\frac{1}{\Lambda^{2}}\beta_{2}(\{u_{\cal B}'\})+\cdots.
\end{gather*}
Then, the large $\Lambda$ behaviour is entirely governed (at least in perturbation theory) by~$\beta_{0}$. To this aim, let us search for a suitable scaling dimension for the bubble interactions in the form}
\begin{gather*}
\delta({\cal B})=\alpha+\beta\kappa({\cal B})+\gamma v({\cal B}).
\end{gather*}

The exponent in the tree-like term is
\begin{gather*}
\delta({\cal B}')+\delta({\cal B}'')-\delta({\cal B})-2=\alpha+\beta+2\gamma-2,
\end{gather*}
while for the loop like term it is
\begin{gather*}
\delta({\cal B}_{c})-\delta({\cal B})-2+d(D-k)-d=-dk-\beta\kappa({\cal B},c)+\beta+2\gamma-2+d(D-1).
\end{gather*}
We recall that $\kappa({\cal B},c)$ is def\/ined as the number of connected components of ${\cal B}$ containing edges of the cut, except for a
{$D$-cut} that increases the number of connected components of ${\cal B}$, in which case $\kappa({\cal B},c)=0$. It obeys $\kappa({\cal B}_{c})=\kappa({\cal B})-\kappa({\cal B},c)+1$.

Setting $\alpha=d(D-1)$, $\beta=-d$, and $2\gamma=-d(D-2)+2$, the exponent of the tree-like term vanishes while the loop-like one it is
$d(\kappa({\cal B},c)-k)$, thus always negative. Therefore, the couplings $u_{\cal B}$ are governed by a f\/low equation that only involves negative powers of $\Lambda$. This leads to a scaling dimension
\begin{gather*}
\delta({\cal B})= d(D-1)-d\kappa({\cal B})- (d(D-2)-2 ) \frac{v({\cal B})}{2}.
\end{gather*}
For connected bubbles, this is the scaling dimension found in \cite{Carrozza:2013wda} by Carrozza, Oriti and Rivasseau using multiscale analysis.

Assuming that the general picture presented in Section~\ref{Wilson:sec} for quantum f\/ield theory remains valid, perturbatively renormalisable interactions correspond to relevant and marginal interactions, or equivalently~$\delta({\cal B})\geq 0$. This leads us to only f\/ive values of $D$ and $d$ for which such interactions can be found. Indeed, if we also take into account the term associated with rescaling of momenta,
\begin{gather*}
\sum_{e}q_{e}\frac{\partial u_{\cal B}(\{q_{e}\})}{\partial q_{e}}
\end{gather*}
that is diagonalised by degree $n$ homogenous polynomials in $q_{e}$, renormalisable terms are characterised by
\begin{gather}
\delta({\cal B},n)=
(D-1)d-\kappa({\cal B}) d- [(D-2)d-2 ]\frac{v({\cal B})}{2}-n\geq 0.
\label{dimension}
\end{gather}
In particular, among all the renormalisable interactions, we always have the mass term $\delta_{\includegraphics[width=0.5cm]{Krajewski-Fig01a},0}=2$ and for the kinetic term $\delta_{\includegraphics[width=0.5cm]{Krajewski-Fig01a},2}=0$.

Let us end by listing all renormalisable interactions in Abelian models with a closure constraint. The condition $\delta({\cal B})\geq 0$ implies $v({\cal B})\leq 2+\frac{4}{d(D-2)-2}$. Non trivial interactions involve bubbles with at most four vertices, so that we are left with only two possibilities: $d(D-2)=4$ and $d(D-2)=3$, which leads to the following 5 renormalisable theories. This list, with connected interactions only, was f\/irst obtained in~\cite{Carrozza:2013wda} using the multiscale analysis of Feynman graphs.
\begin{itemize}\itemsep=0pt
\item $D=3$ and $d=4$ so that $\delta=8-4\kappa-v$. The renormalisable interactions is quartic and melonic
\begin{center}
\parbox{1cm}{\includegraphics[width=1cm]{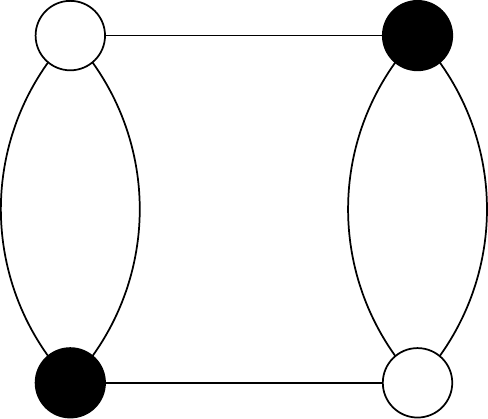}}
\,($\delta=0$).
\end{center}
The f\/ixed point structure of a non-Abelian version of this model has been studied in \cite{Carrozza:2014rya}.
\item $D=4$ and $d=2$ so that $\delta=6-2\kappa-v$. The renormalisable interactions are quartic
\begin{center}
\parbox{1cm}{\includegraphics[width=1cm]{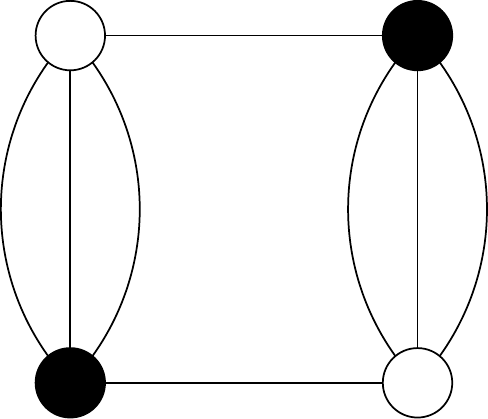}}
\,($\delta=0$),\quad
\parbox{1cm}{\includegraphics[width=1cm]{Krajewski-Fig10b2}}
\,($\delta=0$).
\end{center}
Note that the second interaction is not melonic (called necklace in~\cite{Bonzom:2015axa}).

\item $D=6$ and $d=1$ so that $\delta=5-\kappa-v$. The renormalisable interactions are quartic

\begin{center}
\parbox{1.2cm}{\includegraphics[width=1.2cm]{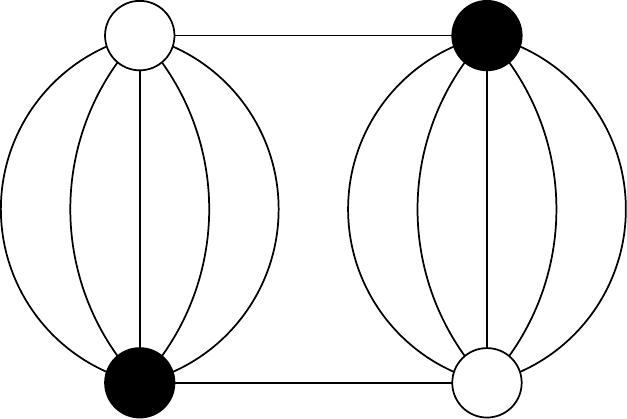}}
\,($\delta=0$),\quad
\parbox{1cm}{\includegraphics[width=1cm]{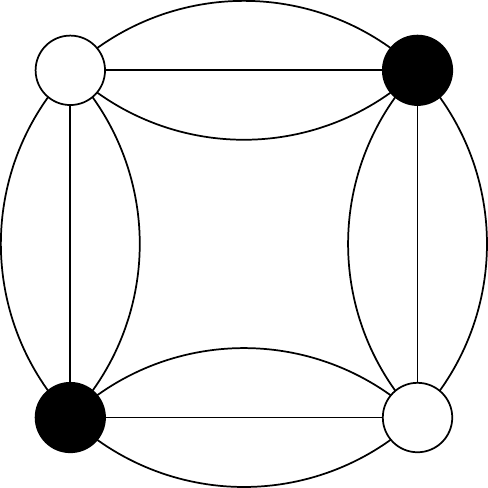}}
\,($\delta=0$),
\quad
\parbox{1cm}{\includegraphics[width=1cm]{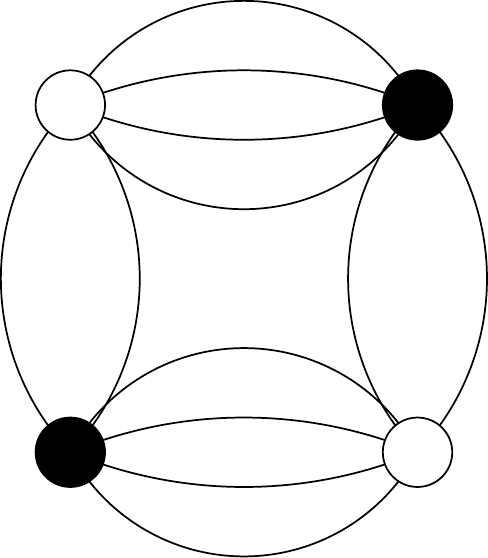}}
\,($\delta=0$).
\end{center}

The last two interactions are not melonic. This model was shown to be renormalisable in~\cite{Samary:2012bw,Lahoche:2015ola} and its f\/ixed point structure was further investigated in~\cite{Benedetti:2015yaa}.

\item $D=3$ and $d=3$ so that $\delta=6-3\kappa-v/2$. The renormalisable interactions are quartic and sextic
\begin{center}
\parbox{1cm}{\includegraphics[width=1cm]{Krajewski-Fig-page23a}}
\,($\delta=1$),\quad
\parbox{1cm}{\includegraphics[width=1cm]{Krajewski-Fig10a2}}
\,($\delta=0$),
\quad
\parbox{1cm}{\includegraphics[width=1cm]{Krajewski-Fig10a3}}
\,($\delta=0$),\quad
\parbox{1cm}{\includegraphics[width=1cm]{Krajewski-Fig10b1}}
\,($\delta=0$).
\end{center}
The f\/irst term is in fact superrenormalisable ($\delta>0$) and the last one is not melonic. Its non-Abelian counterpart is the tensorial $\text{SU(2)}$ group f\/ield theory in $D=3$ which corresponds to three-dimensional Euclidian quantum gravity without the cosmological constant, f\/irst shown to be renormalisable in~\cite{Carrozza:2013wda}.

\item $D=5$ and $d=1$ so that $\delta=4-\kappa-v/2$. The renormalisable interactions are quartic and sextic. We f\/irst have the melonic ones

\begin{center}
\parbox{1.5cm}{\includegraphics[width=1.5cm]{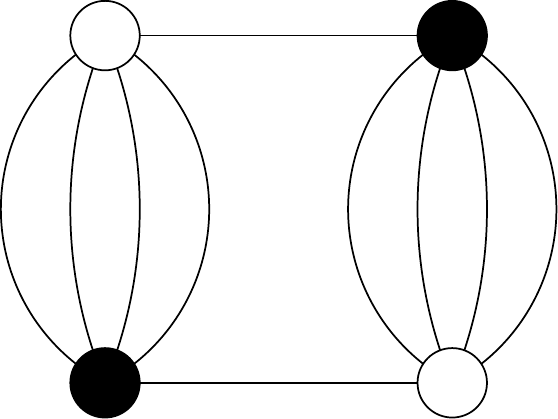}}
\,($\delta=1$),\quad
\parbox{1.5cm}{\includegraphics[width=1.5cm]{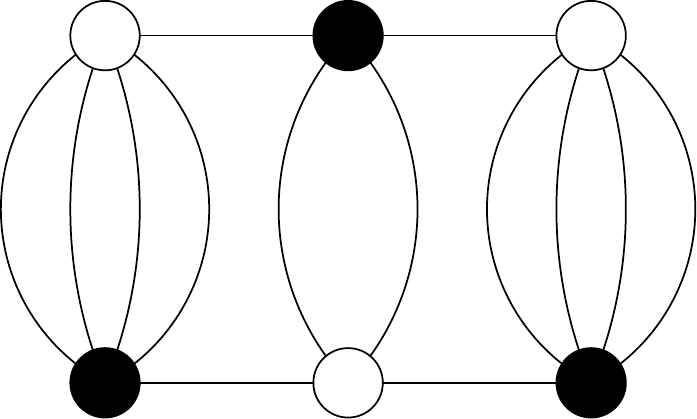}}
\,($\delta=0$),
\quad
\parbox{1cm}{\includegraphics[width=1cm]{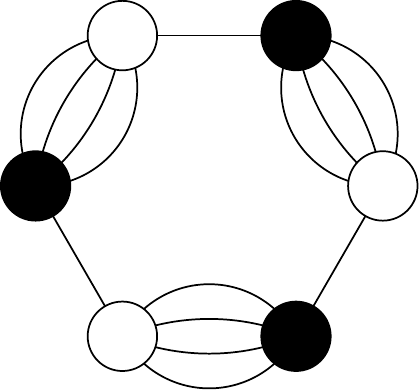}}
\,($\delta=0$),\quad
\parbox{1.5cm}{\includegraphics[width=1.5cm]{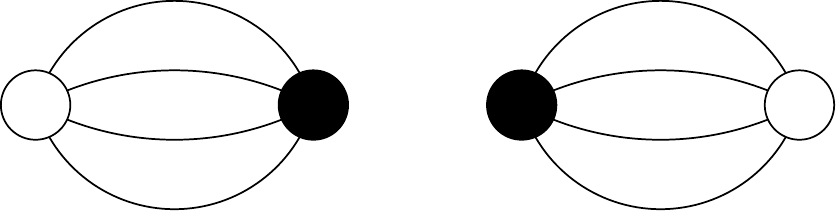}}
\,($\delta=0$).
\end{center}
The last interaction is not connected ($\kappa=2$). This model, with the nonconnected interaction, was shown to be renormalizable in~\cite{Samary:2012bw}. Besides, there are also nonmelonic renormalisable interactions.
\end{itemize}

Note that all these bubble couplings with $v\geq 4$ have dimension~0 or~1, so that there is no possibility of adding a derivative coupling. Indeed, the latter corresponds to a term with $n>0$ in \eqref{dimension}, which would lead to a negative dimension, the case $n=1$ being forbidden by rotational symmetry. Of course, we could always add quadratic terms (mass and kinetic terms) to the list of interaction. Then, the kinetic term can play the role of a derivative coupling.

It is also worthwhile to notice that some of the renormalisable interactions are not necessary melonic. As for unitarily invariant random tensors, nonmelonic interactions appear with a~negative power of $\Lambda$ in the f\/low of renormalisable melonic interactions. Therefore, they do not require renormalisation if the bare theory only contains melonic interactions. However, if the bare theory contains nonmelonic renormalisable interactions, the latter requires renormalisation.

\subsection[$\text{SU}(2)$ models in the spin network basis]{$\boldsymbol{\text{SU}(2)}$ models in the spin network basis}

In the $\text{SU}(2)$ case which is involved in many spin-foam models, it is possible to expand all the bubble couplings on spin networks, see~\cite{Krajewski:2014toa}, rather than just bubbles. For the group $\text{SU}(2)$, spin network observables provide a nice basis. We work with the heat kernel
\begin{gather*}
H\big(\alpha, g\overline{g}^{-1}\big)=
\sum_{j}(2j+1)\exp\big\{{-}\alpha j(j+1)\big\} \sum_{-j\leq m\leq j}{\cal D}^{j}_{mm}\big(g\overline{g}^{-1}\big).
\end{gather*}
The covariance is obtained by averaging over the group
\begin{gather*}
C\big(\Lambda,\Lambda_{0},\big\{g_{i}\overline{g}_{i}^{-1}\big\}\big)=
\sum_{j_{1}\dots j_{D}\,\text{spins}\atop
\iota \,\text{intertwiner}}\Bigg(
\sum_{m_{1},\dots,m_{D}\atop
\overline{m}_{1},\dots,\overline{m}_{D}}
\iota_{m_{1},\dots,m_{D}}
{\cal D}^{j_{1}}_{m_{1}\overline{m}_{1}}\big(g_{1}\overline{g}_{1}^{-1}\big)
\cdots\\
{}\times {\cal D}^{j_{D}}_{m_{D}\overline{m}_{D}}
\big(g_{D}\overline{g}_{D}^{-1}\big)
\overline{\iota}_{\overline{m}_{1},\dots,\overline{m}_{D}}
 d_{j_{1}}\cdots d_{j_{D}}
 \exp\big({-}\alpha\big\{j_{1}(j_{1}+1)+\cdots+ j_{D}(j_{D}+1)\big\}\big)\Bigg),
\end{gather*}
where we performed the group averaging thanks to the relation
\begin{gather*}
\int dh {\cal D}_{m_{1}m'_{1}}^{j_{1}}(h)
\cdots {\cal D}_{m_{D}m'_{D}}^{j_{D}}(h)
=\sum_{\iota}\iota_{m_{1}\dots m_{D}}
\overline{\iota}_{m'_{1}\dots m'_{D}}.
\end{gather*}
Then, the spin network formulation of \eqref{PolchinskiGFTbubble:eq} is
\begin{gather*}
\Lambda\frac{\partial \lambda_{\cal B}(\{j_{e},\iota_{v}\})}{\partial \Lambda}
=
-\frac{2}{\Lambda^{2}}\sum_{0\leq k\leq D
\atop
c\, k\text{-cut }}
\sum_{\iota}
\sum_{\{j_{l}\}_{l\notin c}}
\prod_{l}d_{j_{l}}
\exp\left(-\frac{1}{\Lambda^{2}}\left\{
\sum_{i=1}^{D}j_{i}(j_{i}+1)
\right\}\right)\\
{}\times
\lambda_{B_{c}} (\{j_{e}\}_{e\in{\cal B}},
\{j_{l}\}_{l\notin c},\{i_{v}\}_{v\in{\cal B}},\iota)
+\frac{2}{\Lambda^{2}}\!\!\sum_{D\text{-cut }c\atop
\kappa({\cal B}_{c})>\kappa({\cal B})}\!\!
\sum_{\iota}\!\!\!
\sum_{{\cal B}',{\cal B}''\atop
{{\cal B}_{c}={\cal B}'\cup {\cal B}'', \atop v\in{\cal B}',\, \overline{v}\in{\cal B}''}}\!\!\!\!
\exp\left(-\frac{1}{\Lambda^{2}}\left\{
\sum_{i=1}^{D}j_{i}(j_{i}+1)
\right\}\right)\\
{}\times
\lambda_{\cal B'} ( \{j_{e''}\}_{e''\in {\cal B'}}\{\iota_{v'}\}_{v'\in {\cal B'}},\iota)
\lambda_{\cal B''}(\{j_{e''}\}_{e''\in {\cal B''}}\{\iota_{v''}\}_{v''\in {\cal B''}},\iota).
\end{gather*}
Let us emphasize the analogy with \eqref{AbelianERGE:eq}, the spin playing the role of the momenta. The constraint is implemented by the insertion of the interwiner~$\iota$ at the two vertices created by the cut. Consequently, it acts as a constraint since we can cut some edges only if their spins admit an intertwiner. For $D=3$ this constraint is the triangle inequality. It is also interesting to note the similarity with the tensor model case, with the factor $N^{D-k}$ replaced by the product over the $D-k$ edges not in the cut of the dimensions $d_{j_{l}}$.

\subsection{Wetterich's equation and nontrivial f\/ixed points}\label{truncations:sec}

The search for nontrivial f\/ixed points in group f\/ield theories have been investigated by several authors. We give here a very brief account of the rationale behind these works since they also make use of exact renormalisation group equations. We refer the reader to the original literature \cite{Geloun:2016qyb,Benedetti:2014qsa, Benedetti:2015yaa, Carrozza:2014rya} for the precise calculations and the phase diagrams.

 These works use Wetterich's equation \eqref{Wetterich:eq} expanded as
 \begin{gather}
\frac{\partial\Gamma}{\partial t} =
\operatorname{Tr}\left[C^{-1}KC^{-1}\left(1+C\frac{\partial^{2}\Gamma}{\delta\Phi\delta\overline{\Phi}}C
\right)^{-1}\right]\nonumber\\
\hphantom{\frac{\partial\Gamma}{\partial t}}{} =\sum_{n=0}^{\infty}(-1)^{n}
\operatorname{Tr}\left[C^{-1}KC^{-1}\underbrace{C\frac{\partial^{2}\Gamma}{\delta\Phi\delta\overline{\Phi}}\cdots C\frac{\partial^{2}\Gamma}{\delta\Phi\delta\overline{\Phi}}}_{\text{$n$ times}}
\right].\label{Wetterich:eq+}
\end{gather}
In this framework, a heat kernel type of covariance is used and the ef\/fective average action $\Gamma$ is expanded over bubble couplings as in~\eqref{GFTpotential:eq}. Then, \eqref{Wetterich:eq+} is expanded a~series of colour~$0$ one loop graphs whose vertices are the $D$-bubbles in the expansion of~$\Gamma$.

\begin{figure}[t] \centering
\includegraphics[width=10cm]{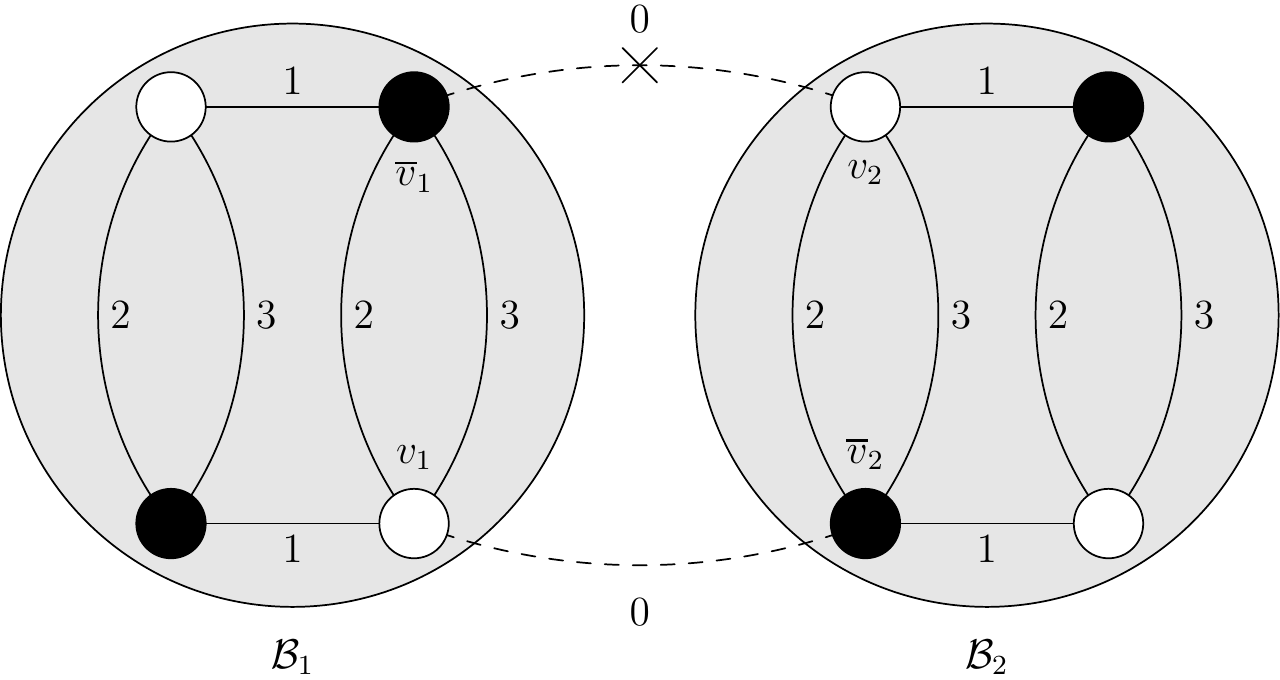}
\caption{A contribution to the evolution of a quartic melon.}
\label{Wetterich1:fig}
\end{figure}

To obtain the contribution to $\frac{\partial \lambda_{\cal B}}{\partial t}$, one has to look for all possibilities of writing ${\cal B}$ as
\begin{gather*}
{\cal B}=\big( {\cal B}_{1}\cup\cdots\cup {\cal B}_{n} \big)\bigg/\big(v_{1}\overline{v}_{2},v_{2}\overline{v}_{3},\dots, v_{n-1}\overline{v}_{n},v_{n}\overline{v}_{1}\big),
\end{gather*}
where $v_{i}$ and $\overline{v}_{i}$ are respectively a white and a black vertex in~${\cal B}_{i}$. For instance, $\frac{\partial}{\partial t} \lambda_{\parbox{0.5cm}{\includegraphics[width=0.5cm]{Krajewski-Fig-page23a}}}$ involves a~term quadratic in $\lambda_{\includegraphics[width=0.5cm]{Krajewski-Fig-page23a}}$, see Fig.~\ref{Wetterich1:fig}.
In order to write down the evolution equation of a coupling pertaining to a given bubble ${\cal B}$ in terms of all the bubbles in the chain ${\cal B}_{1},\dots,{\cal B}_{n}$, it is therefore necessary to f\/ind all such possible chains of bubbles. This is solved by the following algorithm, involving a sequence of $n$ cuts since they are the inverse of inverse operation of a~contraction of a~line of color~0.
\begin{enumerate}\itemsep=0pt
\item The f\/irst cut $c_{1}$ is arbitrary and results in ${\cal B}_{c1}$, with two new vertices $v_{1}$ and $\overline{v}_{1}$. This cut determines the insertion of the derivative of the covariance $K$, materialised by a cross in Fig.~\ref{Wetterich1:fig}.

\item The next cuts $c_{2},\dots,c_{n}$ must always increase the number of connected components (therefore they involve all $D$ colors) with the constraint that each new connected component contains~1 or~2 new vertices $v_{i}$ and $\overline{v}_{j}$. If the f\/irst cut has not increased the number of connected components, all the cuts have to be performed on the connected component of~${\cal B}_{c_{1}}$ containing~$v_{1}$ and $\overline{v}_{1}$; otherwise they may be performed anywhere.

\item Partition the set of connected components of~${\cal B}_{c_{1}\dots c_{n}}$ into $n$ sets, each of which containing one of the white vertices and one of the black vertices previously constructed in such a~way that joining corresponding vertices leads to a cycle of colour~$0$ lines.

\item The $n$ sets of the partition def\/ine the $n$ bubbles ${\cal B}_{1},\dots,{\cal B}_{n}$
\end{enumerate}

As is usual, for a given bubble there is a maximal value of $n$, if we exclude the trivial cuts involving all the $D$ edges incident to a given vertex. Except for the f\/irst cut, this kind of cut simply amounts to an insertion of a chain of dipoles which may be removed by a change in the propagator.

\begin{figure}[t]
\centering
\includegraphics[width=4cm]{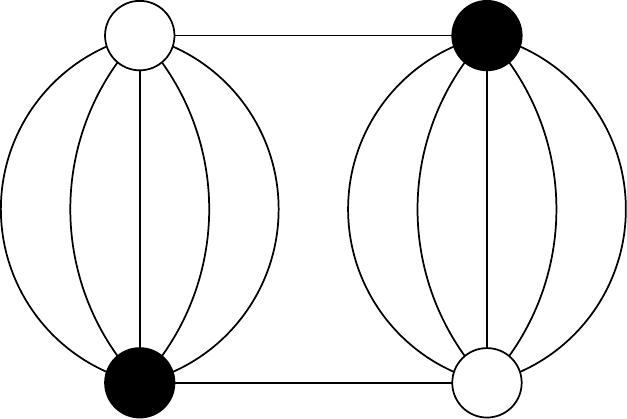}
\caption{The quartic melon in $D=6$.}
\label{quarticmelon:fig}
\end{figure}

To search for f\/ixed points, one usually resorts to truncations. It means that we select a~few couplings, usually including the mass $m^{2}\overline{\Phi}\cdot\Phi$ and the kinetic term $(Z-1)\overline{\Phi}\cdot\Delta\Phi$, both associated to the dipole graph, as well as some interactions, encoded in the bubble couplings, expanded around zero momentum (in Abelian theories). For instance, in the model considered by Benedetti and Lahoche in~\cite{Benedetti:2015yaa} in $D=6$ with the group $\text{U}(1)$, only the interaction associated to the quartic melon given in Fig.~\ref{quarticmelon:fig} has been retained.
 We focus now on this model and review brief\/ly their work. In this case, the f\/low equations reduce to the system
\begin{gather*}
\frac{\partial m^{2}}{\partial t} =\beta_{m^{2}}\big(Z,m^{2},\lambda,t\big),\qquad
\frac{\partial Z}{\partial t} =\beta_{Z}\big(Z,m^{2},\lambda,t\big),\qquad
\frac{ \partial\lambda}{\partial t} =\beta\big(Z,m^{2},\lambda,t\big).
\end{gather*}
The wave function renormalisation is not a coupling and one has to rescale the f\/ields as $\Phi\rightarrow Z^{-1/2}\Phi$ and $\overline{\Phi}\rightarrow Z^{-1/2}\overline{\Phi}$ in order to have f\/ixed points. Second, it is also necessary to rescale~$m$ and $\lambda$ according to their scaling dimension for large~$t$,
\begin{gather*}
{\lambda}=\text{e}^{t(6-D)}Z^{2}\overline{\lambda},\qquad
{m}^{2}=\text{e}^{2t}Z\overline{m}^{2}.
\end{gather*}
Note that $6-D$ is the scaling dimension of the bubble $\parbox{0.5 cm}{\includegraphics[width=0.5cm]{Krajewski-Fig-page23c}}$ found in equation \eqref{dimension}, with $\Lambda\propto \text{e}^{t}$. Finally, determining $Z$ in terms of $\overline{m}$ and $\overline{\lambda}$ leads to a system of equations
\begin{gather*}
\frac{\partial \overline{m}^{2}}{\partial t}=\overline{\beta}_{\overline{m}^{2}}\big(\overline{m}^{2},\overline{\lambda},t\big),\qquad
\frac{ \partial\overline{\lambda}}{\partial t}=\overline{\beta}\big(\overline{m}^{2},\overline{\lambda},t\big).
\end{gather*}
Although nonautonomous (i.e., still involving $t$) because of the f\/inite size of the group, there are f\/ixed points in the limit $t\rightarrow\infty$. Besides the Gau{\ss}ian one with $\overline{m}=0$ and $\overline{\lambda}=0$, there are two others f\/ixed points. One of them cannot be connected to the Gau{\ss}ian one by a trajectory because of the presence of a singularity. The other one has one relevant and one irrelevant direction. It corresponds to a negative value of~$\overline{m}^{2}$, possibly indicating a phase transition. Similar results have been obtained by the other authors mentioned at the beginning of this section.

\section{Loop equations}\label{section4}

 \subsection{Reparametrisation invariance in quantum f\/ield theory}

To begin with, let us consider again a general quantum f\/ield theory whose f\/ields are collectively denoted by~$\phi^{i}$. The basic object of interest is the generating function for correlation functions, which we write symbolically as
\begin{gather*}
Z[J]=\int [D\phi]\exp \{-S[\phi]+J\phi \}.
\end{gather*}
In this section, we consider formal algebraic relations satisf\/ied by the generating function, treating the f\/ield as a f\/inite-dimensional variable for simplicity. This is enough for the heuristic purpose of this following presentation.

 Because it is an integral over all f\/ields, the generating functional $Z[J]$ must be invariant under any change of variables in the space of f\/ields, a property known as reparametrisation invariance. Consider an inf\/initesimal change of variable, which we write as
 \begin{gather*}
 \phi^{i}\rightarrow \phi^{i}+\epsilon X^{i}[\phi].
 \end{gather*}
Invariance of $Z[J]$ under such a change of variable translates into the equation
\begin{gather*}
\int [D\phi] \sum_{i}\bigg(
\frac{\partial X^{i}}{\partial \phi^{i}}
-X^{i}\frac{\partial S}{\partial\phi^{i}}+X^{i}J_{i}
\bigg)\exp \{-S[\phi]+J\phi \}=0.
\end{gather*}
The f\/irst term is the Jacobian of the change of variable and the two others arise from the variation of the argument of the exponential.
Since every function of the f\/ield can be obtained as a derivative acting on the source, we arrive at
\begin{gather}
 \sum_{i}\left\{
\frac{\partial X^{i}}{\partial \phi^{i}}\left(\frac{\partial}{\partial J} \Big)
-X^{i}\Big(\frac{\partial}{\partial J} \right)\frac{\partial S}{\partial\phi^{i}}
\left(\frac{\partial}{\partial J} \right)
+J_{i} X^{i}\left(\frac{\partial}{\partial J} \right)\right\}Z[J]=0.\label{reparametrisation:eq}
\end{gather}
Introducing the dif\/ferential operator
 \begin{gather*}{\cal L}_{X}=
 \sum_{i}\left\{
\frac{\partial X^{i}}{\partial \phi^{i}}\left(\frac{\partial}{\partial J} \right)
-X^{i}\left(\frac{\partial}{\partial J} \right)\frac{\partial S}{\partial\phi^{i}}
\left(\frac{\partial}{\partial J} \right)
+J_{i} X^{i}\left(\frac{\partial}{\partial J} \right)\right\}.
\end{gather*}
 Invariance under reparametrisation takes the following simple form
 \begin{gather*}
 {\cal L}_{X}Z[J]=0.
 \end{gather*}
 Although relatively complicated, these dif\/ferential operators obey a simple commutation relation
 \begin{gather*}
[{\cal L}_{X},{\cal L}_{Y}]={\cal L}_{[X,Y]}
 \end{gather*}
 with $[X,Y]$ the commutator between the two vectors f\/ields $X$ and $Y$ def\/ined on the space of f\/ields
\begin{gather*}
[X,Y]^{i}=\sum_{j} X^{j}\frac{\partial Y^{i}}{\partial \phi^{j}}-Y^{j}\frac{\partial X^{i}}{\partial \phi^{j}}.
\end{gather*}
 $[X,Y]$ is nothing but the inf\/initesimal part of the commutator between the change of variables associated to $X$ and $Y$. Therefore, the operators ${\cal L}$ def\/ine a representation of the Lie algebra of inf\/initesimal change of variables.

 When the vector f\/ield $X$ is independent of $\phi$, the associated change of variables is simply a~translation in the space of f\/ields, $\phi^{i}\rightarrow\phi^{i}+\epsilon^{i}$. Then, \eqref{reparametrisation:eq} takes the form
 \begin{gather*}
\left\{
\frac{\partial S}{\partial\phi^{i}}
\left(\frac{\partial}{\partial J} \right)
+J_{i} \right\}Z[J]=0.
\end{gather*}
 These are the Schwinger--Dyson equations. which are the quantum analogue of the equations of motion. They provide enough information to reconstruct $Z[J]$ in perturbation theory, see~\cite{ZinnJustin:1989mi}.

\subsection{Loop equations for tensor models}

 Let us apply the general framework depicted in the previous section to unitarily invariant random tensor models. These equations where f\/irst derived by Gurau in \cite{Gurau:2011tj} at leading order and at all orders in~\cite{Gurau:2012ix}. We follow the presentation given in~\cite{Krajewski:2012dq}.

 We start with an interaction of the form given in \eqref{invariantpotential:eq}, except that we do not include the combinatorial factor,
 \begin{gather*}
V(T,\overline{T})=\sum_{{\cal B}}\lambda_{\cal B} \operatorname{Tr}_{\cal B}(\overline{T},T).
\end{gather*}
 The inclusion of the combinatorial factor $C_{T}$ is immaterial since it amounts to trading $\lambda_{\cal B}$ for~$\frac{\lambda_{B}}{C_{\cal B}}$. Here, we have included all bubbles, even nonconnected ones and the empty one. The latter is simply a constant $\lambda_{\varnothing}$ that does not couple to the tensors. Then, solving the theory amounts to computing the partition function as a function of all the bubble couplings,
\begin{gather}
Z ( \{\lambda_{\cal B} \} )=
\int d\mu_{C}(T,\overline{T})\exp\{-V(T,\overline{T})\},\label{ZSD:eq}
\end{gather}
with a diagonal covariance $C=t\,\delta_{i_{1}\overline{i}_{1}}\cdots\delta_{i_{D}\overline{i}_{D}}$. This has to be considered as a formal power series in the variables~$\lambda_{\cal B}$, ordered by the number of vertices of the bubbles, in such a way that at a given order there is only a f\/inite number of terms. The expectation value of a bubble is then obtained by
\begin{gather*}
\big\langle
\operatorname{Tr}_{\cal B}(\overline{T},T)
\big\rangle=-
\frac{\partial}{\partial\lambda_{\cal B}}
 (\log Z ).
\end{gather*}
If only a subset of bubble couplings are present in the actual potential, all couplings not present in the potential have to be set to 0 after the derivation.

For unitarily invariant models, there is a preferred set of change of variables parametrised by bubbles with a vertex removed. Indeed, if $\overline{v}$ is a black vertex, then we def\/ine $\operatorname{Tr}_{{\cal B}\backslash \overline{v}}(\overline{T},T)$ as $\operatorname{Tr}_{{\cal B}}(\overline{T},T)$ with the tensor $\overline{T}$ at $\overline{v}$ removed. Treating $T$ and $\overline{T}$ as independent variables, the change of variables associated with ${{\cal B}\backslash \overline{v}}$ is
\begin{gather}
T\rightarrow T+\epsilon_{{\cal B}\backslash \overline{v}} \operatorname{Tr}_{{\cal B}\backslash \overline{v}}(\overline{T},T),\qquad
\overline{T}\rightarrow \overline{T},\label{change:eq}
\end{gather}
with $\epsilon_{{\cal B}\backslash \overline{v}}$ an inf\/initesimal parameter. Performing this change of variables in~\eqref{ZSD:eq} leads to the identity
\begin{gather*}
\int d\mu_{C}(T,\overline{T})\Bigg\{
{-}t^{-1}\operatorname{Tr}_{{\cal B}}(\overline{T},T)+\sum_{v\in{\cal B}\atop \text{white vertex}}N^{l({\cal B},\overline{v},v)}
\operatorname{Tr}_{{\cal B}/\overline{v}v}(\overline{T},T)\\
\qquad{}-\sum_{{\cal B}'\atop
v'\in{\cal B'}\,\text{white vertex}}
\lambda_{{\cal B'}}\operatorname{Tr}_{({\cal B}'\cup{\cal B})/\overline{v}v'}(\overline{T},T)
\Bigg\}\exp\{-V(T,\overline{T})\} =0.
\end{gather*}
For any bubble ${\cal B}$, recall that ${\cal B}/v\overline{v}$ is the bubble with the pair of vertices $v\overline{v}$ contracted (see Fig.~\ref{paircontraction:fig}) and $l({\cal B},v,\overline{v})$ is the number of edges in ${\cal B}$ that connect $v$ and $\overline{v}$. The f\/irst two terms arise from the change of the Gau{\ss}ian measure, the second one being the Jacobian. The third term comes from the change of the potential. Note that the contribution of the covariance is of the same form as that of the dipole, considered as a bubble coupling.

\begin{figure}[t]\centering
\begin{gather*}
\left\langle\parbox{5cm}{\includegraphics[width=5cm]{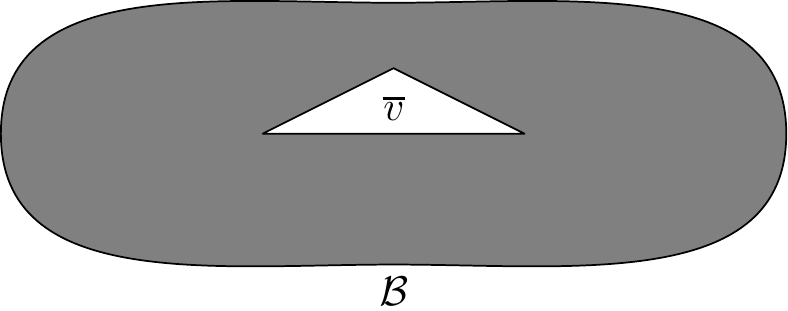}}\right\rangle
 \\
\qquad{}= t\sum_{v\in{\cal B}\atop \text{white vertex}}N^{l({\cal B},\overline{v},v)}
\Bigg\langle\parbox{5cm}{\includegraphics[width=5cm]{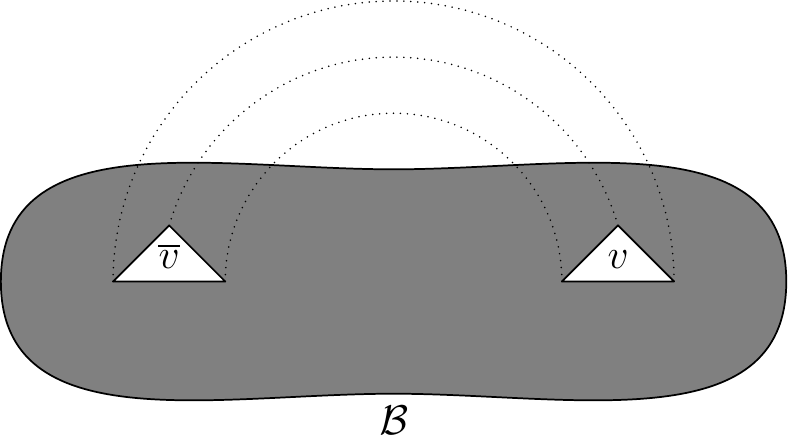}}\Bigg\rangle
\\
\qquad\quad{} -
t\sum_{{\cal B}'\atop
v'\in{\cal B'}\,\text{white vertex}}
\lambda_{{\cal B'}}
\Bigg\langle\parbox{6cm}{
\includegraphics[width=6cm]{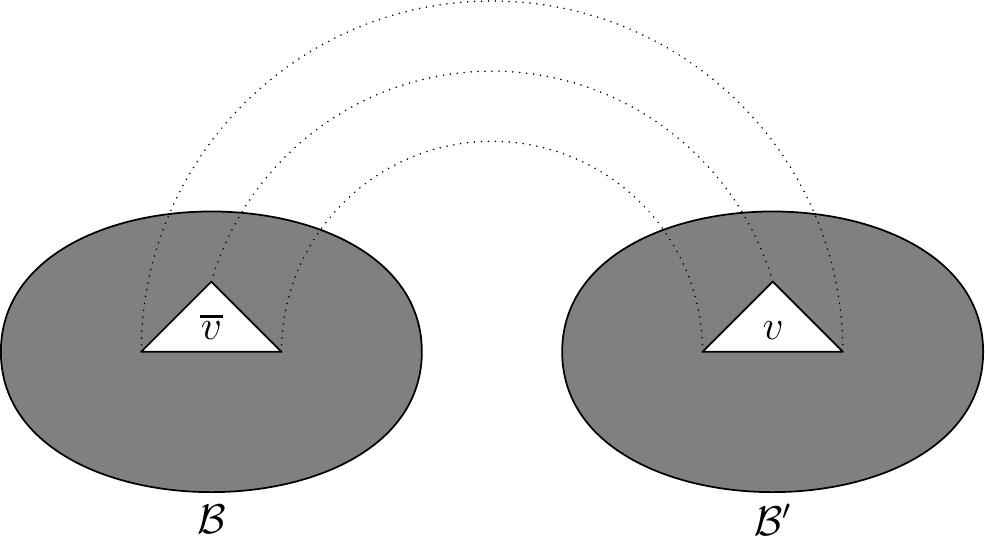}}\Bigg\rangle
\end{gather*}
\caption{Geometrical interpretation of equation \eqref{SDexpectation:eq}.}\label{SD:fig}
\end{figure}

\begin{figure}[t]\centering
\begin{subfigure}[Triangle not sharing an edge.]{
\parbox[b]{6cm}{\includegraphics[width=6cm]{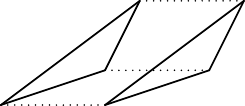}}
}
\end{subfigure}\quad
\begin{subfigure}[Triangle sharing an edge.]{
\parbox[b]{4cm}{\includegraphics[width=4cm]{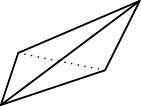}}
\quad \raisebox{10mm}{$\rightarrow$\quad $N$} \quad
}
\end{subfigure}
\caption{Contraction of a pair of triangles.}\label{triangles:fig}
\end{figure}

Writing this equation in terms of bubble expectation values, it reads
\begin{gather}
\big\langle
\operatorname{Tr}_{\cal B}(T,\overline{T})
\big\rangle=
t\!\!\!\sum_{v\in{\cal B}\atop \text{white vertex}} \!\!\!\!\! N^{l({\cal B},\overline{v},v)}
\big\langle\operatorname{Tr}_{{\cal B}/\overline{v}v}(\overline{T},T)\big\rangle
-
t\!\!\!\sum_{{\cal B}'\, v'\in{\cal B'}\atop
\text{white vertex}}\!\!\!\!\!
\lambda_{{\cal B'}}
\big\langle
\operatorname{Tr}_{({\cal B}'\cup{\cal B})/\overline{v}v'}(\overline{T},T)
\big\rangle.\label{SDexpectation:eq}
\end{gather}
Consider a bubble in the potential as a triangulation of a space of dimension $D-1$. Then, \eqref{SDexpectation:eq} has the following interpretation. On a boundary triangulation, consider the simplex of dimension $D-1$ labelled by vertex $v$. Then, the expectation of an observable constructed using such a boundary triangulation is a combination of the expectation values for which this triangle is connected to another triangle on the same boundary triangulation or in a dif\/ferent one, see Fig.~\ref{SD:fig}. The power of $N$ counts the number of simplices of dimension $D-2$ that are shared by the two simplices that we identify, see Fig.~\ref{triangles:fig}. Low powers of $N$ correspond to topology changes, so that the latter are expected to be subdominant.

Since a bubble expectation can be generated by derivation with respect to a bubble coupling, the previous equation can be written as a dif\/ferential operator acting on the partition function
 \begin{gather*}
 {\cal L}_{{\cal B}\backslash\overline{v}} Z=0
 \end{gather*}
 with
 \begin{gather}
 {\cal L}_{{\cal B}\backslash\overline{v}}=
 t^{-1}\frac{\partial}{\partial \lambda_{\cal B}}
-
 \sum_{v\in{\cal B}\atop \text{white vertex}}N^{l({\cal B},\overline{v},v)}
\frac{\partial}{\partial
\lambda_{{\cal B}/\overline{v}v}}
+\sum_{{\cal B}'\atop
v'\in{\cal B'}\,\text{white vertex}}
\lambda_{{\cal B'}}
\frac{\partial}{\partial \lambda_{({\cal B}'\cup{\cal B})/\overline{v}v'}}.\label{Lconnected:eq}
\end{gather}
These equations are constraints on the partition function expressed as dif\/ferential operators with respect to the couplings. They are tensor analogues of the loop equations for matrix models, which are recovered when $D=2$, see Section~\ref{matrixloops:sec}.

 The commutation relation between two such dif\/ferential operators reads
 \begin{gather}
 \big[
 {\cal L}_{{\cal B}_{1}\backslash\overline{v}_{1}},{\cal L}_{{\cal B}_{2}\backslash\overline{v}_{2}}
 \big]=
 \sum_{v'_{1}\neq v_{1}\atop \text{white vertex in ${\cal B}_{1}$}}{\cal L}_{({\cal B}_{1}\cup{\cal B}_{2})/v'_{1}\overline{v}_{2}}
-
 \sum_{v'_{2}\neq v_{2}\atop \text{white vertex in ${\cal B}_{2}$}}{\cal L}_{({\cal B}_{1}\cup{\cal B}_{2})/v'_{2}\overline{v}_{1}}
 \label{commutation:eq}
\end{gather}
and simply reproduces the commutation relation between two changes of variables of the form \eqref{change:eq}. These operators generate a Lie algebra whose structure is related to a Connes--Kreimer algebra, see~\cite{Connes:1999yr}.

By the same token, removing a white vertex on a bubble def\/ines a change of variables for the conjugate tensor
\begin{gather*}
T\rightarrow T,\qquad
\overline{T}\rightarrow \overline{T}+
+\overline{\epsilon}_{{\cal B}\backslash {v}} \operatorname{Tr}_{{\cal B}\backslash {v}}(\overline{T},T).
\end{gather*}
It leads to an analogous equation $\overline{{\cal L}}_{{\cal B}\backslash{v}}\,Z=0$. The operators $\overline{{\cal L}}_{{\cal B}\backslash{v}}$ obey similar commutations relations and commute with ${{\cal L}}_{{\cal B}\backslash\overline{v}}$,
\begin{gather*}
\big[\overline{{\cal L}}_{{\cal B}\backslash{v}},{{\cal L}}_{{\cal B}'\backslash\overline{v}'}\big]=0,
\end{gather*}
because we treat $T$ and $\overline{T}$ as independent variables.

In our derivation of the constraints, we have included all possible bubble couplings in the action, even for nonconnected bubbles and for the empty one. However, most of the authors do not consider such couplings. The empty bubble can only appear as ${\cal B}/v\overline{v}$ when ${\cal B}$ is a dipole. If the empty bubble is not present in the action, we assume that $\frac{\partial }{\partial \lambda_{\varnothing}}$ is the identity. If we start with only connected bubble couplings, nonconnected ones can appear only in the Jacobian. The latter can be generated by higher order derivations on the connected ones. Indeed, let us denote by $({\cal B}/v\overline{v})_{c}$ the connected components of ${\cal B}/v\overline{v}$. Then, we simply have to replace the dif\/ferential operators by
 \begin{gather*}
 {\cal L}_{{\cal B}\backslash\overline{v}}=
 t^{-1}\frac{\partial}{\partial \lambda_{\cal B}}
-
 \sum_{v\in{\cal B}\atop \text{white vertex}}N^{l({\cal B},\overline{v},v)}\hspace{-1cm}\prod_{c\text{ connected}\atop \text{component of ${\cal B}/v\overline{v}$}}\!\!
\left(-\frac{\partial}{\partial
\lambda_{({\cal B}/\overline{v}v)_{c}}}\right)
+\sum_{{\cal B}'\atop
v'\in{\cal B'}\,\text{white vertex}} \hspace{-0.5cm}
\lambda_{{\cal B'}}
\frac{\partial}{\partial \lambda_{({\cal B}'\cup{\cal B})/\overline{v}v'}}. \end{gather*}
This is in general a dif\/ferential operator of order $>1$. Nevertheless, the form of the loop equations and all commutation relations remain unchanged.

\subsection{The case of matrix models}\label{matrixloops:sec}

When $D=2$, coloured tensor models reduce to a theory of complex nonhermitian matrices of size~$N$. In this case, connected bubble are just indexed by a integer $n$ which is the equal number of black and white vertices. Then, connected bubble invariants are just traces of the $n^{\text {th}}$ power of~$M^{\dagger}M$ and the potential is
\begin{gather*}
V(M,M^{\dagger})=\lambda_{0}+\sum_{n=1}^{\infty}\lambda_{n}\operatorname{Tr}\big(M^{\dagger}M\big)^{n},
\end{gather*}
where we have included the empty bubble with $n=0$.
The change of variable associated to a bubble with a black vertex removed (see~\eqref{change:eq}) is {\samepage \begin{gather*}
M\rightarrow M+\epsilon_{n}\,M\big(M^{\dagger}M\big)^{n-1} ,\qquad
M^{\dagger}\rightarrow M^{\dagger},
\end{gather*}
for $n>1$.}

To f\/ind the explicit form of the dif\/ferential operators ${\cal L}_{{\cal B}\backslash \overline{v}}$ \eqref{Lconnected:eq}, f\/irst consider the case of a~dipole, $n=1$. Then, ${\cal B}/v\overline{v}$ is empty and we get
\begin{gather*}
{\cal L}_{1}=t^{-1}\frac{\partial }{\partial \lambda_{1}}-N^{2}\frac{\partial }{\partial \lambda_{0}}+\sum_{k\geq 1}k\lambda_{k}\frac{\partial}{\partial\lambda_{k}}.
\end{gather*}
For $n>1$, ${\cal B}/v\overline{v}$ has either one connected component and a free loop (if~$v$ and $\overline{v}$ are adjacent), or two connected components (if $v$ and $\overline{v}$ are not adjacent). This leads to
\begin{gather*}
{\cal L}_{n}= t^{-1}\frac{\partial }{\partial \lambda_{n}}-2N\frac{\partial }{\partial \lambda_{n-1}}-\sum_{2\leq k\leq n-1}
\frac{\partial ^{2}}{\partial\lambda_{k-1}\partial\lambda_{n-k}}
+
\sum_{k\geq 1}k\lambda_{k}\frac{\partial}{\partial\lambda_{n+k-1}}.
\end{gather*}
In order to compare with the standard form of the loop equations in the literature, it is convenient to def\/ine
\begin{gather*}
\lambda_{0} =N^{2}t_{0},\qquad
\lambda_{1} =N\big(t_{1}-t^{-1}\big),\qquad
\lambda_{n} =Nt_{n}\quad\text{for $n>1$},
\end{gather*}
so that the potential reads
\begin{gather*}
V(M,M^{\dagger})=N\sum_{n=0}^{\infty}t_{n}\operatorname{Tr}\big(M^{\dagger}M\big)^{n}.
\end{gather*}

Then, using $\frac{\partial Z}{\partial t_{0}}=-N^{2}Z$, the dif\/ferential operators take the form
\begin{gather*}
{\cal L}_{n}=
\frac{1}{N^{2}}
\sum_{1\leq k\leq n}
\frac{\partial ^{2}}{\partial t_{k-1}\partial t_{n-k}}
+
\sum_{k\geq 1}kt_{k}\frac{\partial}{\partial t_{n+k-1}}.
\end{gather*}
These operators obey the commutation relation
\begin{gather}
\big[{\cal L}_{m},{\cal L}_{n}\big]=(m-n){\cal L}_{m+n-1}.\label{Witt:eq}
\end{gather}
This is nothing but the commutation relation of the operators $x^{n}\frac{\partial}{\partial x}$, known as the Witt algebra. After a shift $n\rightarrow n+1$ and introducing an extra operator ${\cal L}_{-1}$, these are the Virasoro constraints, see~\cite{Ambjorn:1990ji}.

Finally, it may be interesting to note that the Witt algebra also appears as a subalgebra of the constraint algebra of the tensor models. Indeed, if we def\/ine
\begin{gather*}
{\cal L}_{n}=\sum_{{\cal B}, \, n({\cal B})=n\atop\text{$\overline{v}\in{\cal B}$ black vertex}}{\cal L}_{{\cal B}\backslash\overline{v}},
\end{gather*}
with $n({\cal B})$ the number of white vertices of ${\cal B}$, then the commutation relation~\eqref{commutation:eq} implies those of the Witt algebra~\eqref{Witt:eq}.

 \subsection{An application to melonic dominance} \label{melonicdominance:sec}

\looseness=-1
In Section~\ref{melonic:sec}, we have shown that only melonic couplings contribute at leading order to the partition function, provided each bubble coupling is written as $\lambda_{\cal B}=
N^{D-1}u_{\cal B}$, considering only connected bubbles for simplicity. Then, Schwinger--Dyson equations can be used to evaluate the expectation value of such a bubble, see~\cite{Bonzom:2012cu}. Indeed, with only melonic bubbles~${\cal M}$, \eqref{SDexpectation:eq} reduces to
\begin{gather}
\big\langle
\operatorname{Tr}_{\cal M}(T,\overline{T})
\big\rangle=
t\!\!\!\!\sum_{v\in{\cal M}\atop \text{white vertex}}\!\!\! \!\!\! N^{l({\cal B},\overline{v},v)}
\big\langle\operatorname{Tr}_{{\cal M}/\overline{v}v}(\overline{T},T)\big\rangle
-
t\!\!\!\!\sum_{{\cal M}',\,v'\in{\cal B'}\atop
\text{white vertex}}\!\!\!\!\!\! u_{{\cal M'}}
\big\langle
\operatorname{Tr}_{({\cal M}'\cup{\cal M})/\overline{v}v'}(\overline{T},T)
\big\rangle.\!\!\!\label{meloniclargeSDN:eq}
\end{gather}
Let us make the Ansatz that the expectation value is of Gau{\ss}ian type with a yet to be determined covariance $G(t,\left\{u_{\cal M'}\right\})$ depending on all melonic couplings in $V$,
\begin{gather*}
\big\langle \operatorname{Tr}_{\cal M}(T,\overline{T})\big\rangle
=N G^{n({\cal M})}
 \end{gather*}
 with $n({\cal M})$ the equal number of black or white vertices of~${\cal M}$. We also assume that the integral for a product of melonic bubbles factorizes in the large~$N$ limit. Then~\eqref{meloniclargeSDN:eq} reduces, for $D>2$, to
 \begin{gather*}
-t^{-1}G^{n({\cal M})}+G^{n({\cal M})-1}
-\sum_{{\cal M}'}n({\cal M}')u_{\cal M}
G^{n({\cal M}')+n({\cal M})-1}=0.
\end{gather*}
The assumption $D>2$ is crucial here since in this case there is a single $v$ such that the contribution of the second term in~\eqref{meloniclargeSDN:eq} is dominant, with all others subdominant.

This equation is solved if
\begin{gather*}
G=t+t \sum_{{\cal M}'}n({\cal M}')u_{{\cal M}'} G^{n({\cal M}')}=0.
\end{gather*}
If a f\/inite number of bubbles are present in the action, an explicit perturbative expansion of $G$ in terms of the couplings can be obtained by a recursive solution, starting with $G=t$ at lowest order.

 \section{Conclusion and outlook}\label{section5}

 In this survey, we have presented the exact renormalisation group equations and the loop equations for random tensors and group f\/ield theories. Exact renormalisation group equations are powerful devices that allow to classify the interactions according to their degree of relevancy in a specif\/ied limit. For unitarily invariant tensor models, the most relevant interactions are the melonic ones. In the group f\/ield theory case in the limit of large cut-of\/f, the theory is governed by renormalisable interactions. On the other side, loop equations are constraints on the partition function arising from reparametrisation invariance.

 It is fair to say that the applications of these equations to tensor models and group f\/ield theories have not yet been completely explored. Recent works by various groups suggest the following research directions.

 \begin{itemize}\itemsep=0pt
 \item The group f\/ield theories we have presented are based on abelian groups. However, powerful techniques based on the heat kernel have been developed by Carrozza, Oriti and Rivasseau in \cite{Carrozza:2013mna,Carrozza:2014rba, Carrozza:2012uv} in the context of multiscale analysis. It would certainly be very useful to adapt these techniques to f\/low equations. This would also be particularly useful in the renormalisation of $D=4$ group f\/ield theories related to general relativity.

\item Most of the works presented here is of perturbative nature. However, exact renorma\-li\-sation group equations, especially Wetterich's one, can be used in the nonperturbative regime. Ben Geloun, Benedetti, Carrozza, Lahoche and Oriti have already investigated the existence of f\/ixed points using truncations, see \cite{Geloun:2016qyb,Benedetti:2014qsa,Benedetti:2015yaa, Carrozza:2014rya} and the review~\cite{Carrozza:2016vsq}.

\item Melonic interactions turn out to be the most relevant ones. Nevertheless, the ef\/fects of nonmelonic ones have not yet been understood. This is particularly true in group f\/ield theories where some models exhibit nonmelonic interaction of scaling dimension 0.

 \item Loop equations play a fundamental role in the theory of random matrices, establishing the relation with conformal f\/ield theory and integrable hierarchies. It would be of high interest to establish similar connections in the case of random tensors.
\end{itemize}

\appendix

\section{ERGE for low order bubble couplings}\label{appendixA}

{\allowdisplaybreaks
In this appendix, a few example of evolutions equation for rescaled couplings are given.

\subsection[Couplings in rank $D=3$ tensor models]{Couplings in rank $\boldsymbol{D=3}$ tensor models}\label{example3:sec}

\vspace{-6mm}

\begin{gather*}
{\partial \over \partial t} u_{\includegraphics[width=0.7cm]{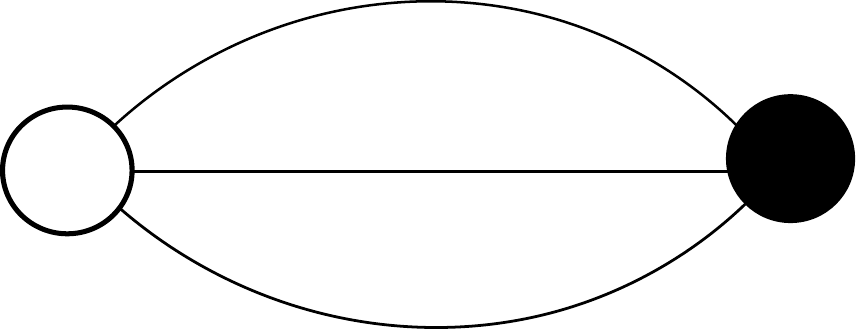}}
 =
\big[
u_{{\includegraphics[width=0.7cm]{Krajewski-Fig-A1-01}} \; {\includegraphics[width=0.7cm]{Krajewski-Fig-A1-01}}}
\big]\big \vert_{0 \, {\rm{cut}}}
+
\big[
3 u_{\includegraphics[width=0.7cm]{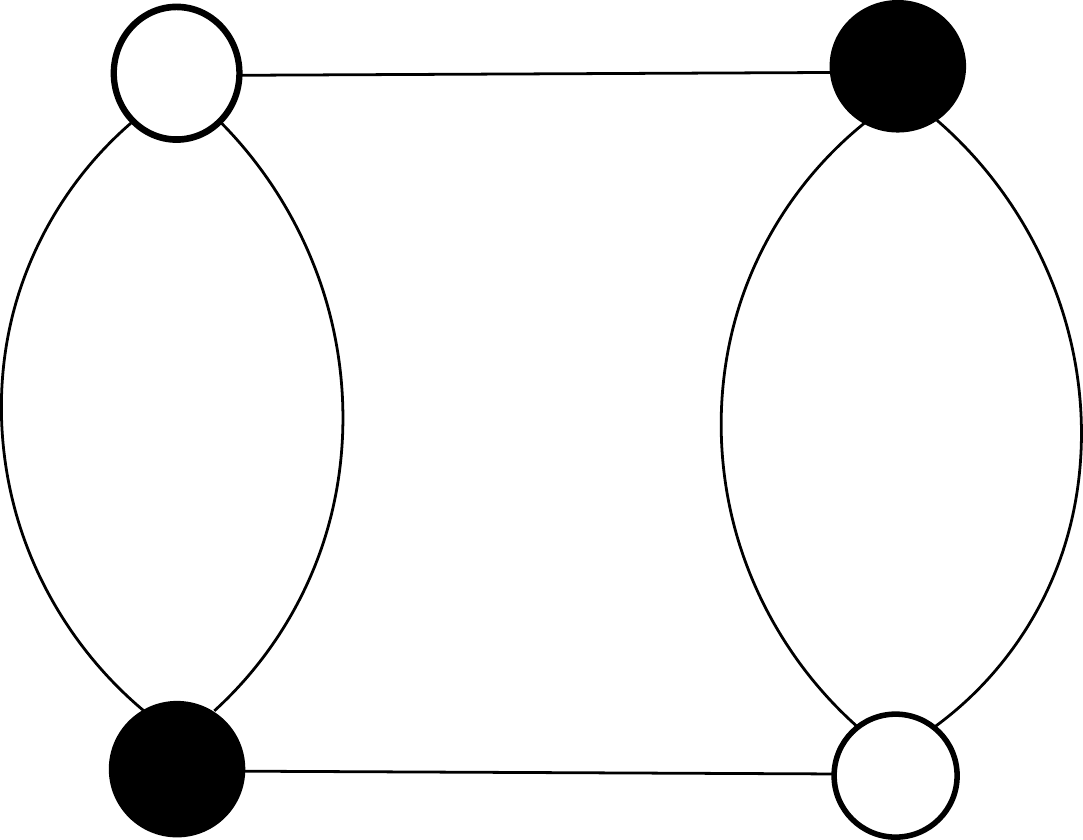}}
\big]\big \vert_{1 \, {\rm{cut}}}
-
\big[
u^2_{\includegraphics[width=0.7cm]{Krajewski-Fig-A1-01}}
\big]\big \vert_{3 \, {\rm{cuts}}}
\nonumber \\
\hphantom{{\partial \over \partial t} u_{\includegraphics[width=0.7cm]{Krajewski-Fig-A1-01}}=}{}
+
{1 \over N}
\big[
3 \,
u_{\includegraphics[width=0.7cm]{Krajewski-Fig-A1-02}}
\big]\big \vert_{2 \, {\rm{cuts}}}
+
{1 \over N^3}
\big[
u_{{\includegraphics[width=0.7cm]{Krajewski-Fig-A1-01}} \; {\includegraphics[width=0.7cm]{Krajewski-Fig-A1-01}}}
\big]\big \vert_{3 \, {\rm{cuts}}},\nonumber
\\
{\partial \over \partial t} u_{\includegraphics[width=0.7cm]{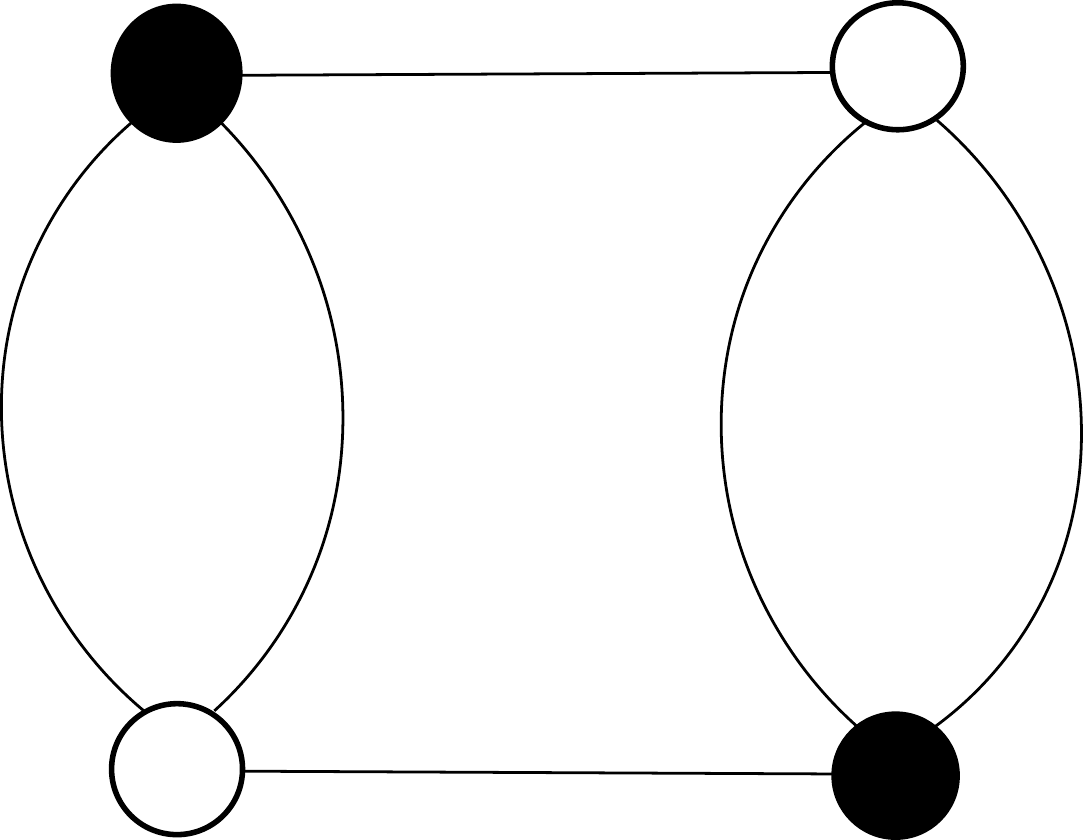}}
 =
\big[
u_{\includegraphics[width=1.4cm]{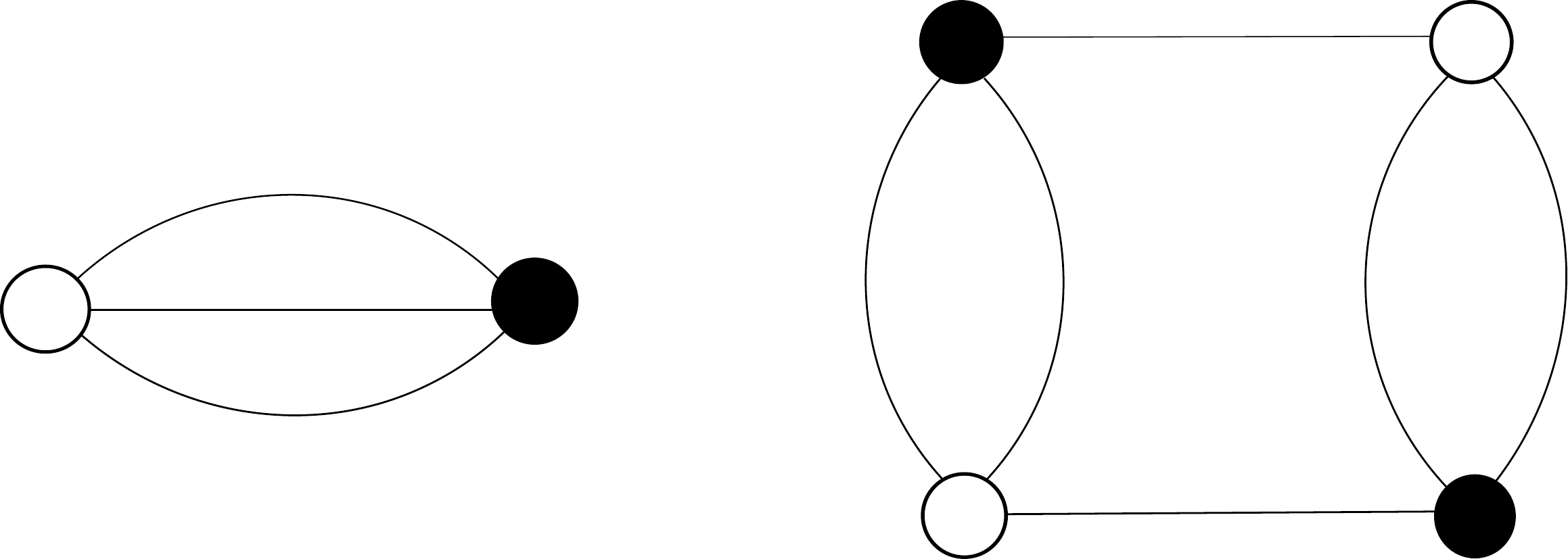}}
\big]\big \vert_{0 \, {\rm{cut}}}
+
\big[
4
u_{\includegraphics[width=0.7cm]{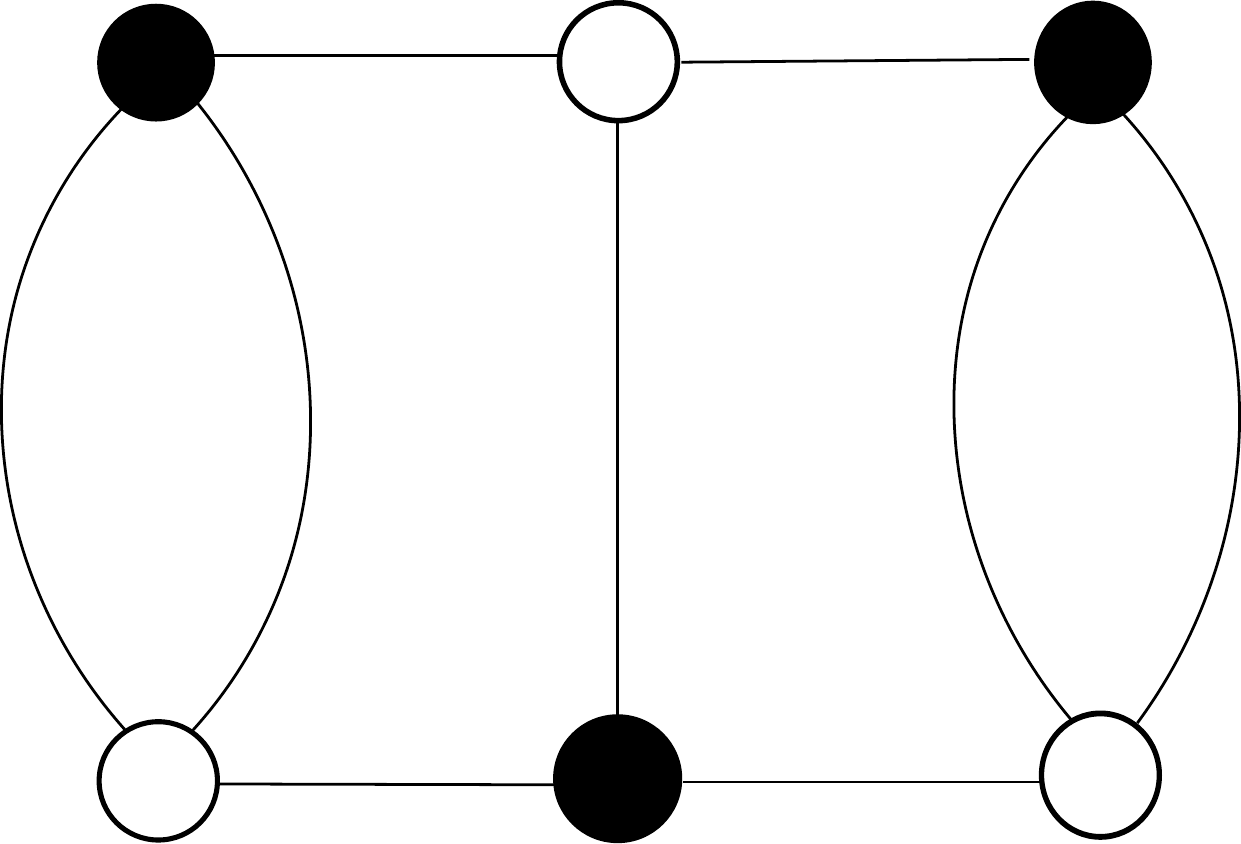}}
+
2
u_{\includegraphics[width=0.7cm]{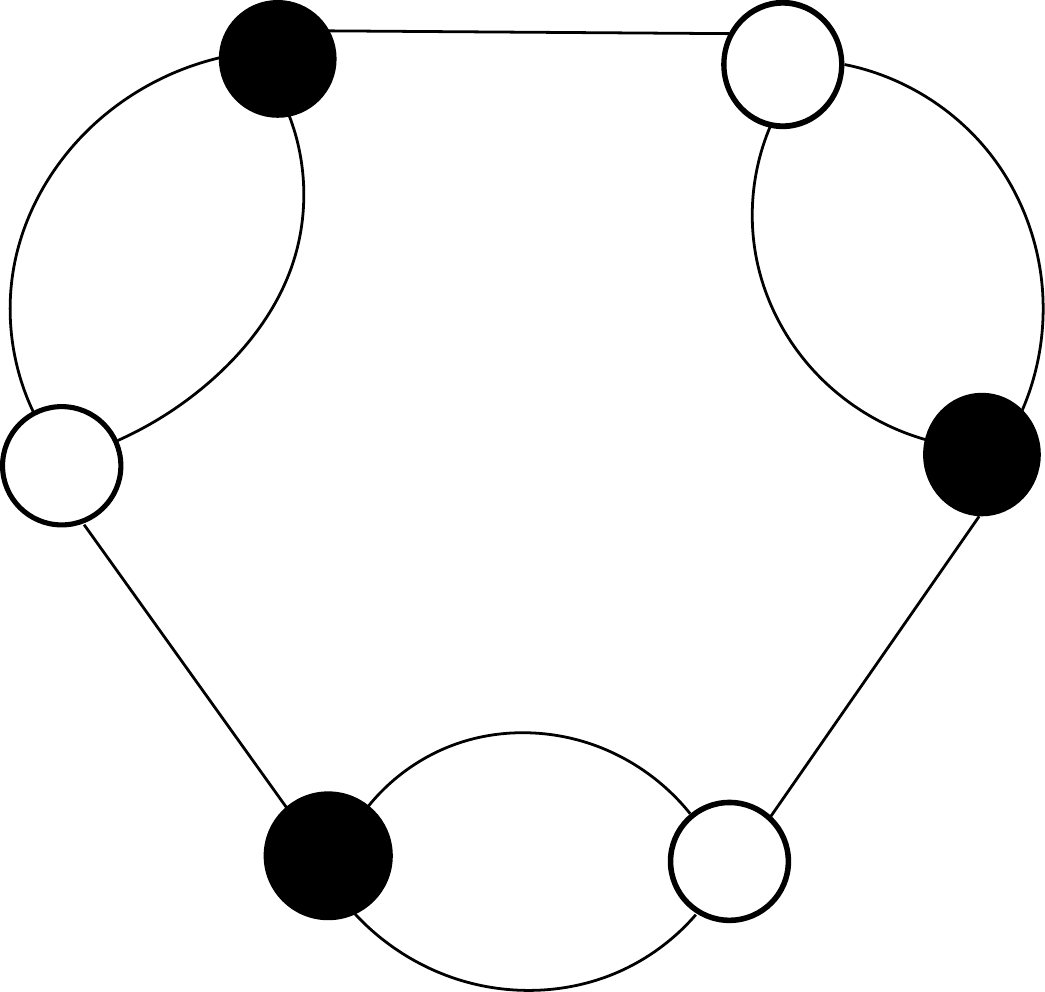}}
\big]\big \vert_{1 \, {\rm{cut}}}
-
\big[
4
u_{\includegraphics[width=0.7cm]{Krajewski-Fig-A1-01}}
u_{\includegraphics[width=0.7cm]{Krajewski-Fig-A1-03}}
\big]\big \vert_{3 \, {\rm{cuts}}}
\nonumber \\
\hphantom{{\partial \over \partial t} u_{\includegraphics[width=0.7cm]{Krajewski-Fig-A1-03}}=}{}
+
{1 \over N}
\big[
8
u_{\includegraphics[width=0.7cm]{Krajewski-Fig-A1-05}}
+
2
u_{\includegraphics[width=0.7cm]{Krajewski-Fig-A1-06}}
+
2
u_{\includegraphics[width=0.7cm]{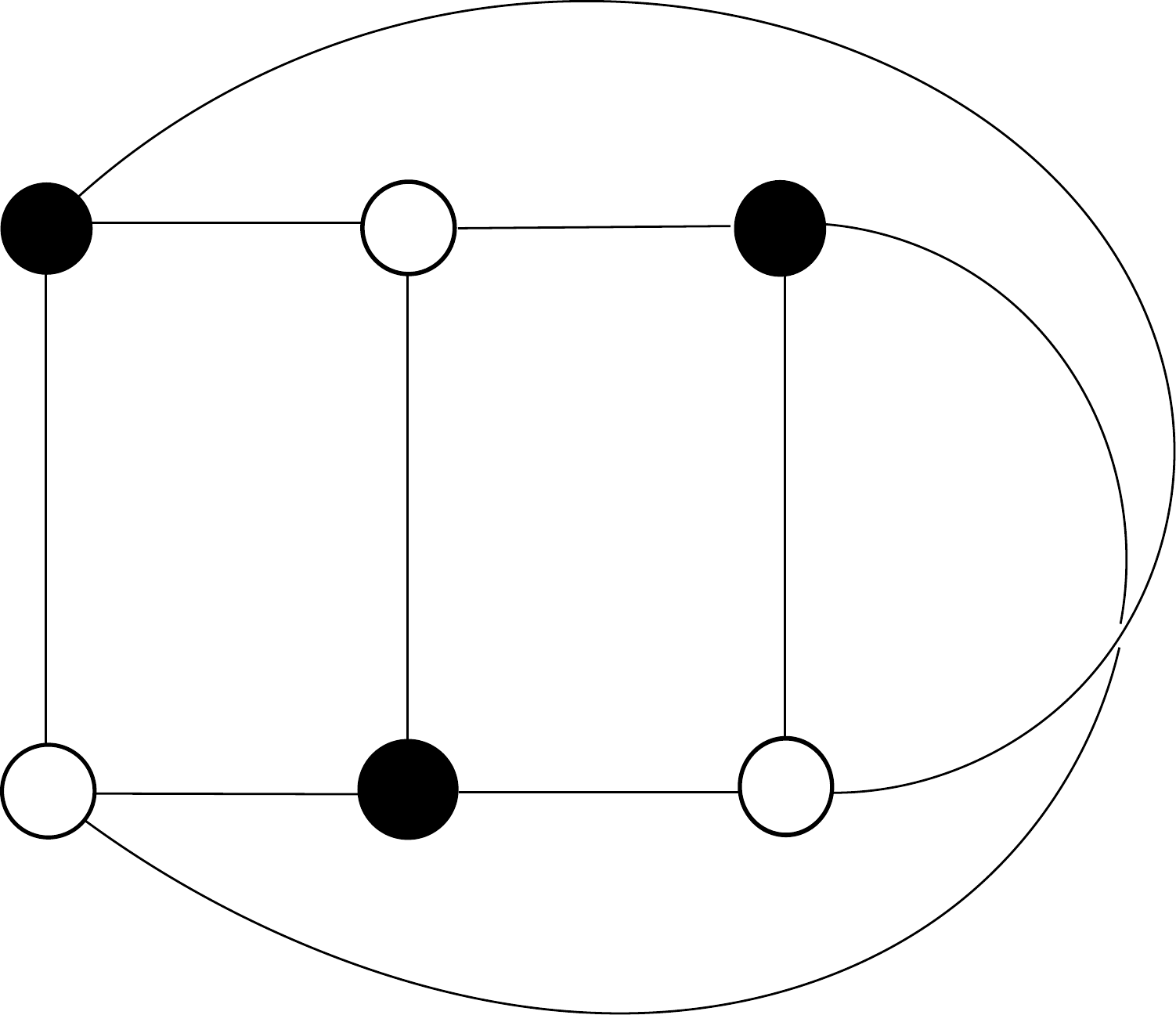}}
\big]\big \vert_{2 \, {\rm{cuts}}}
\nonumber \\
\hphantom{{\partial \over \partial t} u_{\includegraphics[width=0.7cm]{Krajewski-Fig-A1-03}}=}{}
+
{1 \over N^2}
\big[
4 u_{\includegraphics[width=0.7cm]{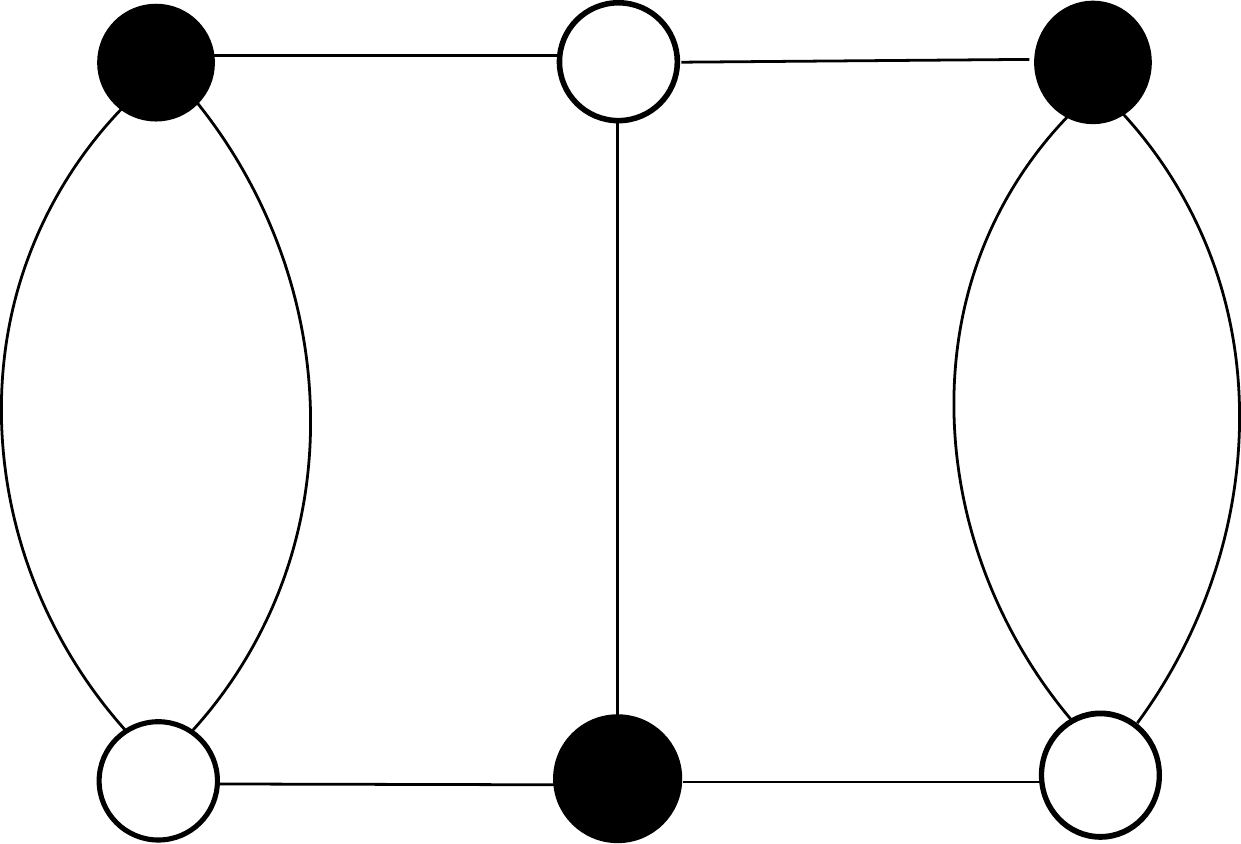}}
\big]\big \vert_{3 \, {\rm{cuts}}}
+
{1 \over N^3}
\big[
4
 u_{\includegraphics[width=1.4cm]{Krajewski-Fig-A1-04}}
\big]\big \vert_{3 \, {\rm{cuts}}} ,\nonumber
\\
{\partial \over \partial t} u_{{\includegraphics[width=0.7cm]{Krajewski-Fig-A1-01}} \; {\includegraphics[width=0.7cm]{Krajewski-Fig-A1-01}}}=
\big[
u_{{\includegraphics[width=0.7cm]{Krajewski-Fig-A1-01}} \; {\includegraphics[width=0.7cm]{Krajewski-Fig-A1-01}}\; {\includegraphics[width=0.7cm]{Krajewski-Fig-A1-01}}}
\big]\big \vert_{0 \, {\rm{cut}}}
+
\big[
6
u_{\includegraphics[width=1.4cm]{Krajewski-Fig-A1-04}}
\big]\big \vert_{1 \, {\rm{cut}}}
+
\big[
6
u_{\includegraphics[width=0.7cm]{Krajewski-Fig-A1-05}}
\big]\big \vert_{2 \, {\rm{cuts}}}\\
\hphantom{{\partial \over \partial t} u_{{\includegraphics[width=0.7cm]{Krajewski-Fig-A1-01}} \; {\includegraphics[width=0.7cm]{Krajewski-Fig-A1-01}}}=}{}
-
\big[
4
u_{{\includegraphics[width=0.7cm]{Krajewski-Fig-A1-01}} \; {\includegraphics[width=0.7cm]{Krajewski-Fig-A1-01}}}
u_{\includegraphics[width=0.7cm]{Krajewski-Fig-A1-01}}
\big]\big \vert_{3 \, {\rm{cuts}}}
+
{1 \over N}
\Big \{
\big[
6
u_{\includegraphics[width=1.4cm]{Krajewski-Fig-A1-04}}
\big]\big \vert_{2 \, {\rm{cuts}}}
+
\big[
6
u_{\includegraphics[width=0.7cm]{Krajewski-Fig-A1-06}}
\big]\big \vert_{3 \, {\rm{cuts}}}
\Big \}\nonumber\\
\hphantom{{\partial \over \partial t} u_{{\includegraphics[width=0.7cm]{Krajewski-Fig-A1-01}} \; {\includegraphics[width=0.7cm]{Krajewski-Fig-A1-01}}}=}{}
+
{1 \over N^3}
\big[
2\;
u_{{\includegraphics[width=0.7cm]{Krajewski-Fig-A1-01}} \; {\includegraphics[width=0.7cm]{Krajewski-Fig-A1-01}}\; {\includegraphics[width=0.7cm]{Krajewski-Fig-A1-01}}}
\big]\big \vert_{3 \, {\rm{cuts}}},\nonumber
\\
{\partial \over \partial t} u_{\includegraphics[width=0.7cm]{Krajewski-Fig-A1-06}}=
\big[
 u_{\includegraphics[width=1.4cm]{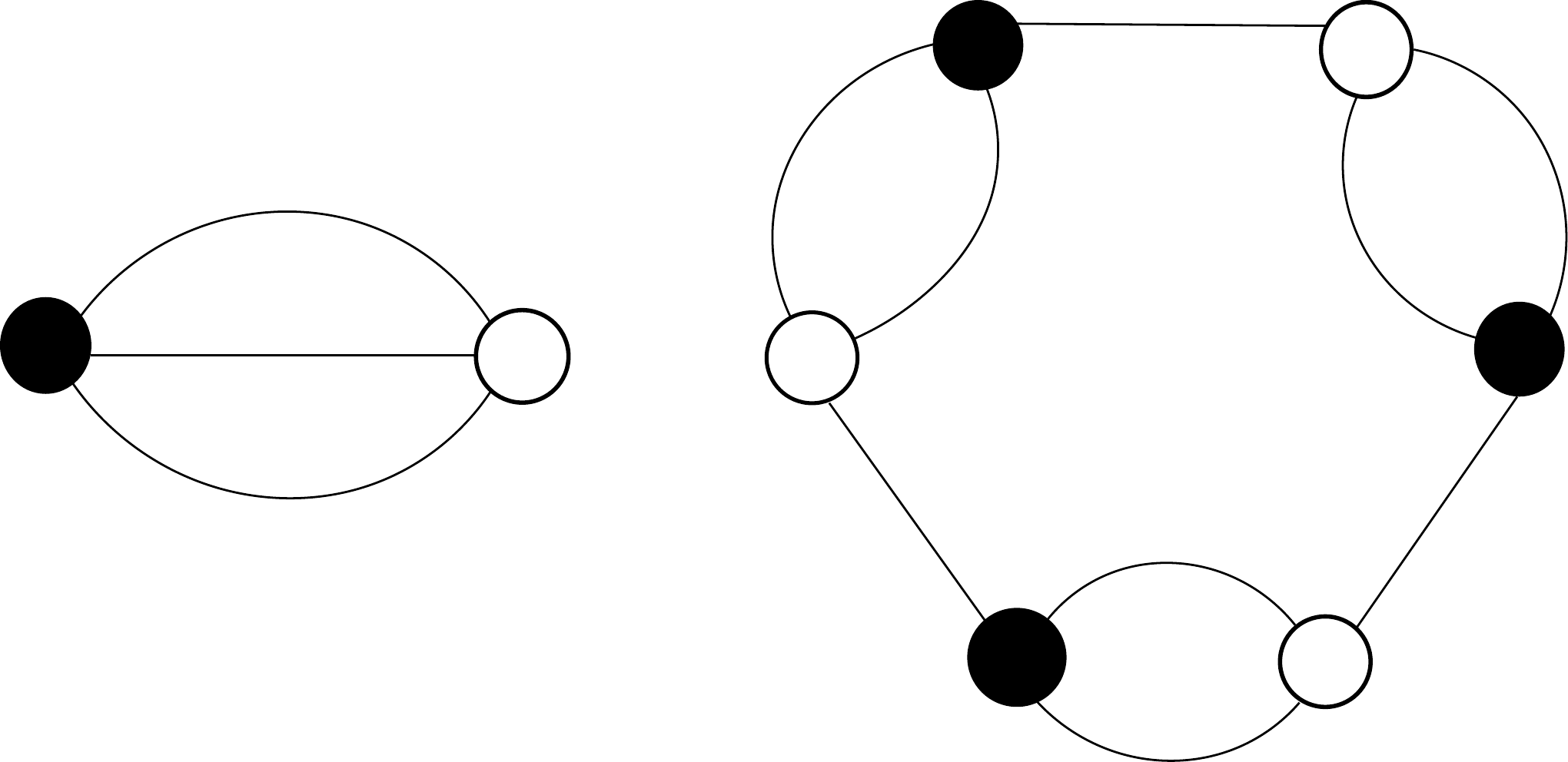}}
\big]\big \vert_{0 \, {\rm{cut}}}
+
\big[
6
u_{\includegraphics[width=1.1cm]{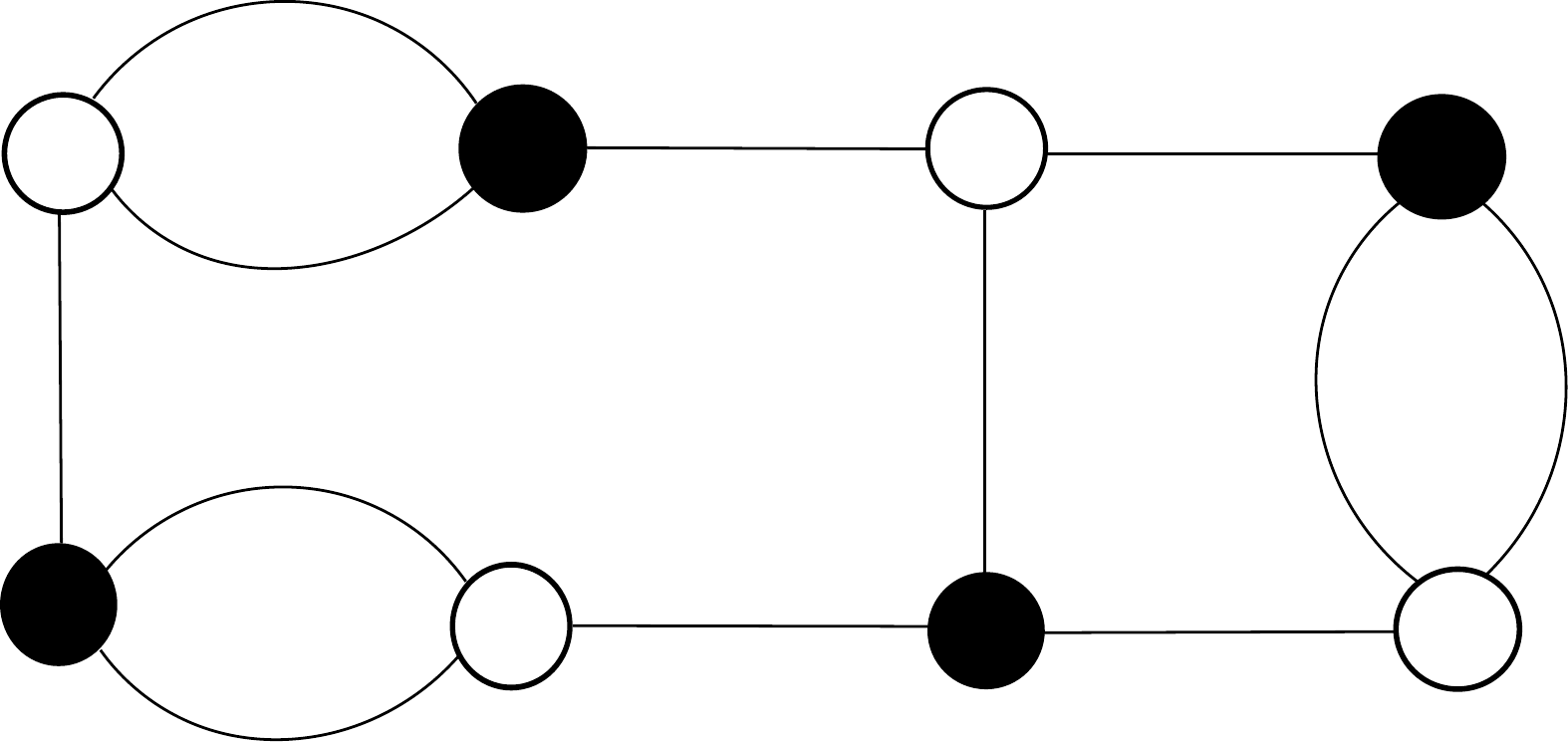}}
+
3
u_{\includegraphics[width=0.7cm]{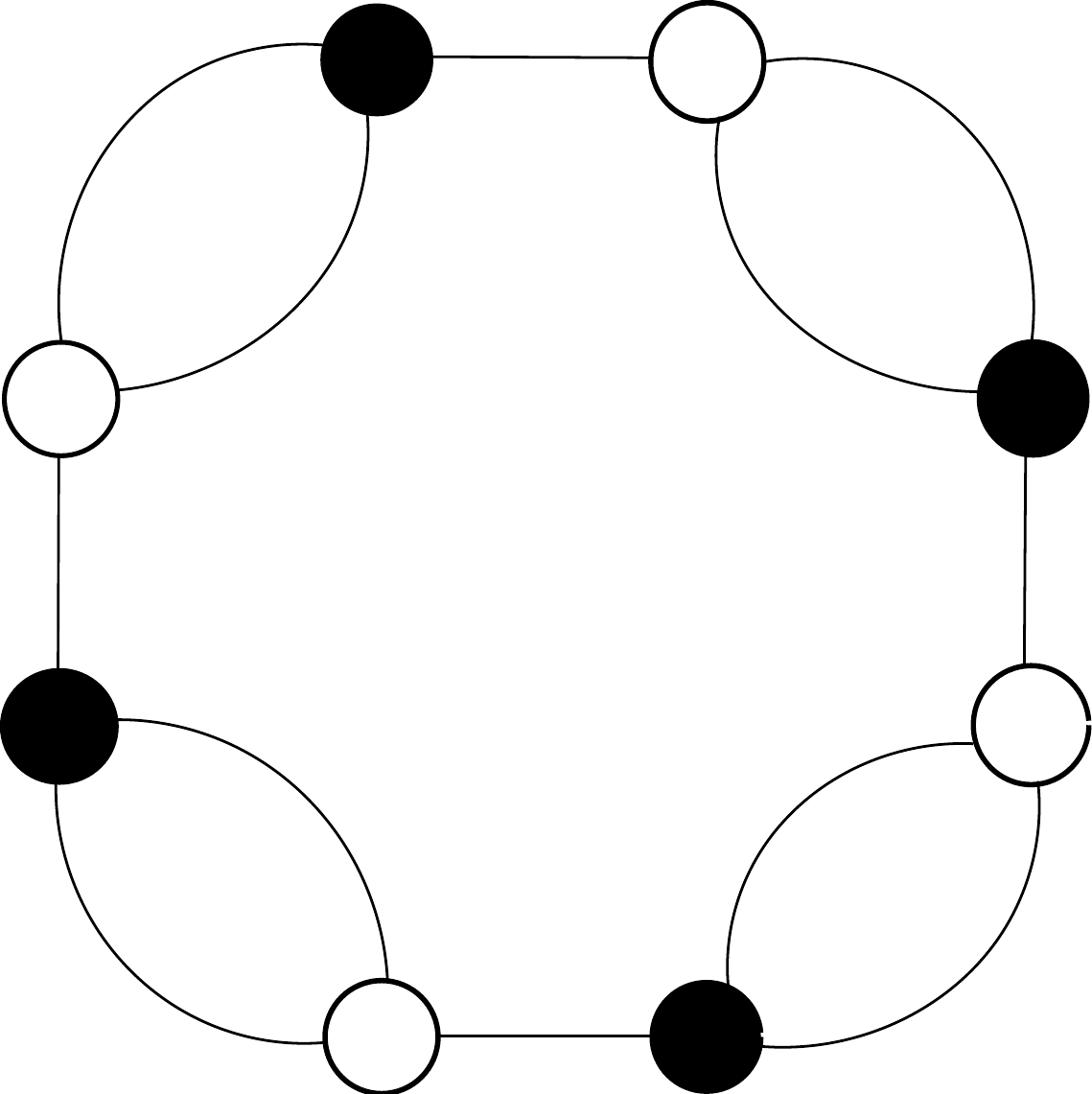}}
\big]\big \vert_{1 \, {\rm{cut}}}
-
\big[
6
u_{\includegraphics[width=0.7cm]{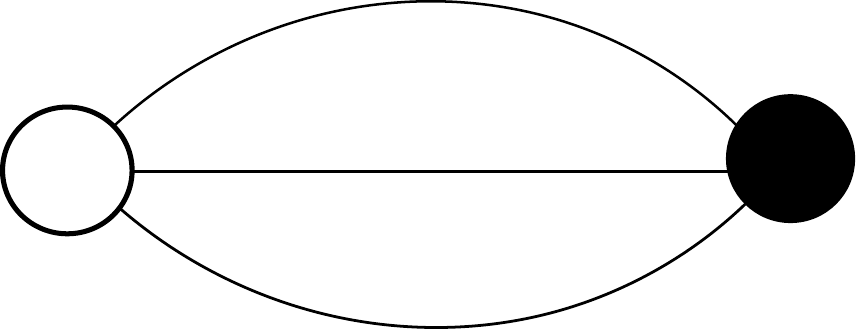}}
u_{\includegraphics[width=0.7cm]{Krajewski-Fig-A1-05}}
+
3
u^2_{\includegraphics[width=0.7cm]{Krajewski-Fig-A1-03}}
\big]\big \vert_{3 \, {\rm{cuts}}}
\nonumber \\
\hphantom{{\partial \over \partial t} u_{\includegraphics[width=0.7cm]{Krajewski-Fig-A1-06}}=}{}
+
{1 \over N}
\big[
3
 u_{\includegraphics[width=0.7cm]{Krajewski-Fig-A1-11}}
+
6
 u_{\includegraphics[width=1.0cm]{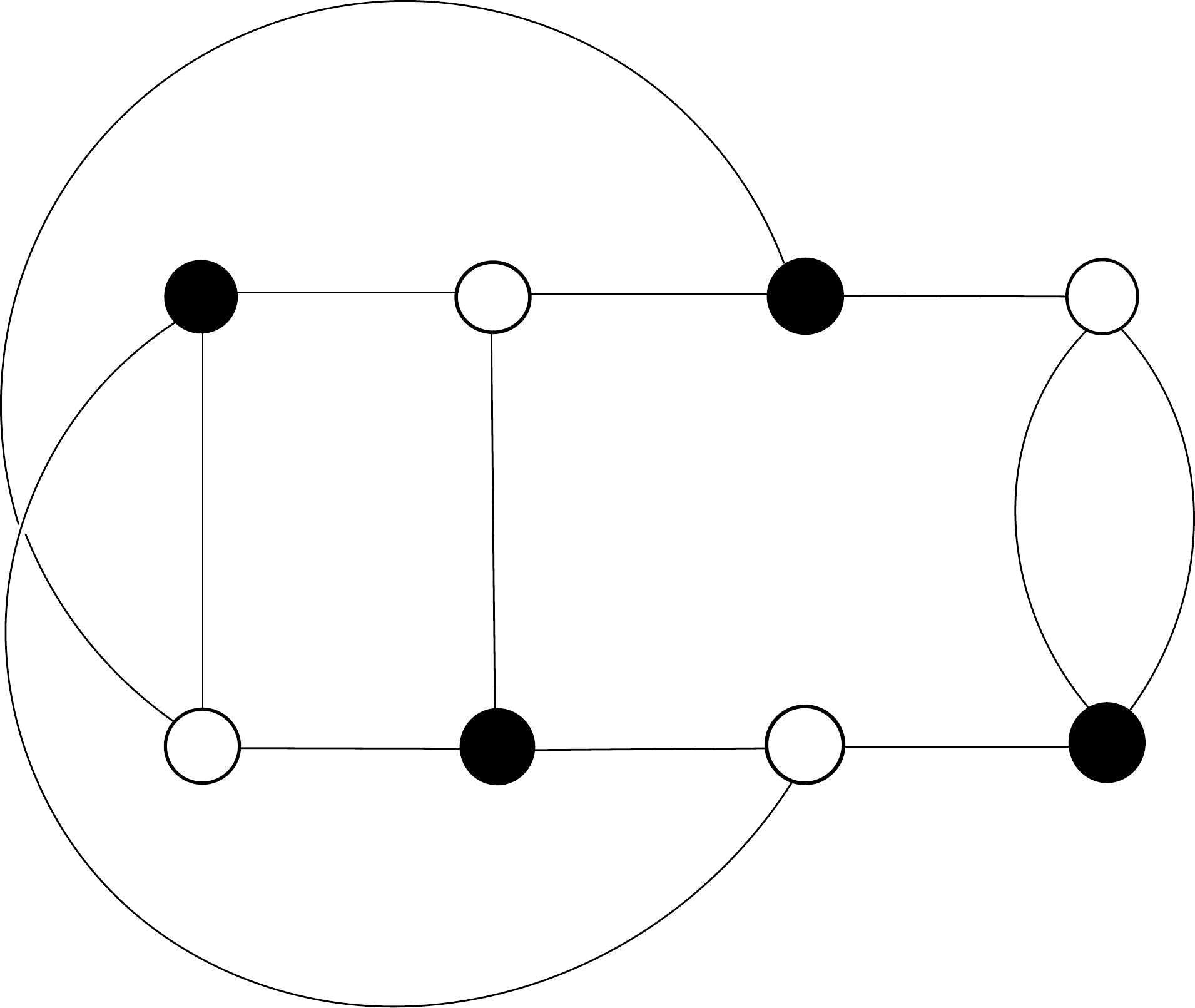}}
+
12
 u_{\includegraphics[width=1.1cm]{Krajewski-Fig-A1-10}}
+
6
 u_{\includegraphics[width=1.0cm]{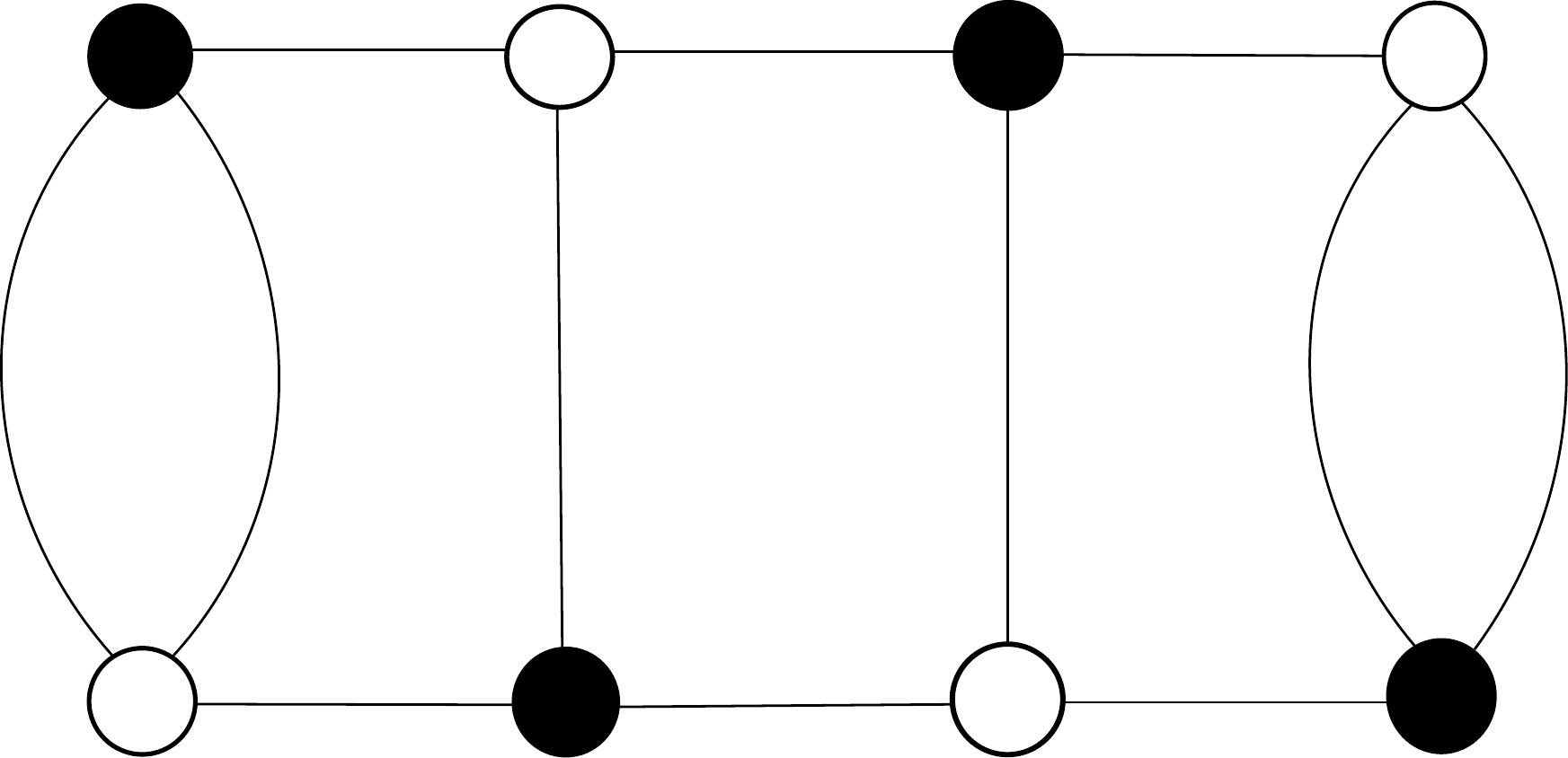}}
\big]\big \vert_{2 \, {\rm{cuts}}}
\nonumber \\
\hphantom{{\partial \over \partial t} u_{\includegraphics[width=0.7cm]{Krajewski-Fig-A1-06}}=}{}
+
{1 \over N^2}
\big[
12
 u_{\includegraphics[width=1.0cm]{Krajewski-Fig-A1-14}}
+
6 u_{\includegraphics[width=0.7cm]{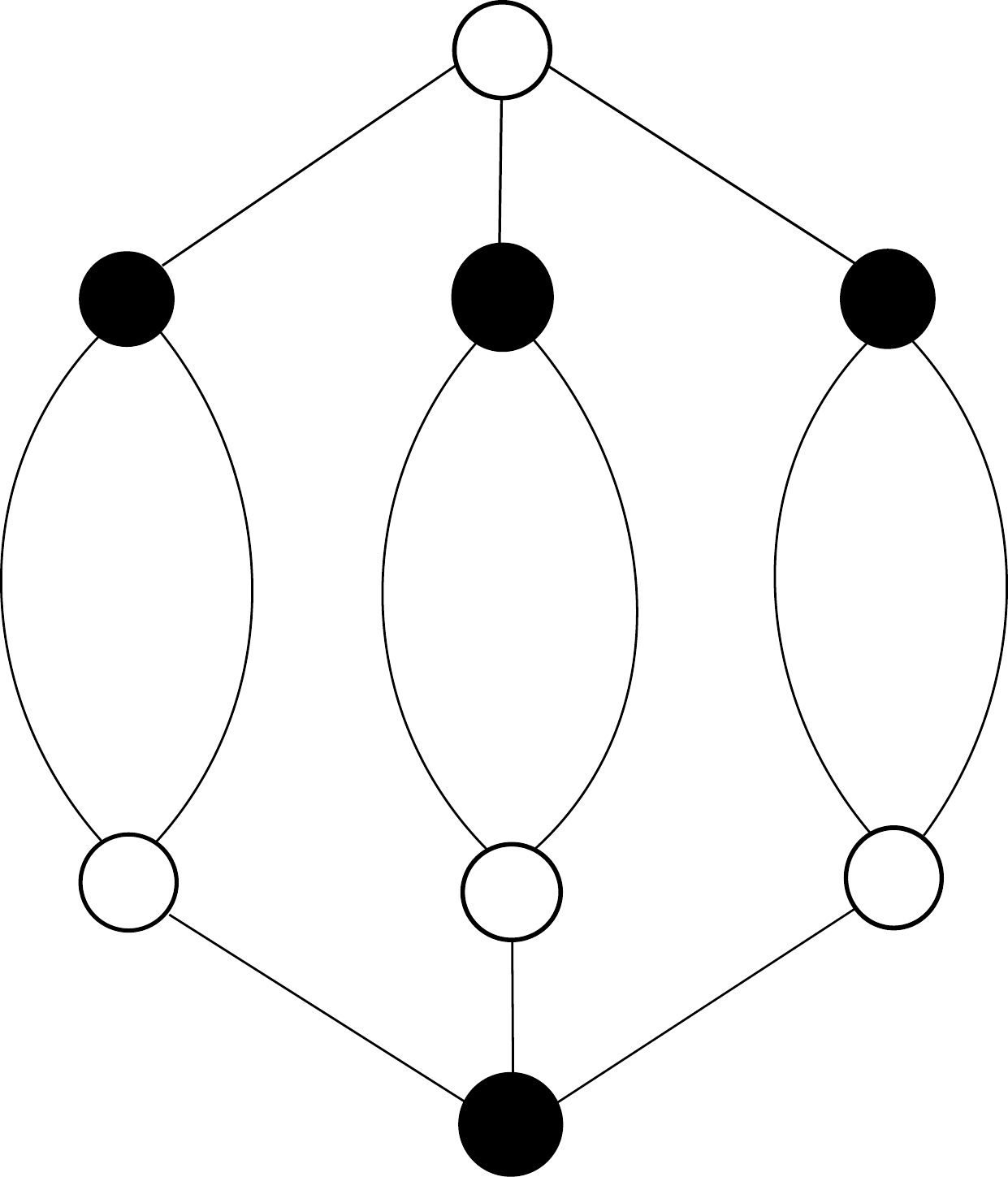}}
\big]\big \vert_{3 \, {\rm{cuts}}}
+
{1 \over N^3}
\big[
6
u_{\includegraphics[width=1.4cm]{Krajewski-Fig-A1-09}}
+
3
u_{\includegraphics[width=1.4cm]{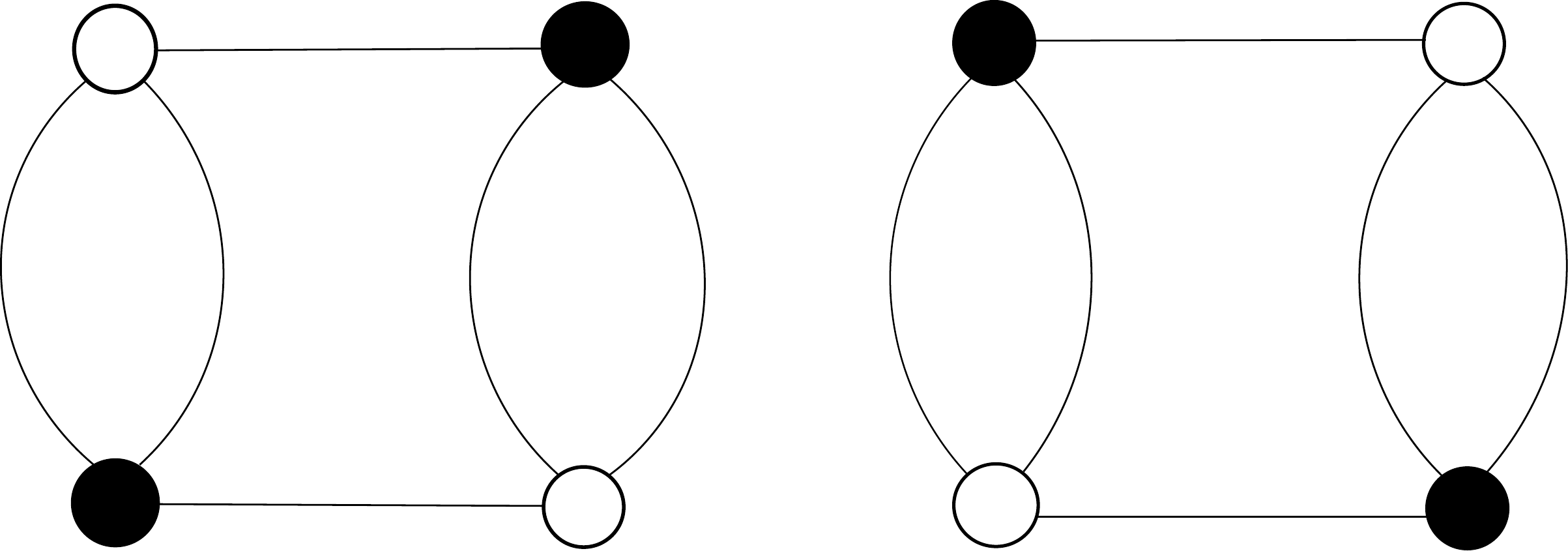}}
\big]\big \vert_{3 \, {\rm{cuts}}},\nonumber
\\
{\partial \over \partial t} u_{\includegraphics[width=0.9cm]{Krajewski-Fig-A1-05}}=
\big[
u_{\includegraphics[width=1.6cm]{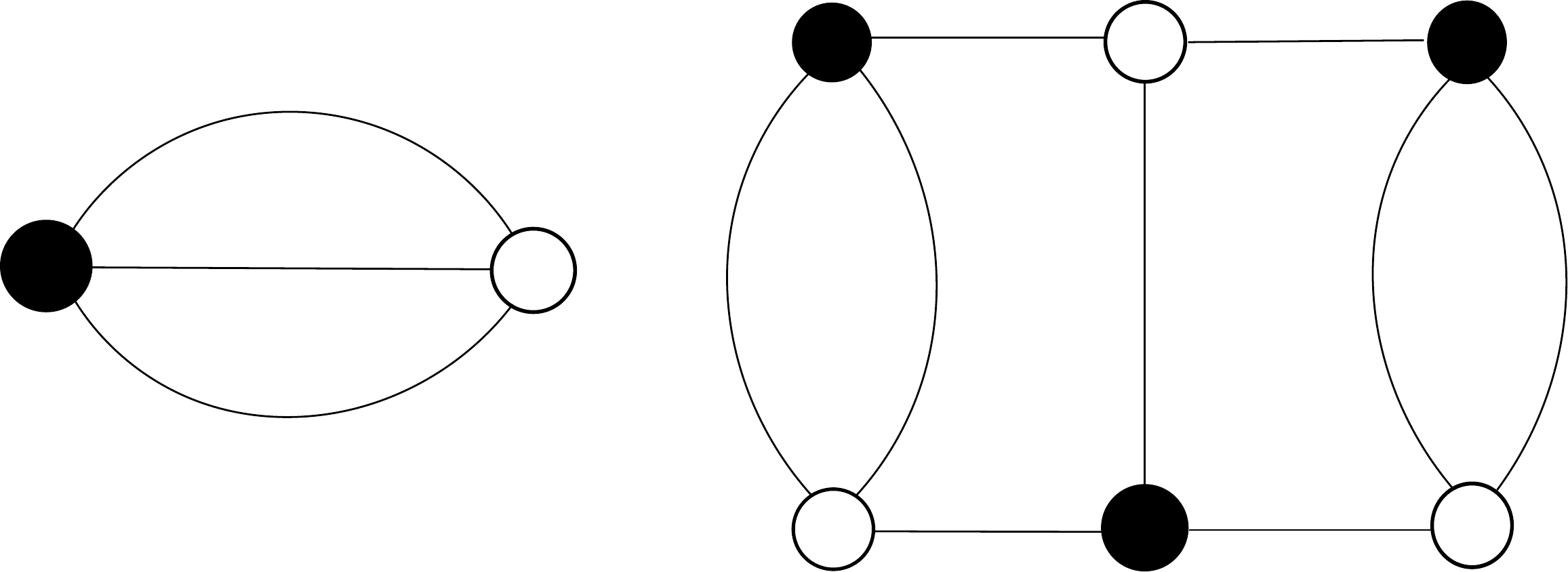}}
\big]\big \vert_{0 \, {\rm{cut}}}
+
\big[
4
u_{\includegraphics[width=1.0cm]{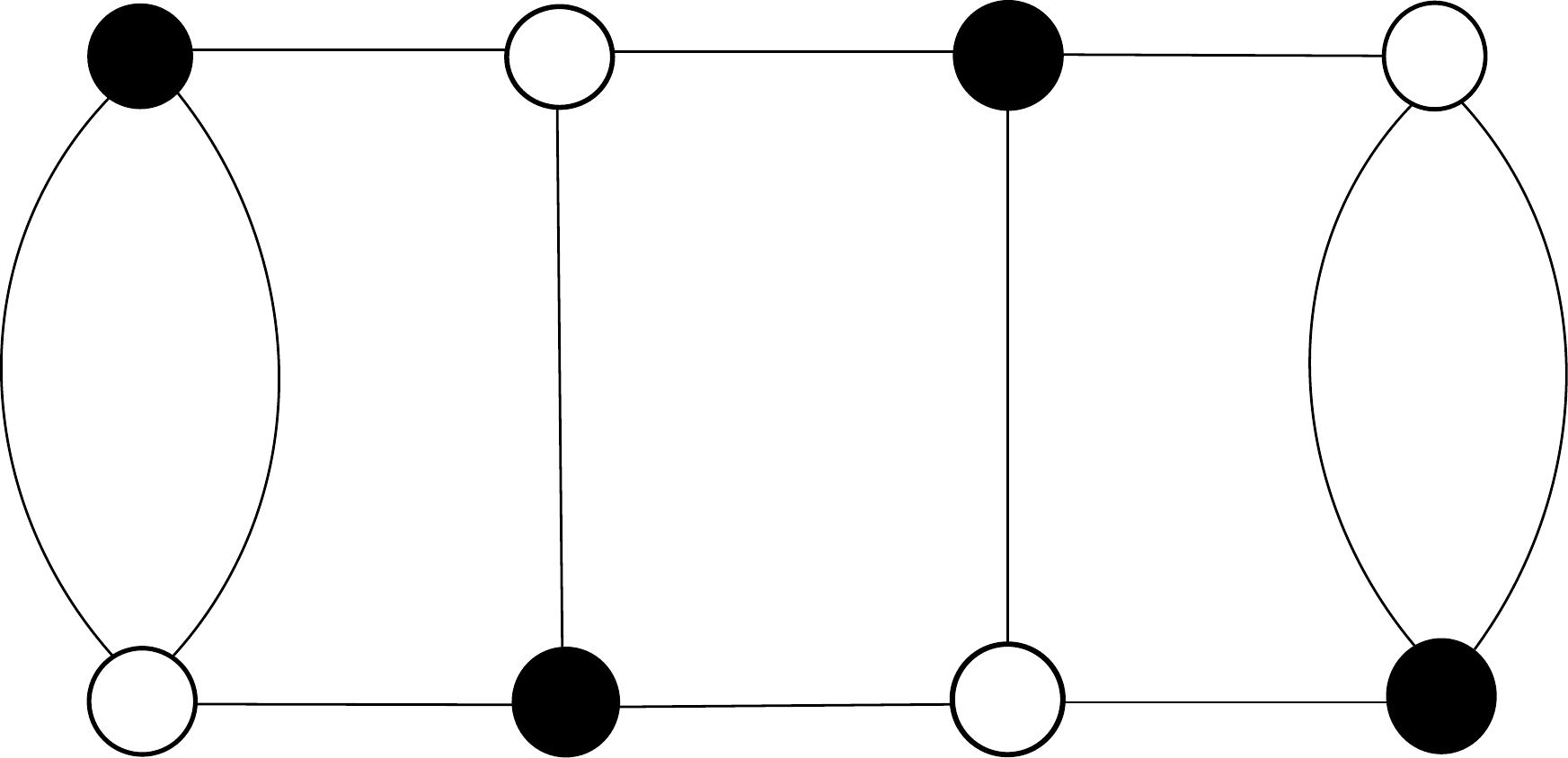}}
+
u_{\includegraphics[width=0.7cm]{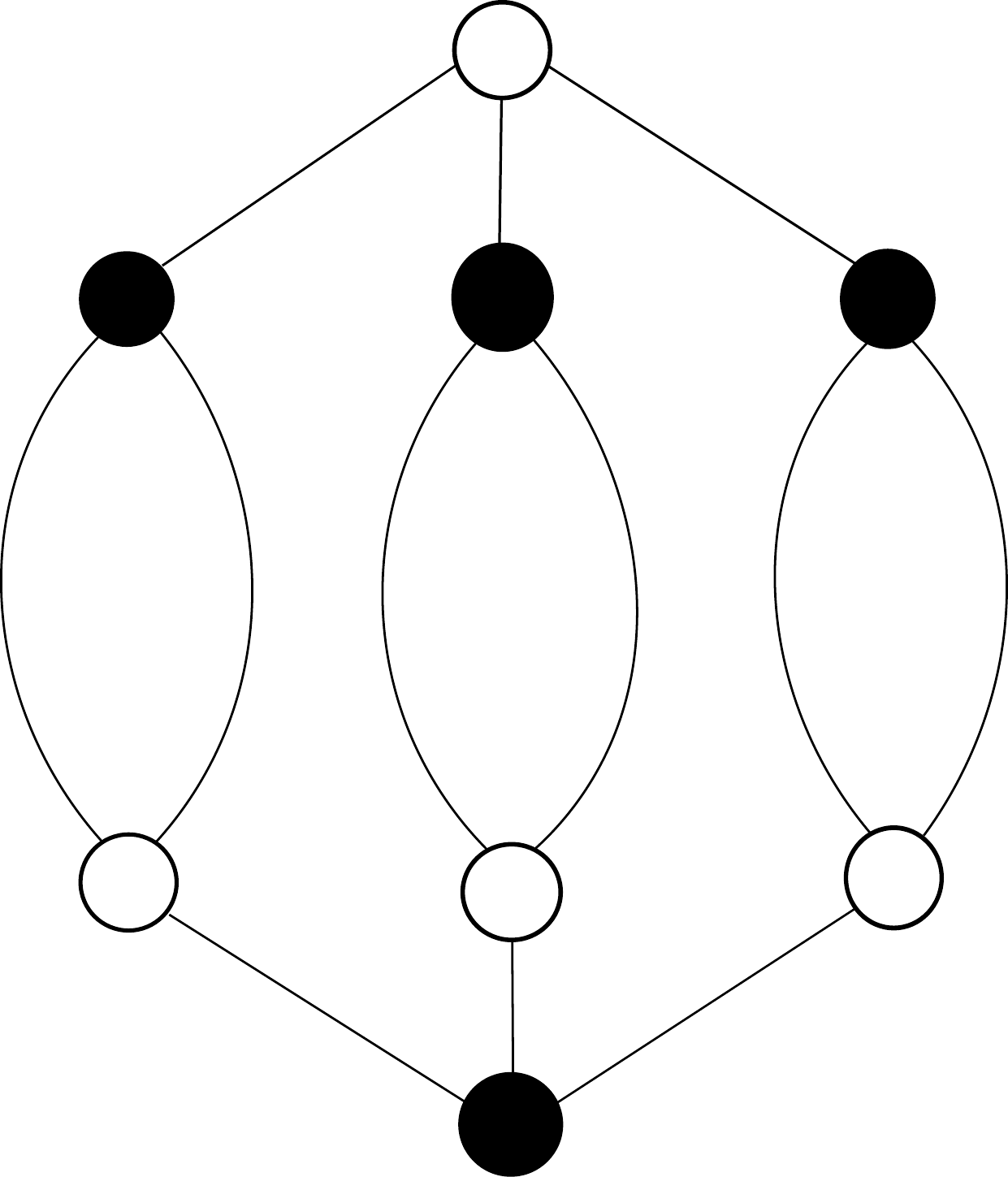}}
+
4
 u_{\includegraphics[width=1.0cm]{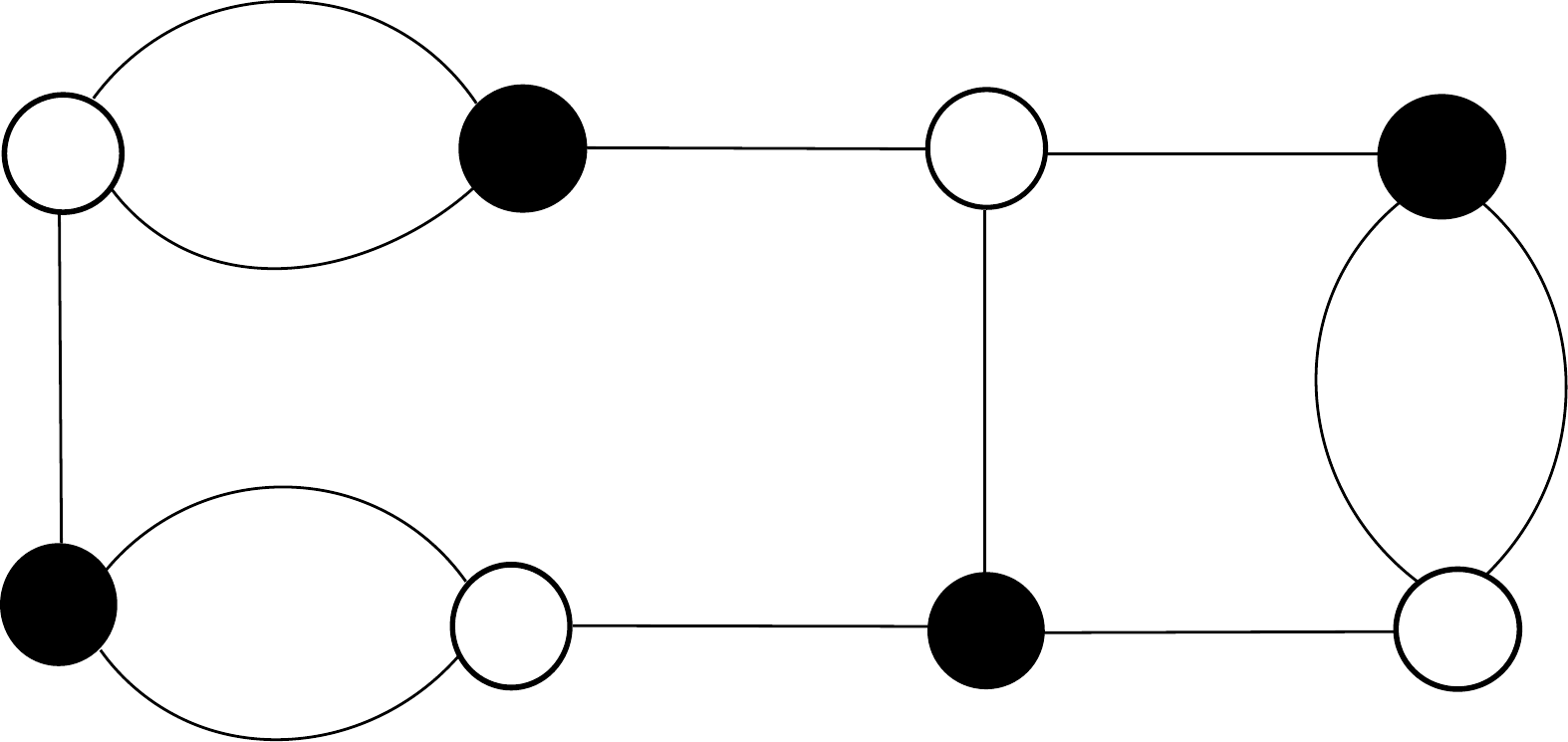}}
\big]\big \vert_{1 \, {\rm{cut}}}
\nonumber \\
\hphantom{{\partial \over \partial t} u_{\includegraphics[width=0.9cm]{Krajewski-Fig-A1-05}}=}{}
-
\big[
6
u_{\includegraphics[width=0.7cm]{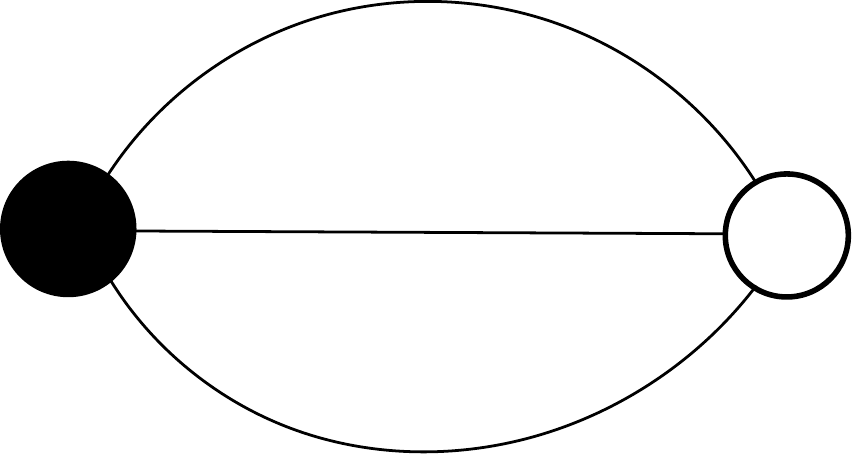}} \;u_{\includegraphics[width=0.7cm]{Krajewski-Fig-A1-05}}
+
2
u^2_{\includegraphics[width=0.7cm]{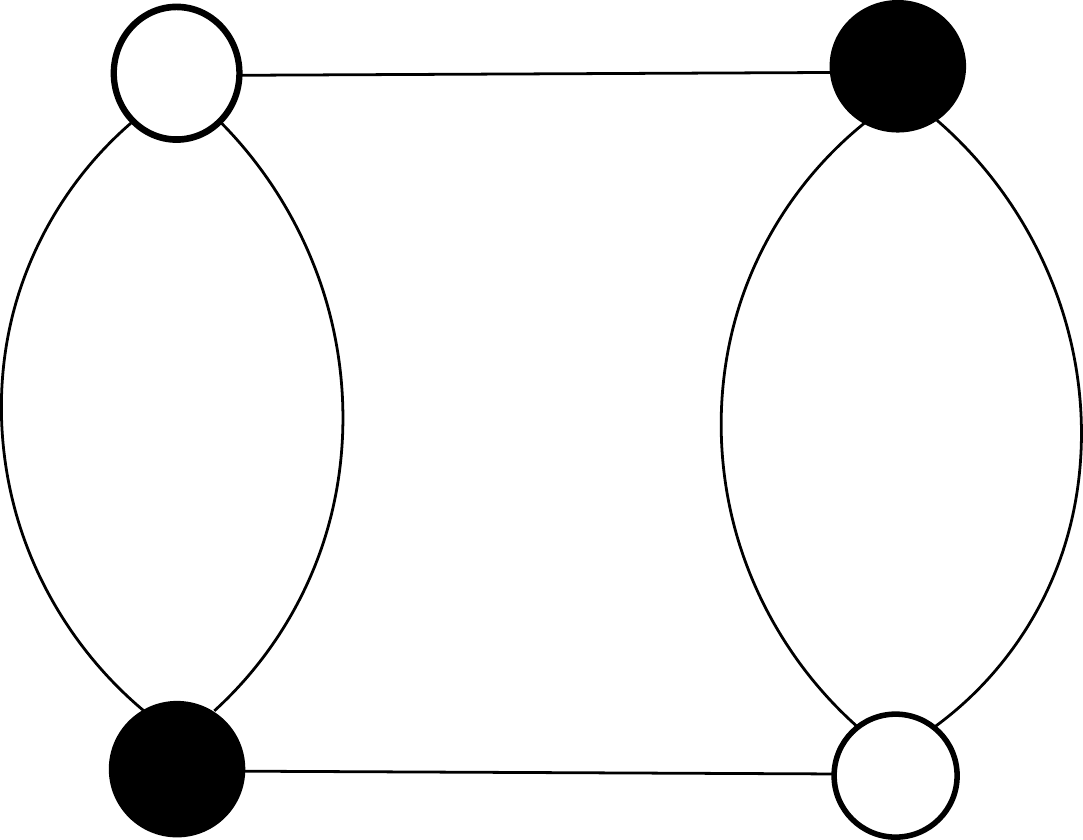}}
\big]\big \vert_{3 \, {\rm{cuts}}}
\nonumber \\
\hphantom{{\partial \over \partial t} u_{\includegraphics[width=0.9cm]{Krajewski-Fig-A1-05}}=}{}
+
{1 \over N}
\big[
6
u_{\includegraphics[width=1.0cm]{Krajewski-Fig-A1-20}}
+
6
u_{\includegraphics[width=1.0cm]{Krajewski-Fig-A1-13}}
+
3
u_{\includegraphics[width=0.9cm]{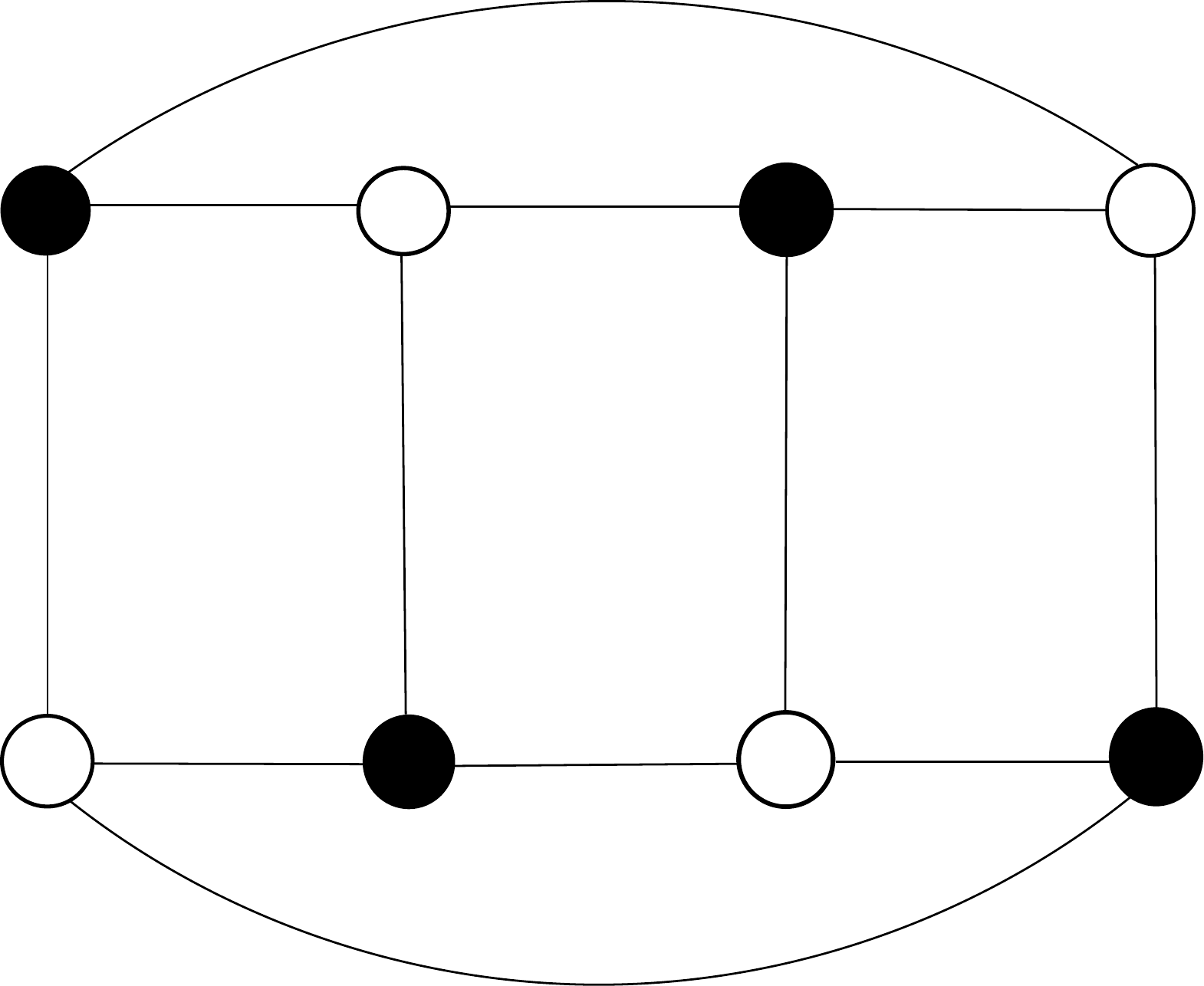}}
+
10
u_{\includegraphics[width=1.0cm]{Krajewski-Fig-A1-18}}
+
2
u_{\includegraphics[width=0.8cm]{Krajewski-Fig-A1-19}}
\big]\big \vert_{2 \, {\rm{cuts}}}
\nonumber \\
\hphantom{{\partial \over \partial t} u_{\includegraphics[width=0.9cm]{Krajewski-Fig-A1-05}}=}{}
+
{1 \over N^2}
\big[
6
u_{\includegraphics[width=1.0cm]{Krajewski-Fig-A1-18}}
+
u_{\includegraphics[width=0.9cm]{Krajewski-Fig-A1-23}}
+
8
u_{\includegraphics[width=1.0cm]{Krajewski-Fig-A1-20}}
\big]\big \vert_{3 \, {\rm{cuts}}}
\nonumber \\
\hphantom{{\partial \over \partial t} u_{\includegraphics[width=0.9cm]{Krajewski-Fig-A1-05}}=}{}
+
{1 \over N^3}
\big[
6
u_{\includegraphics[width=1.6cm]{Krajewski-Fig-A1-17}}
+
2
u_{\includegraphics[width=1.4cm]{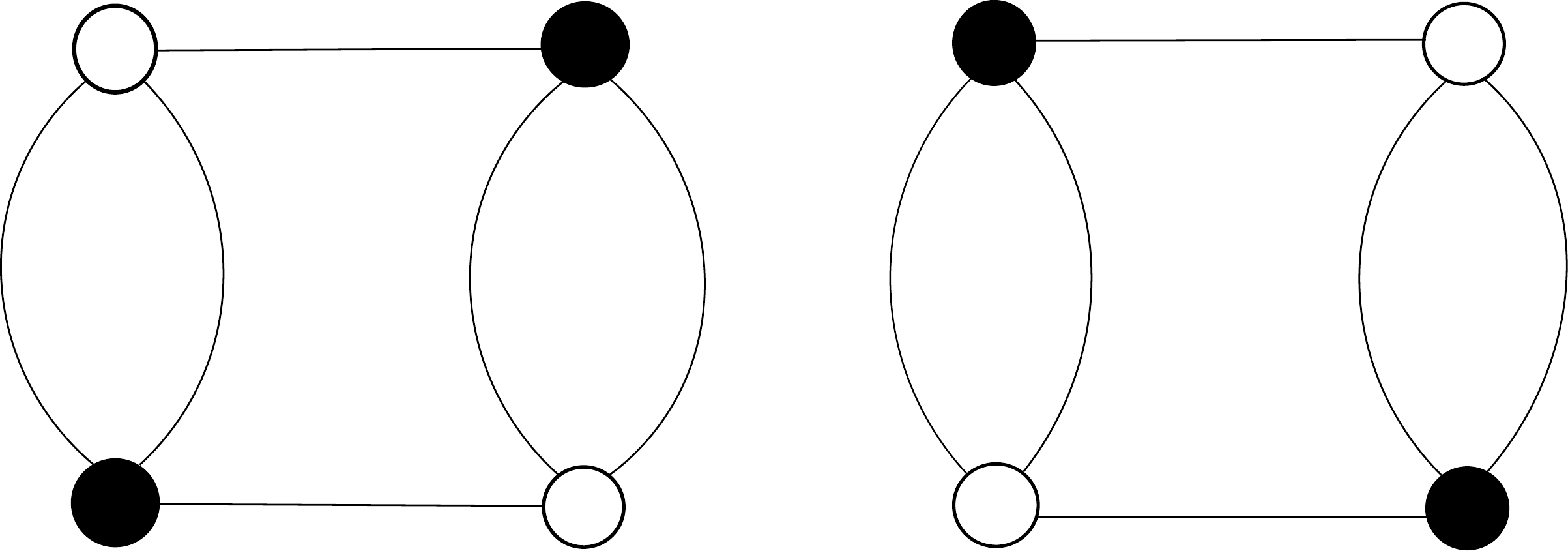}}
\big]\big \vert_{3 \, {\rm{cuts}}},\nonumber \\
{\partial \over \partial t}
u_{{\includegraphics[width=0.7cm]{Krajewski-Fig-A1-01}} \; {\includegraphics[width=0.7cm]{Krajewski-Fig-A1-01}}\; {\includegraphics[width=0.7cm]{Krajewski-Fig-A1-01}}}=
\big[
u_{{\includegraphics[width=0.7cm]{Krajewski-Fig-A1-01}} \; {\includegraphics[width=0.7cm]{Krajewski-Fig-A1-01}}\; {\includegraphics[width=0.7cm]{Krajewski-Fig-A1-01}}\; {\includegraphics[width=0.7cm]{Krajewski-Fig-A1-01}}}
\big]\big \vert_{0 \, {\rm{cut}}}\nonumber\\
\hphantom{{\partial \over \partial t}
u_{{\includegraphics[width=0.7cm]{Krajewski-Fig-A1-01}} \; {\includegraphics[width=0.7cm]{Krajewski-Fig-A1-01}}\; {\includegraphics[width=0.7cm]{Krajewski-Fig-A1-01}}}=}{}
+
\big[
9
u_{\includegraphics[width=2.1cm]{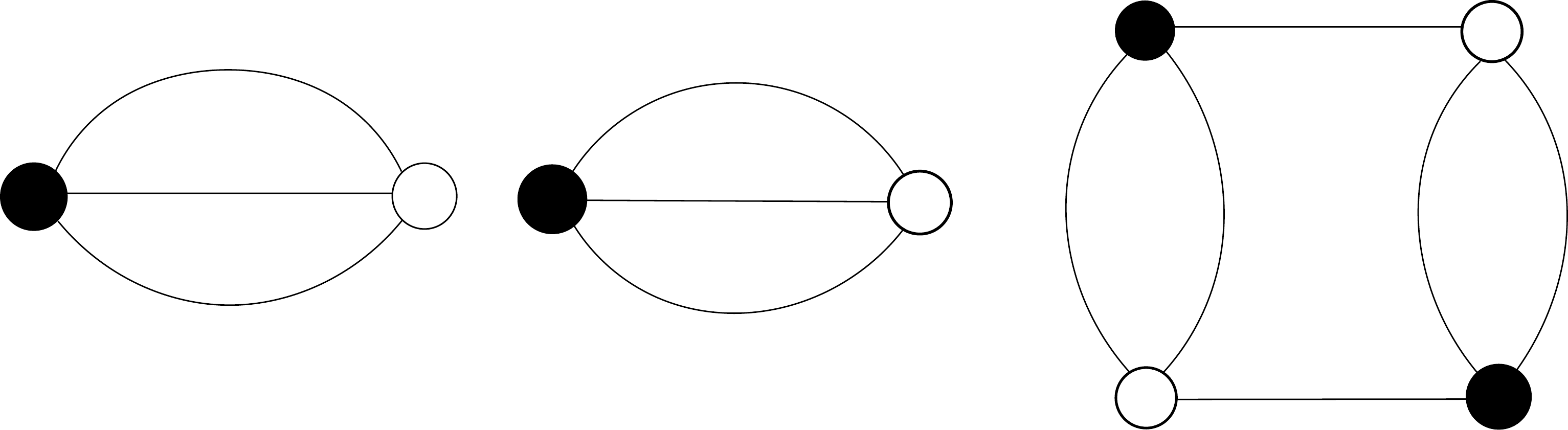}}
\big]\big \vert_{1 \, {\rm{cut}}}
+
\big[
18
u_{\includegraphics[width=1.6cm]{Krajewski-Fig-A1-17}}
\big]\big \vert_{2 \, {\rm{cuts}}}
\nonumber \\
\hphantom{{\partial \over \partial t}
u_{{\includegraphics[width=0.7cm]{Krajewski-Fig-A1-01}} \; {\includegraphics[width=0.7cm]{Krajewski-Fig-A1-01}}\; {\includegraphics[width=0.7cm]{Krajewski-Fig-A1-01}}}=}{}
+
\big[
6
u_{\includegraphics[width=0.7cm]{Krajewski-Fig-A1-15}}
-
6
u_{{\includegraphics[width=0.7cm]{Krajewski-Fig-A1-01}} \; {\includegraphics[width=0.7cm]{Krajewski-Fig-A1-01}}\; {\includegraphics[width=0.7cm]{Krajewski-Fig-A1-01}}}
u_{{\includegraphics[width=0.7cm]{Krajewski-Fig-A1-01}}}
-
6
u^2_{{\includegraphics[width=0.7cm]{Krajewski-Fig-A1-01}} \; {\includegraphics[width=0.7cm]{Krajewski-Fig-A1-01}}}
\big]\big \vert_{3 \, {\rm{cuts}}}
\nonumber \\
\hphantom{{\partial \over \partial t}
u_{{\includegraphics[width=0.7cm]{Krajewski-Fig-A1-01}} \; {\includegraphics[width=0.7cm]{Krajewski-Fig-A1-01}}\; {\includegraphics[width=0.7cm]{Krajewski-Fig-A1-01}}}=}{}
+
{1 \over N}
\Big \{
\big[
9
u_{\includegraphics[width=2.1cm]{Krajewski-Fig-A1-25}}
\big]\big \vert_{2 \, {\rm{cuts}}}
+
\big[
18
u_{\includegraphics[width=1.4cm]{Krajewski-Fig-A1-09}}
\big]\big \vert_{3 \; {\rm{cuts}}}
\Big \}
\nonumber \\
\hphantom{{\partial \over \partial t}
u_{{\includegraphics[width=0.7cm]{Krajewski-Fig-A1-01}} \; {\includegraphics[width=0.7cm]{Krajewski-Fig-A1-01}}\; {\includegraphics[width=0.7cm]{Krajewski-Fig-A1-01}}}=}{}
+
{1 \over N^3}
\big[
3
u_{{\includegraphics[width=0.7cm]{Krajewski-Fig-A1-01}} \; {\includegraphics[width=0.7cm]{Krajewski-Fig-A1-01}}\; {\includegraphics[width=0.7cm]{Krajewski-Fig-A1-01}}\; {\includegraphics[width=0.7cm]{Krajewski-Fig-A1-01}}}
\big]\big \vert_{3 \, {\rm{cuts}}},\nonumber
\\
{\partial \over \partial t} u_{\includegraphics[width=1.4cm]{Krajewski-Fig-A1-04}}=
\big[
u_{\includegraphics[width=2.1cm]{Krajewski-Fig-A1-25}}
\big]\big \vert_{0 \, {\rm{cut}}}
+
\big[
3
u_{\includegraphics[width=1.4cm]{Krajewski-Fig-A1-16}}
+
4
u_{\includegraphics[width=1.4cm]{Krajewski-Fig-A1-17}}
+
2
u_{\includegraphics[width=1.4cm]{Krajewski-Fig-A1-09}}
\big]\big \vert_{1 \, {\rm{cut}}}
\nonumber \\
\hphantom{{\partial \over \partial t} u_{\includegraphics[width=1.4cm]{Krajewski-Fig-A1-04}}=}{}
+
\big[
8
u_{\includegraphics[width=1.0cm]{Krajewski-Fig-A1-14}}
+
4
u_{\includegraphics[width=1.0cm]{Krajewski-Fig-A1-10}}
\big] \big \vert_{2 \, {\rm{cuts}}}\nonumber\\
\hphantom{{\partial \over \partial t} u_{\includegraphics[width=1.4cm]{Krajewski-Fig-A1-04}}=}{}
-
\big[
6
u_{\includegraphics[width=1.4cm]{Krajewski-Fig-A1-04}} \;
u_{\includegraphics[width=0.7cm]{Krajewski-Fig-A1-01}}
+
4
u_{{\includegraphics[width=0.7cm]{Krajewski-Fig-A1-01}} \; {\includegraphics[width=0.7cm]{Krajewski-Fig-A1-01}}} \;
u_{\includegraphics[width=0.7cm]{Krajewski-Fig-A1-03}}
\big]\big \vert_{3 \, {\rm{cuts}}}
\nonumber \\
\hphantom{{\partial \over \partial t} u_{\includegraphics[width=1.4cm]{Krajewski-Fig-A1-04}}=}{}
+
{1 \over N}
\Big \{
\big[
3
u_{\includegraphics[width=1.4cm]{Krajewski-Fig-A1-16}}
+
8
u_{\includegraphics[width=1.4cm]{Krajewski-Fig-A1-17}}
+
2
u_{\includegraphics[width=1.4cm]{Krajewski-Fig-A1-09}}
+
2
u_{\includegraphics[width=1.4cm]{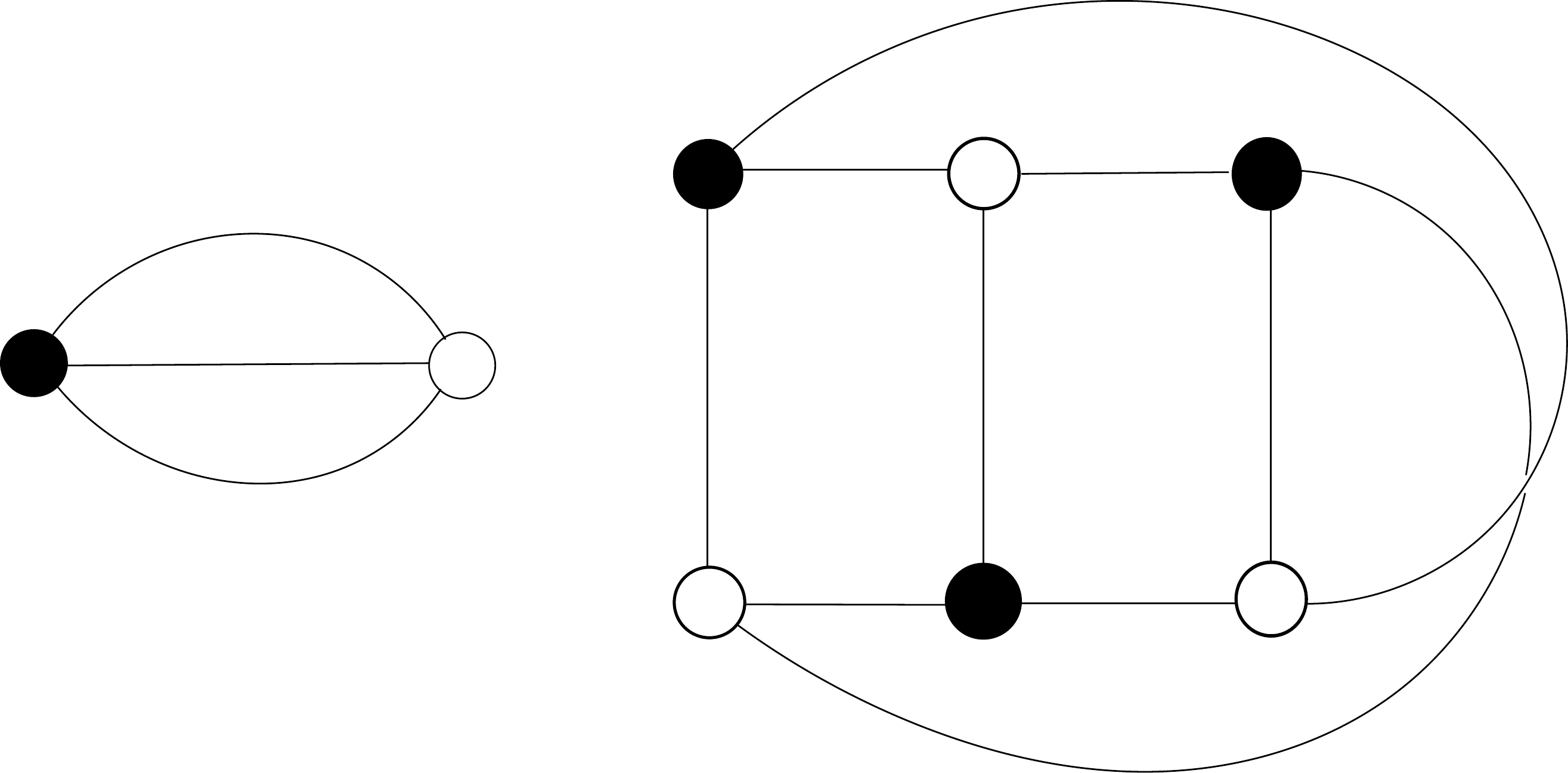}}
\big]\big \vert_{2 \, {\rm{cuts}}}
\nonumber \\
\hphantom{{\partial \over \partial t} u_{\includegraphics[width=1.4cm]{Krajewski-Fig-A1-04}}=}{}
+
\big[
2
 u_{\includegraphics[width=1.0cm]{Krajewski-Fig-A1-13}}
+
8
u_{\includegraphics[width=0.7cm]{Krajewski-Fig-A1-11}}
+
14
u_{\includegraphics[width=1.0cm]{Krajewski-Fig-A1-10}}
\big]\big \vert_{3 \, {\rm{cuts}}}
\Big \}
\nonumber \\
\hphantom{{\partial \over \partial t} u_{\includegraphics[width=1.4cm]{Krajewski-Fig-A1-04}}=}{}
+
{1 \over N^2}
\big[
4
u_{\includegraphics[width=1.4cm]{Krajewski-Fig-A1-17}}
\big]\big \vert_{3 \, {\rm{cuts}}}
+
{1 \over N^3}
\big[
5
u_{\includegraphics[width=2.1cm]{Krajewski-Fig-A1-25}}
\big]\big \vert_{3 \, {\rm{cuts}}}.\nonumber
\end{gather*}

\subsection[Couplings in rank $D=4$ tensor models]{Couplings in rank $\boldsymbol{D=4}$ tensor models}\label{example4:sec}

\vspace{-4mm}

\begin{gather*}
{\partial \over \partial t} u_{\includegraphics[width=0.8cm]{Krajewski-Fig04a}}=
\big[
u_{{\includegraphics[width=0.8cm]{Krajewski-Fig04a}} \; {\includegraphics[width=0.8cm]{Krajewski-Fig04a}}}
\big]\big \vert_{0 \, {\rm{cut}}}
+
\big[
4
u_{\includegraphics[width=0.8cm]{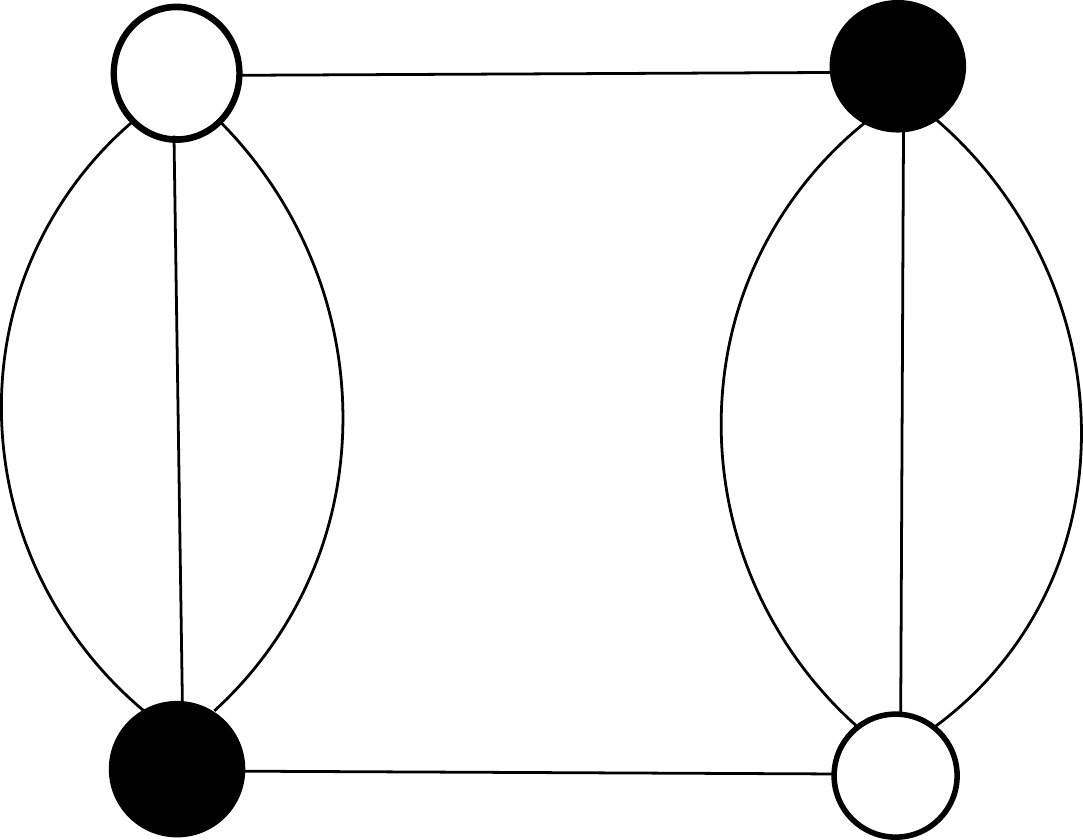}}
\big]\big \vert_{1 \, {\rm{cut}}}
-
\big[
u^2_{\includegraphics[width=0.8cm]{Krajewski-Fig04a}}
\big]\big \vert_{3 \, {\rm{cuts}}}
\nonumber \\
\hphantom{{\partial \over \partial t} u_{\includegraphics[width=0.8cm]{Krajewski-Fig04a}}=}{}
+
{1 \over N}
\big[
6
u_{\includegraphics[width=0.8cm]{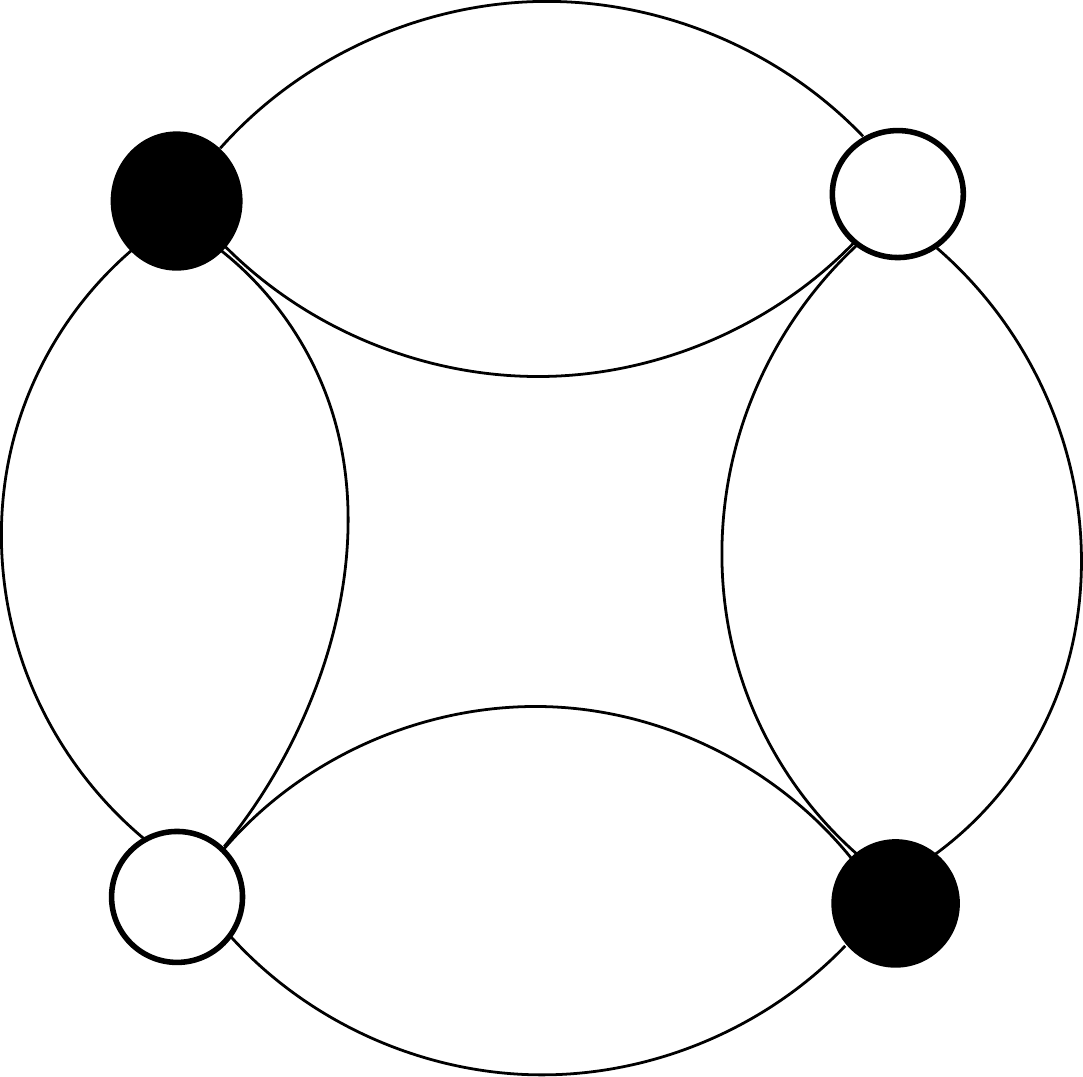}}
\big]\big \vert_{2 \, {\rm{cuts}}}
+
{1 \over N^2}
\big[
4
u_{\includegraphics[width=0.8cm]{Krajewski-Fig-A1-27}}
\big]\big \vert_{3 \, {\rm{cuts}}}
+
{1 \over N^4}
\big[
u_{{\includegraphics[width=0.8cm]{Krajewski-Fig04a}} \; {\includegraphics[width=0.8cm]{Krajewski-Fig04a}}}
\big]\big \vert_{4 \, {\rm{cuts}}},\nonumber
\\
{\partial \over \partial t} u_{\includegraphics[width=0.8cm]{Krajewski-Fig-A1-27}} =
\big[
u_{\includegraphics[width=1.6cm]{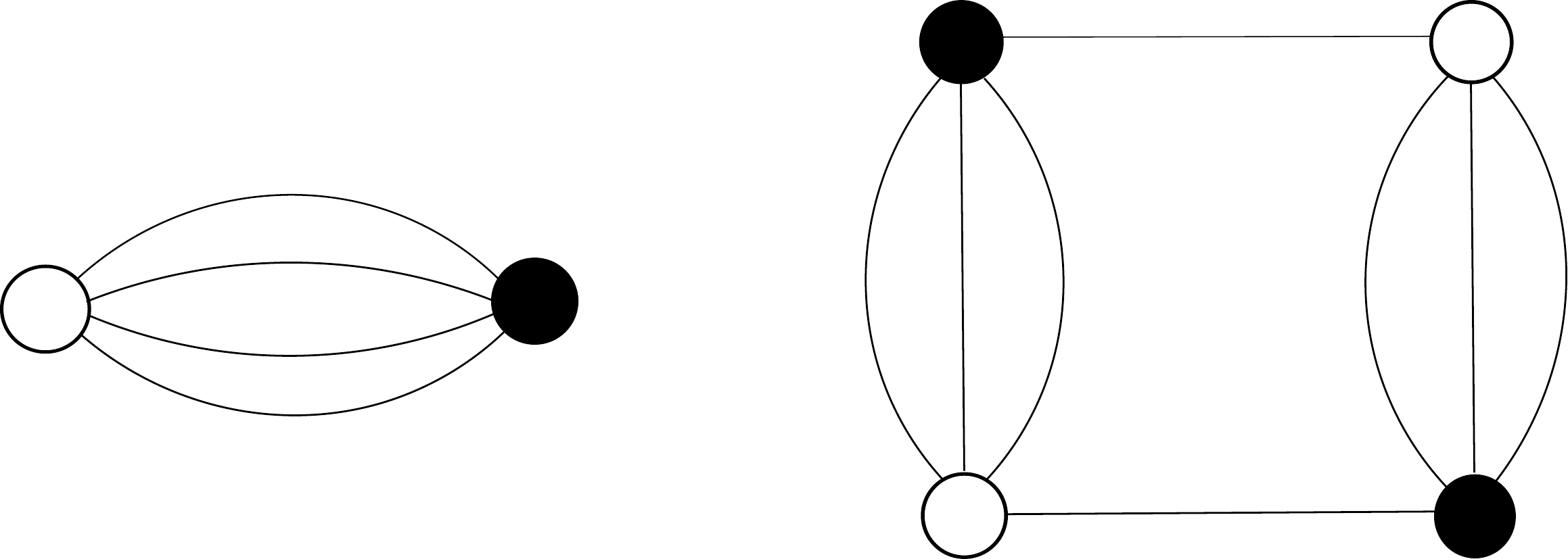}}
\big]\big \vert_{0 \; {\rm{cut}}}
+
\big[
2
u_{\includegraphics[width=0.8cm]{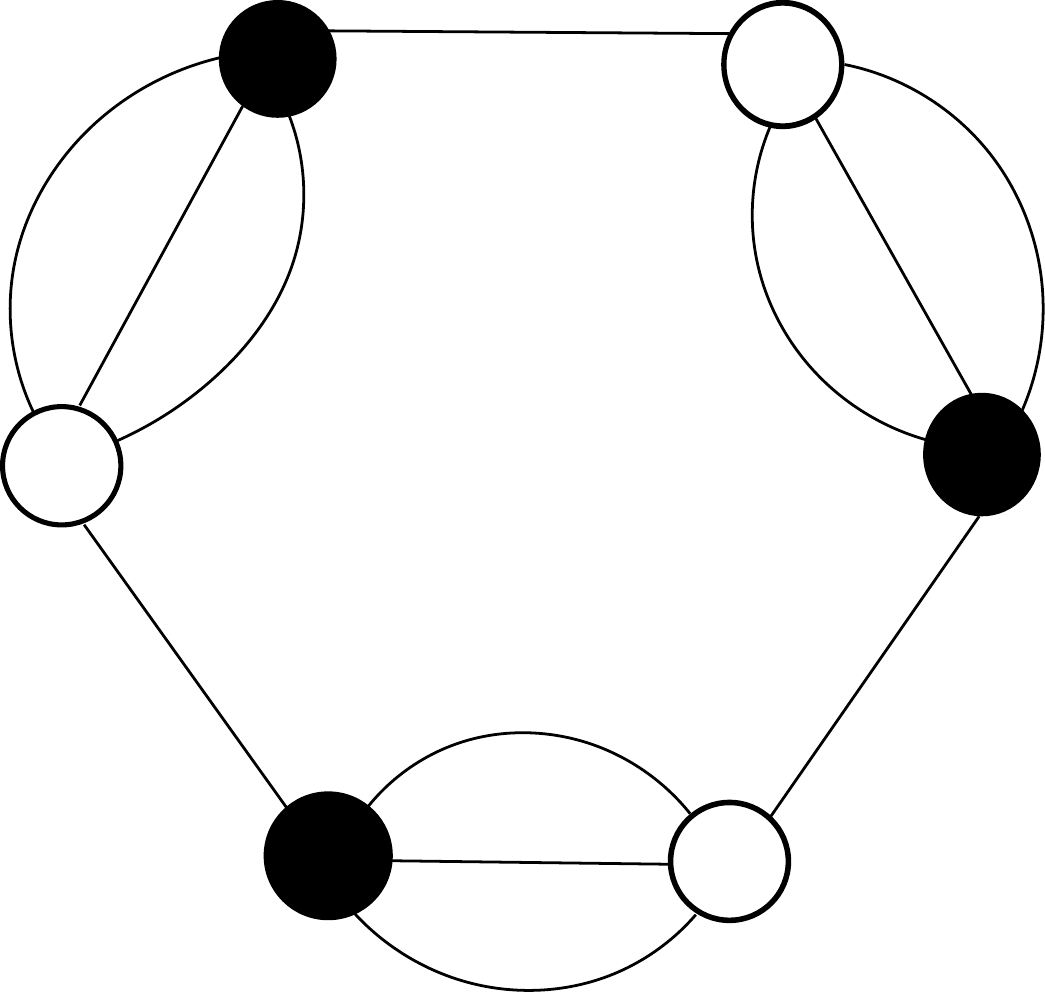}}
\big]\big \vert_{1 \, {\rm{cut}}}
-
\big[
4
u_{\includegraphics[width=0.8cm]{Krajewski-Fig04a}}
u_{\includegraphics[width=0.8cm]{Krajewski-Fig-A1-27}}
\big]\big \vert_{4 \, {\rm{cuts}}}
\nonumber \\
\hphantom{{\partial \over \partial t} u_{\includegraphics[width=0.8cm]{Krajewski-Fig-A1-27}} = }{}
+
{1 \over N}
\Big \{
\big[
6
u_{\includegraphics[width=0.8cm]{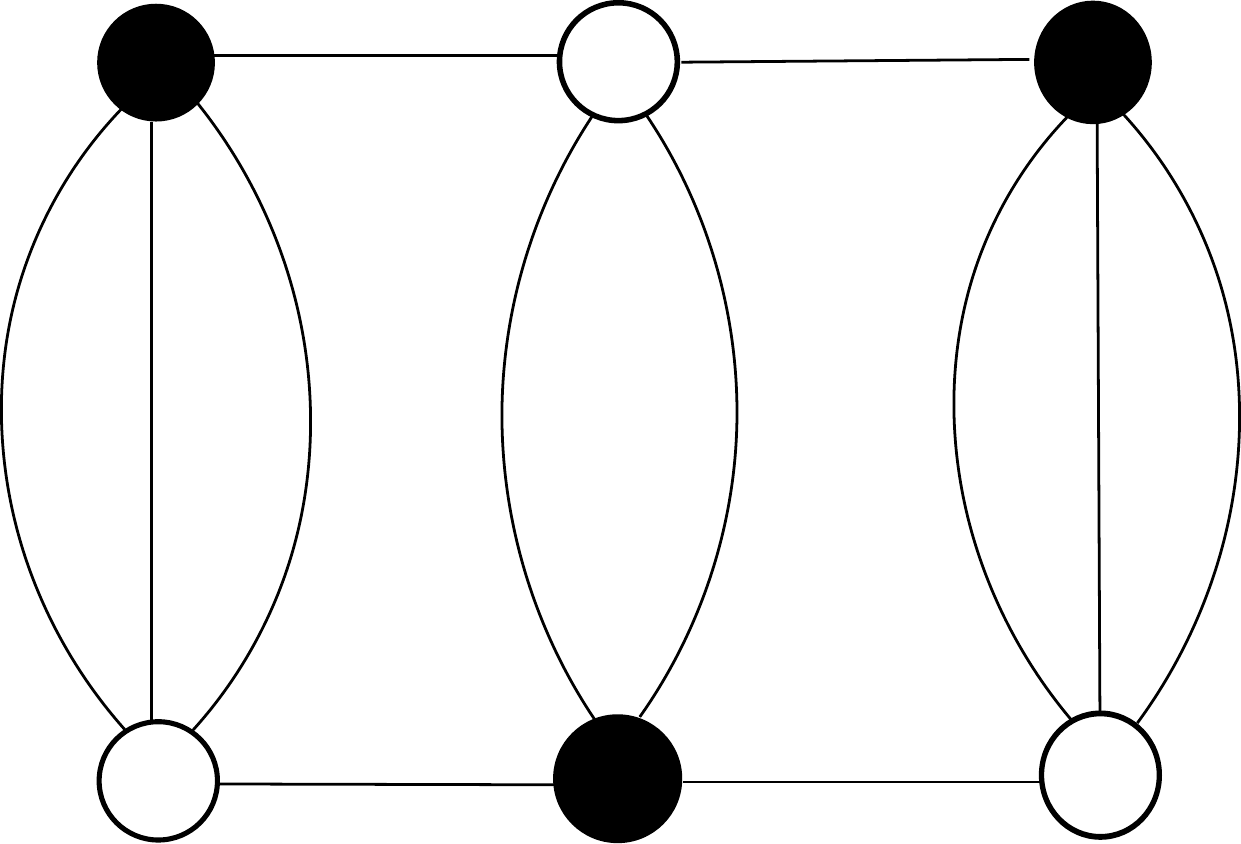}}
\big]\big \vert_{1 \, {\rm{cut}}}
+
\big[
18
u_{\includegraphics[width=0.8cm]{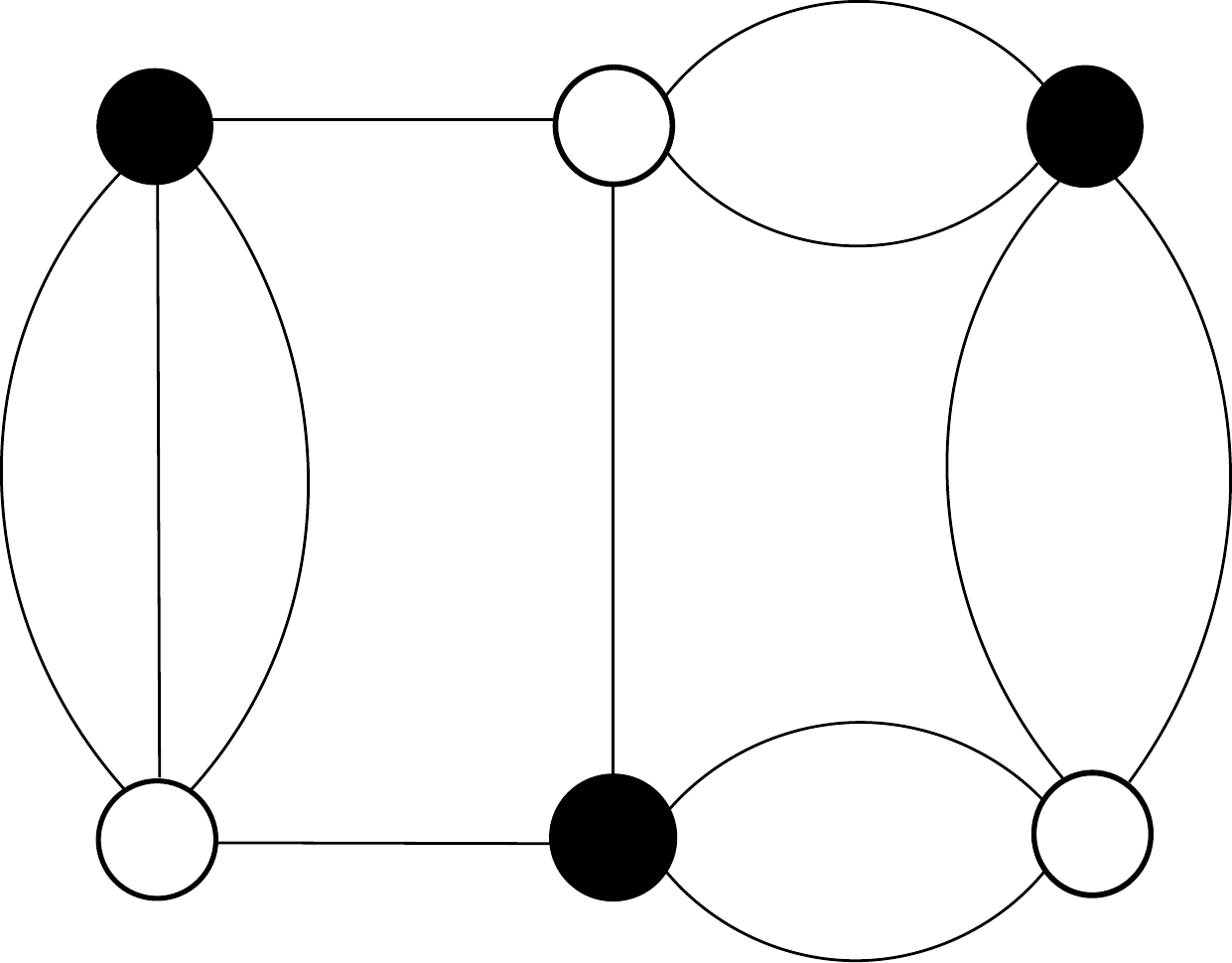}}
+
6
u_{\includegraphics[width=0.8cm]{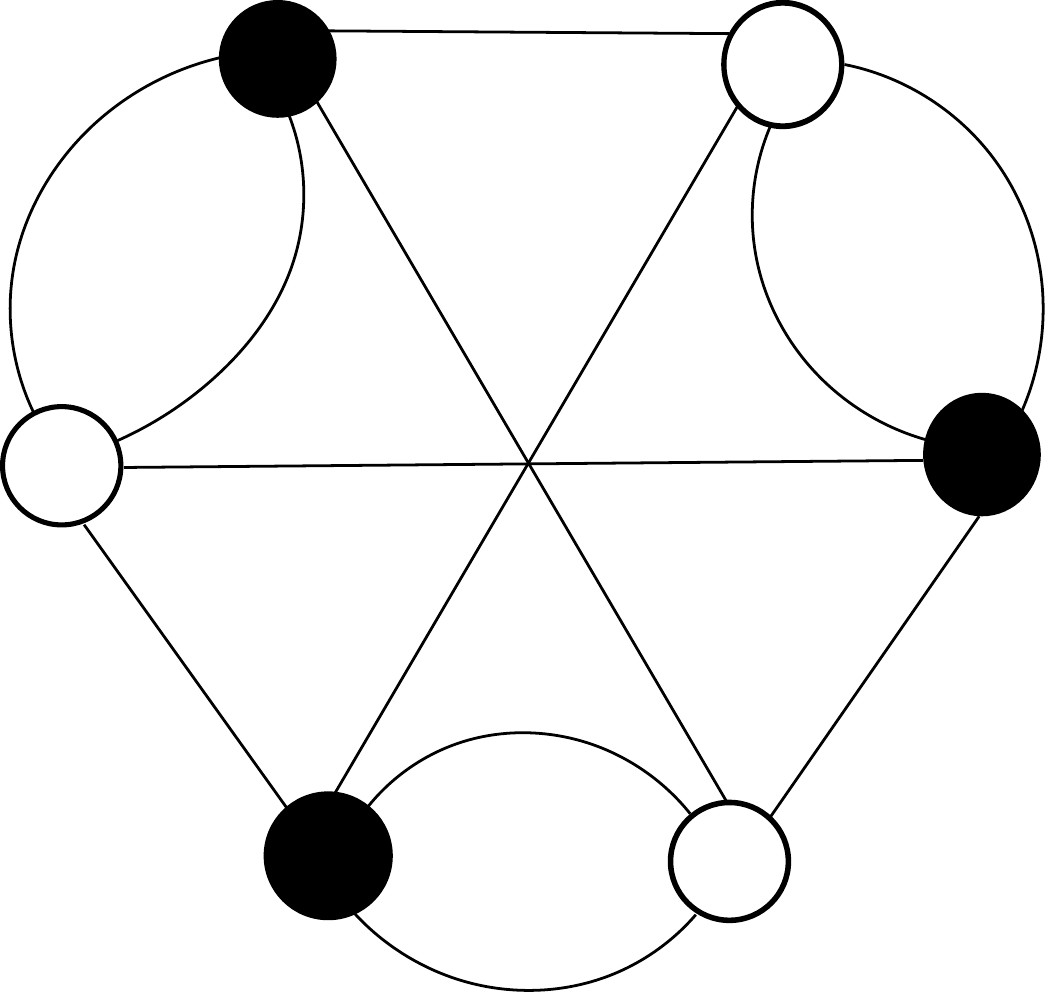}}
\big]\big \vert_{2 \, {\rm{cuts}}}
\Big \}
\nonumber \\
\hphantom{{\partial \over \partial t} u_{\includegraphics[width=0.8cm]{Krajewski-Fig-A1-27}} = }{}
+
{1 \over N^2}
\big[
2
u_{\includegraphics[width=0.8cm]{Krajewski-Fig-A1-30}}
+
12
u_{\includegraphics[width=0.8cm]{Krajewski-Fig-A1-31}}
+
12
u_{\includegraphics[width=0.8cm]{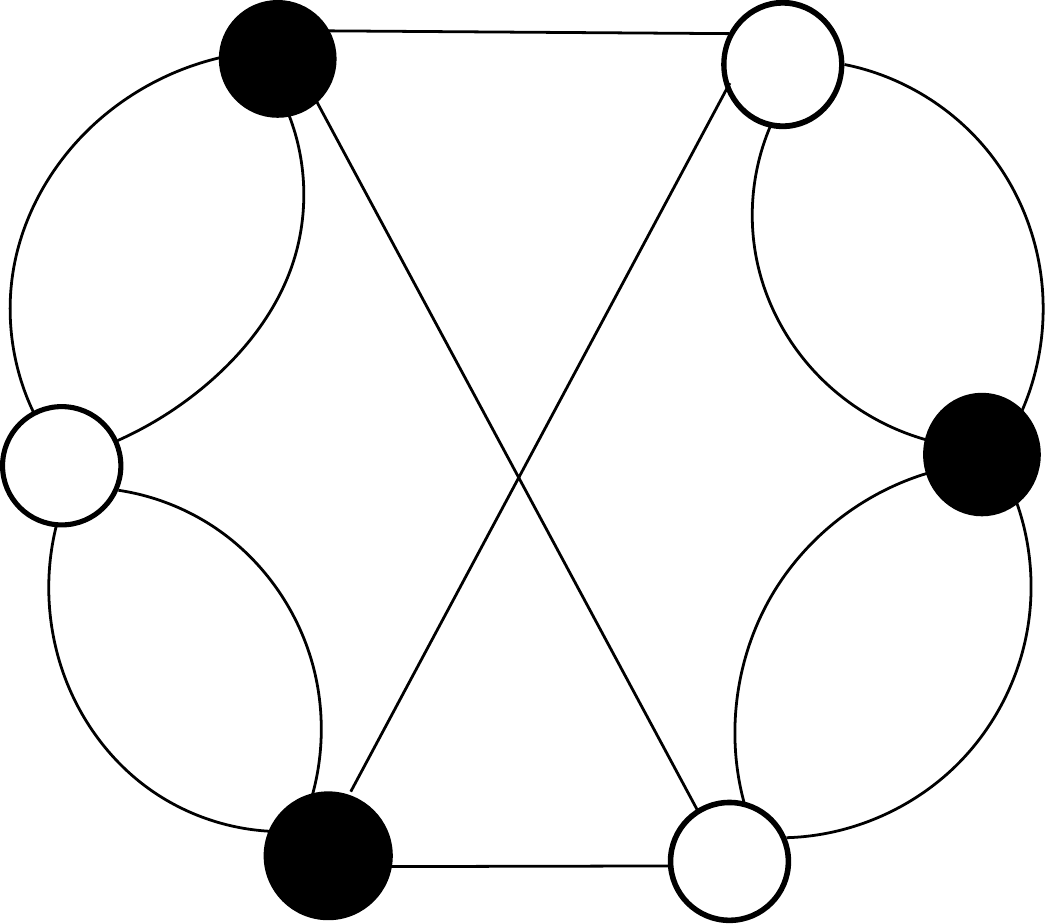}}
+
6
u_{\includegraphics[width=0.8cm]{Krajewski-Fig-A1-33}}
\big]\big \vert_{3 \, {\rm{cuts}}}
\nonumber \\
\hphantom{{\partial \over \partial t} u_{\includegraphics[width=0.8cm]{Krajewski-Fig-A1-27}} = }{}
+
{1 \over N^3}
\big[
12
u_{\includegraphics[width=0.8cm]{Krajewski-Fig-A1-32}}
\big]\big \vert_{4 \, {\rm{cuts}}}
+
{1 \over N^4}
\big[
4
\;u_{\includegraphics[width=1.6cm]{Krajewski-Fig-A1-29}}
\big]\big \vert_{4 \, {\rm{cuts}}},\nonumber
\\
{\partial \over \partial t} u_{\includegraphics[width=0.8cm]{Krajewski-Fig-A1-28}}=
\big[
u_{\includegraphics[width=1.6cm]{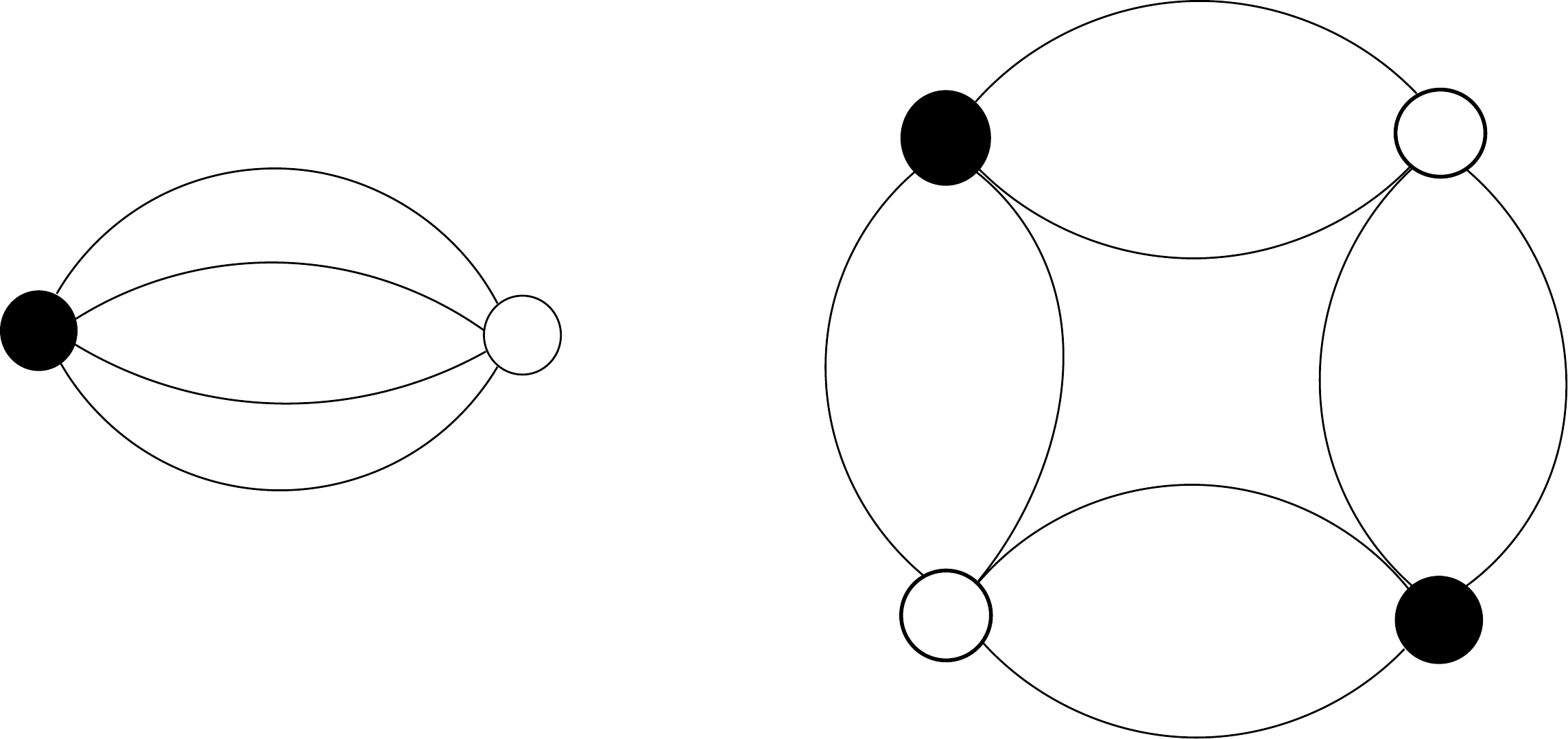}}
\big]\big \vert_{0 \, {\rm{cut}}}
+
\big[
8
u_{\includegraphics[width=0.8cm]{Krajewski-Fig-A1-32}}
\big]\big \vert_{1 \, {\rm{cut}}}
-
\big[
4
u_{\includegraphics[width=0.8cm]{Krajewski-Fig04a}}
u_{\includegraphics[width=0.8cm]{Krajewski-Fig-A1-28}}
\big]\big \vert_{4 \, {\rm{cuts}}}
\nonumber \\
\hphantom{{\partial \over \partial t} u_{\includegraphics[width=0.8cm]{Krajewski-Fig-A1-28}}=}{}
+
{1 \over N}
\big[
4
u_{\includegraphics[width=0.8cm]{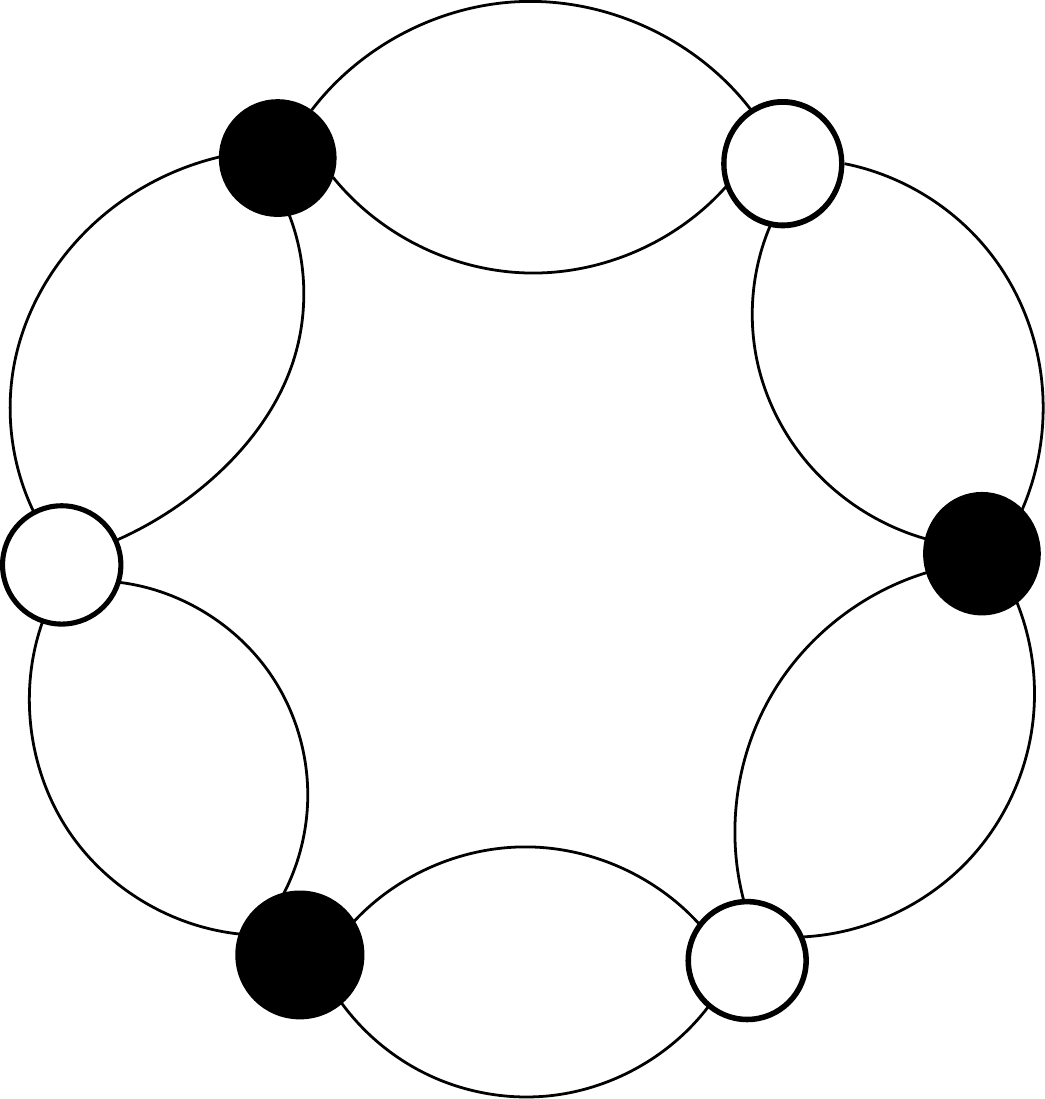}}
+
16
u_{\includegraphics[width=0.8cm]{Krajewski-Fig-A1-34}}
+
4
u_{\includegraphics[width=0.8cm]{Krajewski-Fig-A1-33}}
\big]\big \vert_{2 \, {\rm{cuts}}}
+
{1 \over N^2}
\big[
16
u_{\includegraphics[width=0.8cm]{Krajewski-Fig-A1-32}}
+
16
u_{\includegraphics[width=0.8cm]{Krajewski-Fig-A1-33}}
\big]\big \vert_{3 \, {\rm{cuts}}}
\nonumber \\
\hphantom{{\partial \over \partial t} u_{\includegraphics[width=0.8cm]{Krajewski-Fig-A1-28}}=}{}
+
{1 \over N^3}
\big[
8
u_{\includegraphics[width=0.8cm]{Krajewski-Fig-A1-31}}
+
4
u_{\includegraphics[width=0.8cm]{Krajewski-Fig-A1-34}}
\big]\big \vert_{4 \, {\rm{cuts}}}
+
{1 \over N^4}
\big[
4
u_{\includegraphics[width=1.6cm]{Krajewski-Fig-A1-35}}
\big]\big \vert_{4 \, {\rm{cuts}}},\nonumber
\\
{\partial \over \partial t} u_{{\includegraphics[width=0.8cm]{Krajewski-Fig04a}} \; {\includegraphics[width=0.8cm]{Krajewski-Fig04a}}}=
\big[
u_{{\includegraphics[width=0.8cm]{Krajewski-Fig04a}} \; {\includegraphics[width=0.8cm]{Krajewski-Fig04a}}\; {\includegraphics[width=0.8cm]{Krajewski-Fig04a}}}
\big]\big \vert_{0 \, {\rm{cut}}}
+
\big[
8
u_{\includegraphics[width=1.6cm]{Krajewski-Fig-A1-29}}
\big]\big \vert_{1 \, {\rm{cut}}}
+
\big[
12
u_{\includegraphics[width=0.8cm]{Krajewski-Fig-A1-31}}
\big]\big \vert_{2 \, {\rm{cuts}}}
\nonumber \\
\hphantom{{\partial \over \partial t} u_{{\includegraphics[width=0.8cm]{Krajewski-Fig04a}} \; {\includegraphics[width=0.8cm]{Krajewski-Fig04a}}}=}{}
-
\big[
4
u_{{\includegraphics[width=0.8cm]{Krajewski-Fig04a}} \; {\includegraphics[width=0.8cm]{Krajewski-Fig04a}}}
u_{\includegraphics[width=0.8cm]{Krajewski-Fig04a}}
\big]\big \vert_{4 \, {\rm{cuts}}}
\nonumber \\
\hphantom{{\partial \over \partial t} u_{{\includegraphics[width=0.8cm]{Krajewski-Fig04a}} \; {\includegraphics[width=0.8cm]{Krajewski-Fig04a}}}=}{}
+
{1 \over N}
\Big \{
\big[
12
u_{\includegraphics[width=1.6cm]{Krajewski-Fig-A1-35}}
\big]\big \vert_{2 \, {\rm{cuts}}}
+
\big[
24
u_{\includegraphics[width=0.8cm]{Krajewski-Fig-A1-32}}
\big]\big \vert_{3 \, {\rm{cuts}}}
\Big \}
\nonumber \\
\hphantom{{\partial \over \partial t} u_{{\includegraphics[width=0.8cm]{Krajewski-Fig04a}} \; {\includegraphics[width=0.8cm]{Krajewski-Fig04a}}}=}{}
+
{1 \over N^2}
\Big \{
\big[
8
u_{\includegraphics[width=1.6cm]{Krajewski-Fig-A1-29}}
\big]\big \vert_{3 \, {\rm{cuts}}}
+
\big[
8
u_{\includegraphics[width=0.8cm]{Krajewski-Fig-A1-30}}
+
6
u_{\includegraphics[width=0.8cm]{Krajewski-Fig-A1-36}}
\big]\big \vert_{4 \, {\rm{cuts}}}
\Big \}
\nonumber \\
\hphantom{{\partial \over \partial t} u_{{\includegraphics[width=0.8cm]{Krajewski-Fig04a}} \; {\includegraphics[width=0.8cm]{Krajewski-Fig04a}}}=}{}
+
{1 \over N^4}
\big[
2
u_{{\includegraphics[width=0.8cm]{Krajewski-Fig04a}} \; {\includegraphics[width=0.8cm]{Krajewski-Fig04a}}\; {\includegraphics[width=0.8cm]{Krajewski-Fig04a}}}
\big]\big \vert_{4 \, {\rm{cuts}}}.\nonumber
\end{gather*}

}

\subsection*{Acknowledgements}

T.K.~is partially supported by the grant ``ANR JCJC CombPhysMat2Tens''.

\LastPageEnding

\end{document}